\documentclass[twocolumn,twocolappendix]{aastex701}
\usepackage[dvipsnames]{xcolor}
\usepackage{amsmath}
\usepackage{bbm,bm}
\usepackage{multirow}
\usepackage{colortbl}
\usepackage{preamble}

\graphicspath{{./}{figures-new/}{tables/}}

\newcommand\REV[1]{{#1}}
\shorttitle{\tigresspp: Cosmic Ray Feedback in Galactic Disks}
\shortauthors{C.-G. Kim et al.}

\begin{document}

\title{Cosmic Ray Feedback in Galactic Disks: Star Formation,  Cosmic Ray Transport, and Multiphase Outflows in \tigresspp\ Simulations}

\author[0000-0003-2896-3725]{Chang-Goo Kim}
\affiliation{Department of Astrophysical Sciences, Princeton University, 4 Ivy Lane, Princeton, NJ 08544, USA}
\email[show]{changgoo@princeton.edu}

\author[0000-0002-5708-1927]{Lucia Armillotta}
\email{lucia.armillotta@unifi.it}
\affiliation{University of Florence, Department of Physics and Astronomy, via G. Sansone 1, 50019, Sesto Fiorentino, Firenze, Italy}
\affiliation{Department of Astrophysical Sciences, Princeton University, 4 Ivy Lane, Princeton, NJ 08544, USA}

\author[0000-0002-0509-9113]{Eve C. Ostriker}
\affiliation{Department of Astrophysical Sciences, Princeton University, 4 Ivy Lane, Princeton, NJ 08544, USA}
\email{eco@astro.princeton.edu}

\author[0000-0002-6302-0485]{Sanghyuk Moon}
\affiliation{Korea Astronomy and Space Science Institute, 776 Daedeok-daero, Yuseong-gu, Daejeon 34055, Republic of Korea}
\affiliation{Department of Astrophysical Sciences, Princeton University, 4 Ivy Lane, Princeton, NJ 08544, USA}
\email{smoon@kasi.re.kr}

\author[0000-0002-0041-4356]{Lachlan Lancaster}
\affiliation{Center for Computational Astrophysics, Flatiron Institute, 162 5th Avenue, New York, NY 10010, USA}
\affiliation{Department of Astronomy, Columbia University,  550 W 120th St, New York, NY 10025, USA}
\email{ltl2125@columbia.edu}

\author[0000-0001-6228-8634]{Jeong-Gyu Kim}
\affiliation{Quantum Universe Center, Korea Institute for Advanced Study, Hoegiro 85, Seoul 02455, Republic of Korea}
\affiliation{Division of Science, National Astronomical Observatory of Japan, Mitaka, Tokyo 181-0015, Japan}
\email{jeonggyukim@kias.re.kr}

\author[0000-0001-8840-2538]{Nora B. Linzer}
\affiliation{Department of Astrophysical Sciences, Princeton University, 4 Ivy Lane, Princeton, NJ 08544, USA}
\email{nlinzer@princeton.edu}

\author[0000-0002-7047-3730]{Ronan N. Hix}
\affiliation{Department of Astrophysical Sciences, Princeton University, 4 Ivy Lane, Princeton, NJ 08544, USA}
\email{ronanhix@princeton.edu}

\begin{abstract}
We present new simulations of local star-forming disks that self-consistently evolve cosmic rays (CRs) and multiphase gas using \tigresspp.
To isolate the role of CRs, we conduct paired simulations under solar-neighborhood conditions:
a magnetohydrodynamics (MHD) model following the standard \classic\ framework with FUV heating and supernova (SN) feedback from star clusters formed via gravitational collapse; and a CRMHD model in which an additional 10\% of each SN's energy is injected as CRs.
These CRs are transported anisotropically along magnetic field lines via a two-moment solver, with the CR scattering rate set by balancing Alfv\'en-wave growth and damping based on the self-confinement paradigm.
The CRMHD model develops a characteristic two-zone vertical CR profile:
uniform pressure in the diffusion-dominated, high-density midplane gas, and an exponential atmosphere shaped primarily by advection and streaming in low-density extraplanar gas.
The CR pressure is comparable to the total thermal gas pressure in the midplane,
but is too uniform to affect gas dynamics, leaving SFRs unchanged.
In contrast, the vertical CR pressure gradient at $|z|\gtrsim 1\kpc$  accelerates warm outflowing gas, resulting in an approximately 4 times higher mass loading factor than in the MHD model.
CR--gas interactions increase CR energy near the midplane through compressional work, while CR streaming heats low-density warm-hot gas.
\tigresspp\ opens
a path toward investigating CR transport and CR-regulated ISM and outflow dynamics at high resolution across diverse galactic environments.
\end{abstract}


\section{Introduction}\label{sec:intro}

Understanding the mechanisms that regulate star formation and drive galactic outflows remains one of the central challenges in galaxy formation \citep{2015ARA&A..53...51S,2017ARA&A..55...59N}.
The interstellar medium (ISM) is a complex, multiphase gaseous system governed by gravity, turbulence, and magnetic fields that requires sophisticated numerical methods to model. A realistic treatment of stellar feedback is an essential component of contemporary studies \citep[e.g.][]{2017ApJ...846..133K,2023ApJ...946....3K,2023MNRAS.522.1843R,2021ApJ...920...44H,2025ApJ...984..142G}. In addition to thermal gas, the ISM is permeated by a relativistic component, cosmic rays (CRs).
Observations suggest that the energy density of CRs in the Milky Way is roughly in equipartition with the thermal gas and magnetic fields, at approximately $1 \eV\pcc$ \citep[e.g.,][]{1990ApJ...365..544B,Grenier+15,2001SSRv...99..243B}.
CR pressure may contribute to maintaining vertical hydrostatic balance in the ISM disk and to driving galactic winds, depending on how CRs are transported away from the regions where they are accelerated \citep[e.g.,][]{2017PhPl...24e5402Z,2023A&ARv..31....4R}.

Based on recent magnetohydrodynamic (MHD) simulations with resolved supernova (SN) feedback, multiphase galactic outflows driven by clustered SNe are characterized by energy-loaded hot ($T\sim10^6-10^7\Kel$) winds and mass-loaded warm-cold ($T<10^4\Kel$) fountains  \citep[e.g.,][]{2018ApJ...853..173K,2020ApJ...894...12V}.
This general conclusion holds in self-consistent local simulations with varying environments \citep[e.g.,][]{2020ApJ...900...61K,2023MNRAS.522.1843R} and global simulations of dwarf galaxies \citep[e.g.,][]{2019MNRAS.483.3363H,2022arXiv220509774S,2024ApJ...960..100S}.

Multiphase outflows can therefore regulate the growth of galaxies in two different ways: energy-loaded hot outflows heat the circumgalactic medium and may prevent it from cooling and flowing inward (preventative feedback), while mass-loaded cool outflows remove material from the disk that could otherwise form stars (ejective feedback).
However, the ability of ejective feedback to limit the stellar-to-halo mass in galaxy formation models often demands unrealistically high mass loading factors, inconsistent with recent observations \citep[e.g.,][]{2019ApJ...886...74M,2024ApJ...966..129K} and with simulations resolving the emergence of outflows \citep[e.g.,][]{2018ApJ...853..173K,2020ApJ...900...61K,2021MNRAS.508.2979P}.
At the same time, resolved simulations find it difficult to launch hot wind with sufficiently high specific energy \citep[energy loading/mass loading $> 0.1$, ; e.g.,][]{2018ApJ...853..173K,2020ApJ...900...61K,2024ApJ...960..100S,2019MNRAS.483.3363H,2021MNRAS.506.3882S} for preventative feedback to be solely responsible for star formation regulation \citep{2023ApJ...949...21C,2024MNRAS.527.1216S,2025MNRAS.543.1456B}.
Energy loading factors this high are typically only achieved in idealized starbursts with strongly clustered SNe \citep[e.g.,][]{2018MNRAS.481.3325F,2020ApJ...895...43S}.
These shortcomings highlight the need to quantify effects of additional feedback mechanisms beyond ``thermal'' SN feedback alone, pointing to a careful study of the role of CRs.

Incorporating CR feedback into ISM and galaxy simulations has traditionally relied on simplified transport models (e.g., CR streaming is often neglected; but see \citealt{2017ApJ...834..208R,2019MNRAS.488.3716C}). The most common approach assumes CRs propagate via isotropic \citep[e.g.,][]{2012MNRAS.423.2374U,2016ApJ...827L..29S} or anisotropic diffusion \citep[e.g.,][]{2019MNRAS.488.3716C,2020MNRAS.492.3465H,2023MNRAS.520.4621S} with a constant diffusion coefficient, typically $\kappa_{\parallel} \sim 10^{28-29} \diffunit$.
While these models demonstrate that CRs can thicken galactic disks and launch winds, the results depend sensitively on the chosen value of $\kappa_{\parallel}$ \citep[e.g.,][]{2014MNRAS.437.3312S,2019MNRAS.488.3716C,2020A&A...638A.123D,2021MNRAS.501.4184H}. This sensitivity is useful in empirically constraining the effective diffusion coefficient \citep{2021MNRAS.501.4184H}, but the predictive power of such simulations for the role of CR feedback is limited. Furthermore, the adoption of constant $\kappa_{\parallel}$ neglects the role of local plasma conditions in setting the rate at which CR particles scatter, which may result in an inaccurate local  coupling between CRs and thermal gas.
In particular, because ion-neutral damping of the waves that scatter CRs is much stronger in dense, neutral midplane gas than elsewhere in the ISM (see below),  a constant diffusion coefficient risks overestimating CR-gas coupling in this region, which might have serious consequences for dynamics.

In this work, we use a scattering rate determined by the self-confinement paradigm \citep[][see also \citealt{2022MNRAS.517.5413H,2025A&A...698A.104T}]{2021ApJ...922...11A}.
In this picture, CRs are scattered by Alfv\'en waves that they themselves generate via the streaming instability \citep{1969ApJ...156..445K}. The scattering rate is set by a local steady-state balance between the wave growth rate (driven by the streaming instability) and wave damping rates. We explicitly account for non-linear Landau damping (dominant in low density, ionized gas) and ion-neutral damping (dominant in high density, neutral gas). By coupling these microphysical processes to the macrophysical cosmic ray + magnetohydrodynamic (CRMHD) fluid model, our simulation naturally captures the (huge) variations in $\kappa_\parallel$ and other quantities that affect  CR transport in multiphase galactic gas.  The multiphase, star-forming ISM is itself modeled by the \tigress\ framework \citep{2017ApJ...846..133K}.
To evolve the CR fluid, we employ a two-moment method to solve the CR transport equations \citep{2018ApJ...854....5J}, evolving both CR energy density and flux, coupled with the MHD equations \citep[see also][]{2019MNRAS.485.2977T,2019MNRAS.488.3716C,2022MNRAS.509.3779H,2026ApJS..283...37Z}.

We have previously applied the model of two-moment CR transport with self-consistent scattering to \tigress\ simulations of the multiphase ISM, investigating CR transport in a ``post-processing'' mode \citep{2021ApJ...922...11A,2022ApJ...929..170A,2025ApJ...989..140A,2025ApJ...994...45H}.
These studies consider a wide range of galactic environments, and demonstrate that CR transport coefficients vary by many orders of magnitude depending on the local ISM phase.
CR pressure in the neutral gas near the midplane  is nearly uniform due to ion-neutral damping efficiently suppressing wave growth and CR scattering.
In the extraplanar region, where wave damping is much weaker, CR transport along field lines is no longer fully diffusive. Rather, dynamical processes --- advection at the bulk gas velocity and streaming at the Alfv\'en wave speed --- combine with diffusive processes set by the scattering rate to control the effective transport velocity of the CR fluid along the magnetic field. While the scattering rate is large enough that the CR scale height and effective transport speed out of the disk are mainly set by dynamical processes for GeV CRs, diffusion is increasingly important for higher-energy CRs \citep{2025ApJ...989..140A}.

In \citet{2024ApJ...964...99A}, we transitioned from post-processing to time-dependent CRMHD simulations in studying the dynamics of the multiphase gas and outflow acceleration in extraplanar regions.  These simulations employed the upper extraplanar cutouts ($z>500\pc$) of \tigress\ outputs for initial conditions, but did not include ongoing star formation and feedback; rather, in these controlled models, fixed fluxes of gas and CRs into the extraplanar region were prescribed.

In this paper, we present our first results from fully integrating CRs within the \tigress\ numerical framework, for the whole ISM including both the midplane region with star formation and feedback,  and the extraplanar region where outflows are accelerated.  Although we have the capability of modeling multiple CR energy groups simultaneously \citep[see][]{2025ApJ...989..140A,2025ApJ...988..214L}, this adds to the numerical expense, and results presented in this work focus on $\sim$GeV CRs because this is the most dynamically important component.
We compare our new self-consistent CRMHD simulation against an otherwise identical MHD simulation without CRs. Our results reveal a nuanced role for CRs. As uniform CR pressure in the neutral gas does not contribute to vertical support, the self-regulation of the global SFR remains largely governed by the balance of total pressure of the thermal gas and weight, similar to the MHD case \citep{2022ApJ...936..137O}.
In the CRMHD simulation, the CR pressure gradient in the extraplanar region drives outflows that are cooler, slower, and more mass-loaded than those in the MHD simulation, in qualitative agreement with previous work either with a simplified CR transport \citep[e.g.,][]{2016ApJ...816L..19G,2018MNRAS.479.3042G,2016ApJ...827L..29S,2021MNRAS.504.1039R} or with a more self-consistent transport model similar to this work \citep[e.g.,][]{2025ApJ...987..204S,2025A&A...698A.104T}.

The rest of the paper is structured as follows. \autoref{sec:method} describes our numerical methods, which we use to run two models with and without CRs (\autoref{sec:model}).
We compare results from these simulations, focusing on the effects of CR feedback on star formation rates (\autoref{sec:sfr}), multiphase ISM and CR transport within it (\autoref{sec:ism_cr_trans}), and multiphase outflows (\autoref{sec:wind}). We discuss our results in comparison to our previous post-processing models (\autoref{sec:diss_pp}) and other simulations (\autoref{sec:diss_sim}). \autoref{sec:conclusion} closes with a summary and future perspectives.

\section{Methods}\label{sec:method}

\tigresspp\ is an extension of the \tigress\ numerical framework\footnote{The original \tigress\ framework referred to here is broadly defined including a version with a simpler cooling and heating, called \classic\ \citep{2017ApJ...846..133K}, and a version with explicit radiation transfer and photochemistry solutions, called \ncr\  \citep{2023ApJ...946....3K,2023ApJS..264...10K}. Both are implemented within the C version of the \athena\ code \citep{2008ApJS..178..137S,2009NewA...14..139S}.} implemented within the C++ version of the \athena\ code, called \athenapp\ \citep{2020ApJS..249....4S}. In addition to overall performance improvement using task-based parallelization, and to new capabilities such as adaptive-mesh refinement, the migration of \tigress\ to \athenapp\ enables us to couple evolution of CRs with the MHD solver, using a two-moment method for CR fluids \citep{2018ApJ...854....5J}. The version of the CR solver used in \tigresspp\ adopts a
CR scattering coefficient parallel to the magnetic field direction based on the self-confinement model \citep[e.g.,][]{2013PhPl...20e5501Z,2017PhPl...24e5402Z}, as developed by \citet{2021ApJ...922...11A}. This CR transport model has been applied to the \classic\ outputs in a post-processing mode with frozen gas properties and magnetic fields \citep{2021ApJ...922...11A,2022ApJ...929..170A} or for a brief dynamically coupled evolution \citep{2024ApJ...964...99A,2025ApJ...994...45H}.

In this paper, we present the first results from
our \tigresspp\ framework, which allows for fully coupled dynamical evolution of thermal gas with CR transport.  For the present simulations,
we adopt the \classic\ approach for multiphase ISM thermodynamics implemented in \athenapp.   In this section, we summarize the governing equations and numerical algorithms, including describing key modifications to \classic\ relative to \citet{2017ApJ...846..133K}, and our new algorithm for injection of CR energy from SN feedback. In a companion paper, we will present a comprehensive description of the \tigresspp\ methods, including  the updated version of \ncr\ and the multi-energy group CR transport model \citep[][]{2025ApJ...988..214L, 2025ApJ...989..140A}.

\subsection{Governing Equations}
The governing equations for the CRMHD evolution model consist of three main parts: the gas conservation equations with gravitational, radiative, and CR source terms; the magnetic induction equation; and the two-moment CR transport equations including source terms.

The conservation equations for gas mass, momentum, and total energy
including source terms
are, respectively,
\begin{equation}\label{eq:mass_con}
    \pderiv{\rho}{t} + \divergence{\rho\vel} = 0,
\end{equation}
\begin{eqnarray}
    \label{eq:mom_con}
    \pderiv{(\rho\vel)}{t} &+&\divergence[\sbrackets]{\rho \vel\vel + \rbrackets{\Pth+\Pmag}\rttensor{I} - \frac{\Bvec\Bvec}{4\pi}}
    \nonumber\\
 &=&- \rho \nabla \Phi +\sigmatot \cdot\sbrackets{ \Fcr - \vel \cdot \rbrackets{\Pcrtensor + \ecr \rttensor{I}}} ,
\end{eqnarray}
and
\begin{eqnarray}\label{eq:energy_con}
    \pderiv{\etot}{t} &+&
    \divergence[\sbrackets]{\vel(\etot+\Pth+\Pmag)- \frac{\Bvec(\Bvec\cdot\vel)}{4\pi}}
    \nonumber\\
    &=&\mathcal{G} - \mathcal{L} - (\rho\vel) \cdot\nabla{\Phi}  \nonumber\\
    &+& \rbrackets{\vel + \vel_\mathrm{s}} \cdot
    \sigmatot \cdot
    \sbrackets{  \Fcr - \vel \cdot \rbrackets{\Pcrtensor + \ecr \rttensor{I}} }.
\end{eqnarray}
\autoref{eq:mom_con} and \autoref{eq:energy_con} do not explicitly include the momentum and energy injected in the gas from SNe as we adopt a prescription that changes from energy to momentum injection depending on the condition of the feedback region; this is discussed in \autoref{sec:feedback} below.
The magnetic field evolution is governed by the induction equation without explicit resistivity (ideal MHD):
\begin{equation}\label{eq:induction}
    \pderiv{\Bvec}{t}=\curl{\vel\times\Bvec},
\end{equation}
with the divergence-free constraint
\begin{equation}\label{eq:divzero}
\nabla\cdot \Bvec =0.
\end{equation}
The transport of CR fluid is described by the two-moment formalism \citep{2018ApJ...854....5J,2021ApJ...922...11A}:
\begin{eqnarray}\label{eq:ecr}
    \pderiv{\ecr}{t} +
    \divergence[]{\Fcr} =
    &-& \rbrackets{\vel + \vel_\mathrm{s}} \cdot
    \sigmatot \cdot
    \sbrackets{  \Fcr - \vel \cdot \rbrackets{\Pcrtensor + \ecr \rttensor{I}} }
    \nonumber\\
     &-& \Lambda_\mathrm{coll} n_\mathrm{H} \ecr + \dot{e}_{\rm c,SN},
\label{eq:CRenergy}
\end{eqnarray}
\begin{eqnarray}\label{eq:Fcr}
\frac{1}{\vmax^2} \pderiv{\Fcr}{t} + \divergence[]{\Pcrtensor}= &-& \sigmatot \cdot \sbrackets{\Fcr - \vel \cdot \rbrackets{\Pcrtensor + \ecr \rttensor{I}}}\nonumber \\
&-& \frac{\Lambda_\mathrm{coll} n_\mathrm{H}}{v_\mathrm{p}^2} \Fcr
\label{eq:CRflux}
\end{eqnarray}
where the tensor $\sigmatot$ includes both explicit scattering terms and a term that captures the effect of streaming at the Alfv\'en speed when the scattering rate is large (see \autoref{sec:cr_method}).
\autoref{eq:ecr} includes as $\dot{e}_{\rm c,SN}$ the term representing CR energy injection by SNe; this is accomplished via a passive scalar variable as detailed in \autoref{sec:crinjection}.

In the above, $\rho = \muH m_{\rm H}\nH$ is the gas density, $\nH$ the  number density of hydrogen nuclei, $\muH$  the mean molecular weight per H nucleus, and
$m_{\rm H}$  the mass of a hydrogen atom;
$\vel$ and $\Bvec$ are velocity and magnetic field vectors, respectively;
$\Pth$ and $\Pmag=\Bvec\cdot\Bvec/(8\pi)$ are thermal and magnetic pressure, respectively;
$\etot=\rho \vel\cdot\vel/2 + \Pth/(\gamma -1) +
\Pmag$ is the total energy density of thermal gas, where $\gamma=5/3$ is the adiabatic index of the thermal gas.
$\ecr$ is the CR energy density, $\Fcr$ is the CR energy flux, and $\Pcrtensor$ is the CR pressure tensor. We assume approximately isotropic pressure in the CR streaming frame, such that $\Pcrtensor\equiv \Pcr\rttensor{I}$, with $\Pcr =(\gammacr - 1)\ecr = \ecr/3$ and $\gammacr = 4/3$ is the adiabatic index of the CR fluid, assumed to be relativistic. $\rttensor{I}$ is the identity tensor. The speed $\vmax$ is the maximum CR transport speed, which in principle is close to the speed of light but is reduced here for computational efficiency.
The collisional CR loss terms $\Lambda_{\rm coll}$ on the RHS of \autoref{eq:CRenergy} and \autoref{eq:CRflux} will be detailed in \autoref{sec:cr_method}, along with the terms in $\sigmatot$ that represent effects of interactions of CRs with Alfv\'en waves. $v_{\rm p}$ is the proton velocity.
Strictly speaking, these collisional loss terms would appear as source terms for gas on the RHS of \autoref{eq:mom_con} and \autoref{eq:energy_con}, but these collisional source terms are generally quite small compared to other MHD source terms.

The total gravitational potential $\Phi = \Phi_{\rm sg} + \Phi_{\rm ext}(z)$ includes the self-gravitational potential obtained as the solution of Poisson's equation (including contributions from both gas and young star clusters, represented numerically as sink/star particles),
\begin{equation}\label{eq:poisson}
    \nabla^2 \Phi_{\rm sg} = 4\pi G (\rho + \rho_{\rm sp}),
\end{equation}
and the fixed external gravitational potential in the vertical direction. We adopt an identical functional form to that in \classic\ \citep{2017ApJ...846..133K},
\begin{equation}
\begin{split}
\Phi_{\mathrm{ext}}(z) = \; & 2 \pi G \Sigma_*z_* \sbrackets{ \rbrackets{ 1 + \frac{z^2}{z_*^2}}^{1/2} -1}\\
&+ 2 \pi G \rho_\mathrm{dm} R_0^2 \,\ln\rbrackets{ 1 + \frac{z^2}{R_0^2}}\,,
\end{split}
\label{eqn:pot}
\end{equation}
where $\Sigma_* = 42$ M$_{\odot}$ pc$^{-2}$, $z_*=245$ pc, $\rho_{\rm{dm}} = 0.0064$ M$_{\odot}$ pc$^{-3}$ and $R_0 = 8$ kpc represent solar neighborhood-like conditions (the ``R8'' model in \classic\ and \ncr) and are held fixed throughout this paper.

The radiative heating ($\mathcal{G}$) and cooling ($\mathcal{L}$) terms are also identical to those of \classic. We adopt a tabulated cooling coefficient that depends only on gas temperature, $\mathcal{L}=\nH^2\Lambda(T)$. Our table combines a simple fitting formula from \citet{2002ApJ...564L..97K} and a table from \citet{1993ApJS...88..253S} for collisional ionization equilibrium at solar metallicity. We adopt $\mu_{\rm H}=1.4271$ consistent with \citet{1993ApJS...88..253S}. The heating rate $\mathcal{G}=\nH \Gamma$ represents the photoelectric effect on small grains which scales with the far-ultraviolet (FUV) radiation field. Adopting a simple plane-parallel global attenuation model for FUV radiation emitted from star particles, the heating rate per hydrogen nucleon can be written as
\begin{equation}\label{eq:Gamma}
    \Gamma = \Gamma_0 \rbrackets{\frac{\Sigma_{\rm FUV}}{4\pi J_{\rm FUV,0}}f_\tau + 0.0024},
\end{equation}
where the reference values of $\Gamma_0=2\times10^{-26}\ergs\,{\rm H^{-1}}$ and $J_{\rm FUV,0}=2.1\times10^{-4}\ergs\cm^{-2}{\rm sr^{-1}}$ are for solar neighborhood conditions. The unattenuated FUV luminosity per unit area $\Sigma_{\rm FUV}$ is calculated by summing over all star particles' contribution (see \autoref{sec:feedback}).
Here we adopt the same attenuation function, $f_\tau=(1-E_2(\tau_\perp/2))/\tau_\perp$ with $\tau_\perp=\kappa_{\rm FUV}\Sigma_{\rm gas}$ and $\kappa_{\rm FUV} = 10^3 \cm^2\,{\rm g}^{-1}$, as in our previous set of \classic\ simulations \citep{2020ApJ...900...61K}, which derives from a uniform slab model \citep{2010ApJ...721..975O}; \citet{2024ApJ...975..173L} show that this is a reasonable approximation to the \ncr\ full radiative transfer results. The temporal variation of $\Gamma$ is mainly driven by the time dependence of $\Sigma_{\rm FUV}$, as the number and ages of star particles vary in time. The heating rate is floored by the second term in \autoref{eq:Gamma}, which represents the minimum heating from meta-galactic UV \citep{2002ApJS..143..419S}.

As we shall explain in \autoref{sec:cr_method}, we consider locally varying CR ionization rate scaled to $\ecr$ in calculating the ionization fraction of gas at $T<2\times10^4\Kel$, which affects the CR scattering coefficient calculation.  However, we do not explicitly include heating from CR ionization in the current simulations. We defer fully consistent coupling of the CR fluid with both thermodynamic and chemical state evolution of the gas to future development within the \ncr\ framework, in which CR ionization enters into both heating and ionization of hydrogen and other species \citep{2023ApJS..264...10K}.

Note that all our code implementations are compatible with the shearing-box formalism implemented in \athenapp, but in this paper we focus on results from non-shearing-box simulations. Thus, unlike \classic\ and \ncr, we do not include Coriolis or tidal gravity terms in \autoref{eq:mom_con}, and boundary conditions are periodic in both horizontal directions.

\subsection{MHD and Gravity Solvers}
\athenapp\ solves the ideal MHD equations using a high-order Godunov method with a constrained transport scheme. This evolves the volume-averaged density, momentum, and total energy using the corresponding area and time averaged fluxes computed by a Riemann solver. To calculate fluxes, we use piece-wise linear (second order) reconstruction and the HLLD Riemann solver.

To update the area-averaged magnetic fields defined on cell faces, we need to evaluate the line-averaged electro motive force $-\vel \times \Bvec$ at cell corners. Rather than the original \athenapp\ constrained transport method based on \citet{2008JCoPh.227.4123G}, we adopt the UCT-HLLD algorithm proposed by \citet{2004JCoPh.195...17L} and extended in \citet{2021JCoPh.42409748M}. The time integration is done with a second-order Runge-Kutta (RK2) time integrator consisting of two sub-steps \citep{2020ApJS..249....4S}.

The gravitational force and work terms are included as source terms at each sub-step within the time integrator. Poisson's equation is solved by using a fast Fourier transform method
identically to \citet{2017ApJ...846..133K}, which allows for open boundary conditions in the vertical direction. In our usual MHD simulations, we solve Poisson's equation at every sub step. For CRMHD simulations with the time step constrained by $\vmax$, we sparsely update the gravitational potential by solving Poisson's equation at every MHD time step determined by the maximum speed of MHD waves, which is typically $\sim 5-10$ times smaller than $\vmax$.

For the combined radiative cooling and heating energy source term, we update only temperature (keeping density fixed) with subcycling at the end of the time integrator (i.e., the cooling solver is called after the MHD integrator). Each subcycle evolves the internal energy $e_{\rm int}$ for 10\% of the instantaneous cooling time $|t_{\rm cool}| \equiv e_{\rm int}/|\mathcal{L}-\mathcal{G}|$ until the accumulated evolution time reaches the time step of the main integrator.

After the transport step (applying flux divergence), if the high-order fluxes yield density or pressure smaller than the preset floors, we recalculate the fluxes of the problematic cells using first-order reconstruction and the local Lax-Friedrich (LLF) Riemann solver.
At the end of every integration step (after applying source terms), we check the density and pressure again and correct the problematic cells by taking an average from a local 6-cell stencil. We additionally check the flow, Alfv\'en, and sound speeds and raise the cell density if any speed exceeds a velocity ceiling. We use the density floor $n_{\rm H, floor}=7\times 10^{-6}\pcc$, pressure floor $P_{\rm floor}= 10^{-4}k_B\pcc\Kel$, and velocity ceiling $v_{\rm ceil}=5\times10^3\kms$.

\subsection{CR transport}\label{sec:cr_method}

The two-moment CR transport method implemented by \citet{2018ApJ...854....5J} advances \autoref{eq:CRenergy} and \autoref{eq:CRflux} in time using a finite-volume solver similar to the MHD solver, splitting transport (LHS) and source (RHS) terms.
An HLLE Riemann solver with a second-order piecewise linear reconstruction scheme is used for the flux calculations.

The source term is calculated in the frame where the $x$-direction is parallel to $\Bvec$ with rotated flow velocity $\vel$ and CR flux $\Fcr$.
In this rotated frame, the CR-wave interaction
coefficient is a diagonal
tensor $\sigmatot$ consisting of one component $\sigma_{\rm tot,\parallel}$ parallel and two components  $\sigma_{\rm tot,\perp}=\sigma_\perp$
perpendicular to the direction of the magnetic field.
For the parallel component, there are two terms,
\begin{equation}\label{eq:sigma_def}
    \sigma_{\rm tot,\parallel}^{-1}= \sigma_\parallel^{-1} + \frac{v_{A,i}}{|\hat \Bvec \cdot \nabla \Pcr |} (\Pcr+\ecr).
\end{equation}
The first is the inverse of the explicit wave scattering term parallel to the magnetic field, $\sigma_\parallel$, and in this work, we always include the perpendicular scattering as $\sigma_\perp = 10\sigma_\parallel$.
The parallel component of the scattering coefficient $\sigma_\parallel$ is derived based on the self-confinement picture as detailed in Section 2.2.3 of \citet{2021ApJ...922...11A}.  In this approach, the scattering rate is computed by assuming that the growth rate of
the streaming instability balances the Alfv\'en wave damping rate, including terms for both ion-neutral damping and nonlinear Landau damping.  The former is more important in denser, low-ionization gas, while the latter is more important in more diffuse, high-ionization gas.

The local Alfv\'en speed of ions is
$v_{A,i}\equiv B/(4\pi \rho_i)^{1/2}$, for $\rho_i$ the ion density, so that the  CR streaming velocity is defined by
\begin{equation}
\vel_\mathrm{s} = -v_{A,i}\frac{\hat{\Bvec}\cdot\nabla \Pcr}{| \hat{\Bvec}\cdot\nabla \Pcr |} \hat{\Bvec};
\end{equation}
i.e. $\vel_\mathrm{s}$ points along the magnetic field down the CR pressure gradient.  With \autoref{eq:sigma_def}, in the limit of large $\vmax$ (i.e. steady state for the CRs) and for our assumption of nearly-isotropic, relativistic CRs, with small collisional losses, it is straightforward to show that
\autoref{eq:CRflux} becomes
\begin{equation}\label{eq:steady_flux}
\nabla \Pcr \approx -\rttensor{\sigma}\cdot \sbrackets{\Fcr - (\vel+\vel_\mathrm{s})(4\Pcr)}.
\end{equation}
Physically, this means that scattering of individual CRs occurs at a rate $c^2 \sigma$ in the frame moving at the sum of the gas velocity and the streaming velocity, i.e. the frame of Alfv\'en waves. In the steady-state limit, the first source term on the RHS of \autoref{eq:CRenergy} becomes
\begin{equation}\label{eq:source_steady}
(\vel+\vel_\mathrm{s})\cdot \nabla \Pcr.
\end{equation}
It is evident from \autoref{eq:steady_flux} that in regions where the scattering rate is large, the flux will be $\Fcr \rightarrow (\vel + \vel_\mathrm{s})(4 \Pcr)$, such that the effective transport speed of the CR fluid will be the sum of the gas flow velocity and the Alfv\'en speed.

The ion mass density is $\rho_i = \mu_i n_i  m_\mathrm{H}$, for $n_i=x_i n_\mathrm{H}$ the number density and $\mu_i$  the ion mean molecular weight (see Eq.~28 in \citealt{2025ApJ...988..214L}).
In the simulations, the ionization fraction $x_i$ at $T<2\times10^4\Kel$ is obtained from the balance
between ionization and recombination processes, including CR ionization, collisional ionization, radiative recombination, and grain-assisted recombination for hydrogen. We also include contribution of ions from metals with ionization potential $<13.6\eV$ (mainly C$^+$). This treatment follows \citet{2021ApJ...922...11A}, but additionally accounts for collisional ionization to ensure a smooth transition of $x_i$ and $\mu_i$ at $T \sim 2 \times 10^4 \Kel$.
The local CR ionization rate is calculated by assuming a functional form for the CR spectrum from \citet{2018A&A...614A.111P} with the high energy slope of $-2.7$ and the low energy slope of $\delta =-0.35$. At $T>2\times10^4\Kel$, $x_i$ is simply taken from the tabulated values in \citet{1993ApJS...88..253S}.

The CR loss term consists of ionization and hadronic losses $\Lambda_{\rm coll}=\Lambda_{\rm coll,ion} +\Lambda_{\rm coll, pion}$ which we calculate using the expression derived in Eq.~9 of \citet{2021ApJ...922...11A}, with the loss functions taken from \citet{2018A&A...614A.111P}. Evaluating at 1 GeV, we have $\Lambda_{\rm coll} = 1.62\times10^{-16}\,{\rm cm^{3}\,s^{-1}}$ with $\Lambda_{\rm coll,pion}=9\times10^{-17}\,{\rm cm^{3}\,s^{-1}}$ and $\Lambda_{\rm coll,ion}=7.2\times10^{-17}\,{\rm cm^{3}\,s^{-1}}$.

We apply bad cell mitigation steps similar to those used in the MHD part. We recalculate the CR fluxes using first-order reconstruction and the LLF Riemann solver if the CR energy density is below a floor after applying the transport term. At the end of the integration step, we apply the same neighbor averaging for CR energy density. We use a floor value of $e_{\rm c, floor}=10^{-20}\eV\pcc$.

\subsection{Modifications from \classic}\label{sec:mod_classic}
The star formation and stellar feedback models in this paper are philosophically similar to \classic\ as detailed in \citet{2017ApJ...846..133K,2020ApJ...900...61K}. Here, we summarize a few implementation details for sink particles and SN feedback.

\subsubsection{Sink Particles}\label{sec:sink}

We utilize sink particles to model star cluster formation and stellar feedback. From a numerical point of view, the introduction of sink particles is necessary to avoid unresolved runaway gravitational collapse.
The main implementation follows \citet{2013ApJS..204....8G} with modifications introduced in subsequent papers \citep{2017ApJ...846..133K,2020ApJ...900...61K}. Here, we summarize the key features that differ from the original implementations.

\paragraph{Sink Creation} We first test the density threshold. We use locally isothermal Larson-Penston density at $r=\Delta x /2$
\begin{equation}
    \rho_{\rm thr} = \frac{8.86}{\pi}\frac{c_s^2}{G\Delta x},
\end{equation}
where the isothermal sound speed is defined locally as $c_s^2\equiv \Pth/\rho$. If the cell density exceeds $\rho_{\rm thr}$, we subsequently conduct a local potential minimum test and a converging flow test. The gravitational potential of the target cell should be the minimum in the control volume, defined by the surrounding $3^3$ cells. We then check the strict converging flow condition; i.e., the mass fluxes averaged over the control volume faces (each face consists of $3^2$ cell faces) should be convergent in all three directions. The particle creation task is called after the main integrator step in a separate task list. When the creation condition is met, we introduce a zero-mass particle at the position of the target cell. We then apply \emph{Sink Merging} to check any overlap with existing particles and \emph{Sink Accretion} to assign the actual mass and other properties as described below.

\paragraph{Sink Merging} We check the overlap of the control volumes of every pair of particles. If two overlapping particles are both new, zero mass particles introduced by \emph{Sink Creation}, we remove both particles and introduce a new zero mass particle at the geometrical center of the two. If one overlapping particle is new, we simply remove the new particle expecting 
that the corresponding density peak shall be accreted into the existing particle in the subsequent \emph{Sink Accretion} phase.
If both overlapping particles are not new, we merge two particles at the center of mass position. Since more than two particles can overlap, \emph{Sink Merging} is performed
iteratively until there are no overlapping particles.

\paragraph{Sink Accretion} For each particle, we again check the strict converging flow condition. If this check returns false, any existing sink does not accrete, and any new particle is removed, i.e., creation is canceled. If the condition is satisfied, all $3^3$ control volume cells are
reset using the average values of exterior adjacent cells (face-sharing cells). This include the cell where the particle is contained, which is reset by an average over 6 face-sharing cells. This averaging is applied to gas density, momentum, and internal energy. The magnetic field and CR variables are untouched. The momentum and total energy are recalculated based on the reset primitive variables. For passive scalars (including CR injection scalar, see below), we take averages of their conservative variables (i.e., species density) rather than the specific scalars (i.e., concentration). The mass difference between before and after reset $\Delta M_{\rm reset}\equiv M_{\rm ctrl,after} - M_{\rm ctrl,before}$ is usually negative, implying that the mass is removed from the control volume, which is a defining property of accretion. The removed mass after reset is added to sink $\Delta\Msink = -\Delta M_{\rm reset}$. See \emph{Sink Mass Partitioning} for partitioning of the added mass into stars and gas within the sink. The sink's momentum and metal mass is similarly updated. When $\Delta M_{\rm reset}>0$, we simply cancel the accretion and restore the fluid variables in the control volume to their original values before the reset.
The accretion rate determined by the above procedure is identical to the mass inflow rate into the control volume, modulo the rate of change in the remaining gas mass present in the control volume (the latter is zero for steady accretion, see Equation (55) in \citealt{2025ApJ...987...78M}).

\paragraph{Sink Integration}
Each particle experiences the same gravitational field as the gas (i.e., both self and external gravity described by the total potential). The force is interpolated onto the particles' positions. We use a kick-drift-kick leapfrog integrator coupled within the RK2 integrator. We also have an option to use a drift-kick-drift leap frog integrator for the van Leer predictor-corrector integrator \citep{2009NewA...14..139S}. The two different choices are made to guarantee the total (gas + sink) momentum conservation.

\paragraph{Sink Mass Partitioning}
In the original \classic\ framework, all mass locked into sink particles is treated as stars. In other words, it adopts an instantaneous, 100\% star formation model. To relax this somewhat extreme assumption, we introduce a model that treats a sink particle as a star cluster with a gas reservoir. We anticipate the development of a more sophisticated sink subgrid model motivated by cloud scale simulations including radiation and stellar wind feedback \citep[e.g.,][]{2018ApJ...859...68K,2021ApJ...911..128K,2021ApJ...922L...3L}. In this work, we take a simple version with an instantaneous partitioning with a constant gas fraction $\fgas=0.7$; i.e., $\Delta \Mgas = \Delta M_{\rm sink} \fgas$ and $\Delta \Mstar = \Delta M_{\rm sink} (1-\fgas)$.

\paragraph{Sink Mass Return}
Besides mass gain through \emph{Sink Accretion} and mass loss through SN feedback, which we shall discuss in \autoref{sec:feedback}, we implement an additional mass loss process, representing ionizing outflows from GMCs by photo-evaporation or similar gas release from sink gas reservoirs. We plan to implement a more physically motivated model based on small-scale simulations in the future and investigate its impact more deeply. In the present work, however, we adopt one simple choice, as we are primarily interested in the effects of CRs while keeping other treatments fixed.
We determine the return mass $\Delta \Mret = \Mgas(\Delta t_{\rm ret}/t_{\rm ret})$ at every $\Delta t_{\rm ret}=1\Myr$ for a fixed time scale of $t_{\rm ret}=5\Myr$ over $2t_{\rm ret}$ (i.e., 10 mass return events per sink over 10 Myr).
For each mass return event, we spread $\Delta \Mret$ uniformly into a spherical region centered at the sink with a radius of $r_{\rm ret}=100\pc$.
We then subtract $\Delta M_{\rm ret}$ from $\Mgas$. We adjust the return mass not to exceed $\Mgas$.\footnote{In R. Hix et al (2026, in prep), we explore effects of varying parameters for sink mass return ($\fgas$, $r_{\rm ret}$, $t_{\rm ret}$).}

\subsubsection{FUV and Supernova Feedback}\label{sec:feedback}

For each sink particle, we keep track of its stellar mass $\Mstar$, age since birth $t_{\rm age}$, and mass-weighted mean age $t_{\rm mage}$. Both $t_{\rm age}$ and $t_{\rm mage}$ increases in increments of the simulation time step $\Delta t$. When two sinks merge, $t_{\rm age}$ is taken from the older particle and $t_{\rm mage}$ is set by the mass weighted mean value. When accretion occurs (i.e., adding zero age stars), $t_{\rm mage}$ is adjusted by
\begin{equation}
    t_{\rm mage} = \frac{\Mstar}{\Mstar+\Delta \Msink(1-\fgas)}t_{\rm mage}.
\end{equation}

We adopt the same population synthesis model from {\tt STARBURST99} as in \classic. In particular, we adopt Geneva non-rotating tracks, solar metallicity, and a fully sampled Kroupa IMF. We tabulate a luminosity-to-mass ratio $\Psi_{\rm FUV}(t)$ for FUV radiation and calculate the FUV luminosity $L_{\rm FUV} = \Psi_{\rm FUV}(t_{\rm mage})\Mstar$. Each star particle's FUV luminosity is then summed up and divided by the horizontal area of the simulation domain to get $\Sigma_{\rm FUV}$, and then to set $\Gamma$ using \autoref{eq:Gamma}.

We also tabulate the specific SN rate $\xi (t)$. We assign a SN for the sink when the expectation value of SN $\mathcal{N}_{\rm SN} = \xi_{\rm SN}(t_{\rm age}) \Mstar \Delta t$ exceeds a uniform random number in $[0,1)$. Note that in the present work we do not include SNe from runaway OB stars nor type Ia SNe (although these capabilities have been implemented).

When a SN event occurs, we assign a spherical region with radius $r_{\rm inj}=3\Delta x$ around the event site as a feedback volume. We calculate the mean gas properties within the feedback volume. If the mass within the feedback volume $M_{\rm fb}$ (including mass to be injected, $M_{\rm inj}$; see below) is small enough to resolve the energy conserving stage of SN remnant evolution, based on the criterion $M_{\rm fb}<0.1M_{\rm sf}$, a total energy $E_{\rm SN}$ is injected within the feedback volume, with the thermal to kinetic fraction consistent with that of the Sedov-Taylor solution, i.e., $E_{\rm th}:E_{\rm kin}=0.72:0.28$.
Here, we use the shell formation mass in the two-phase medium $M_{\rm sf}=1344\Msun E_{\rm SN,51}^{0.87} n_{{\rm H},0}^{-0.26}$ as calibrated by \citet{2015ApJ...802...99K}, where $E_{\rm SN,51}=E_{\rm SN}/10^{51}\erg$ and $n_{\rm H,0} = \nH /1\pcc$.
In future work, this prescription can be updated with metallicity dependence based on the results of \citet{2023ApJS..264...10K}. When $M_{\rm fb}>0.1M_{\rm sf}$, we instead inject terminal momentum calibrated by \citet{2015ApJ...802...99K}:
\begin{equation}
    p_{\rm final} =2.8\times10^5\Msun\kms n_{{\rm H},0}^{-0.17}.
\end{equation}

Each feedback event injects a total energy of $E_{\rm SN}=10^{51}\erg$ (or the terminal momentum, which gives total energy less than that) coupled with the injection mass $M_{\rm inj}$ and the mass in the feedback volume. The injection mass consists of the SN ejecta mass $M_{\rm ej}$ plus additional mass representing swept-up material from the unresolved gas reservoir. We subtract $M_{\rm ej}$ from $\Mstar$ and $M_{\rm inj}-M_{\rm ej}$ from $\Mgas$. For the present work we adopt $M_{\rm inj}=40\Msun$ and $M_{\rm ej}=10\Msun$.

Note that here we set the feedback volume differently from \classic, where for resolved SN, we
adjust the
feedback volume so that the mass is close to $0.1M_{\rm sf}$. The \classic\ approach is somewhat equivalent to the Lagrangian way of coupling SN energy (where the mass resolution is fixed), but the current implementation is more natural for an Eulerian code (where the volume resolution is fixed).
Depending on the adopted value of $M_{\rm inj}$ and the parameters of the mass return model described in \autoref{sec:sink}, we have found that the choice of the feedback volume can strongly affect the driving of hot outflows. We will present a comprehensive study of this in future work. Our results suggest that it will be important to test sensitivity of wind driving to mass resolution in simulations using Lagrangian codes as well.
For this work, we choose feedback model parameters that produce results for MHD-only simulations that are similar to those obtained with our \classic\ implementation, and explore the role of CRs.

\subsection{CR injection} \label{sec:crinjection}
In CRMHD simulations, we additionally inject a fraction $\fcr$ of SN energy as CR energy.
This means that the total energy injected to the simulation from each resolved SN event is $(1+\fcr)\Esn$.\footnote{\REV{In principle, one could keep the total injected energy fixed at $\Esn$ and partition it into thermal ($(1-\fcr)\Esn$) and CR ($\fcr\Esn$) energy. We opt for the current approach and the full exploration of the other choice with varying $\fcr$ will be presented in a companion paper.}}
Extensive testing has shown that instantaneous CR injection within the feedback volume causes numerical difficulties, especially when subsequent SNe explode within pre-cleared interiors of superbubbles.
Moreover, in reality CRs are accelerated at SN shocks over the course of SN remnant evolution rather than instantaneously at the moment of explosion.
\REV{Motivated by these numerical and physical reasons, we first inject CR energy as a passive scalar and convert it into actual CR energy  over a scalar decay time scale of $t_{\rm dec}$. This way, the expanding velocity field within a superbubble interior quickly advects the newly-injected CR scalar to the superbubble boundary, such that most of CR energy is effectively deposited near the interface.}

In practice, at the time of each SN event, we add a CR passive scalar $\scr = \fcr E_{\rm SN}/V_{\rm inj}$ uniformly within the feedback volume, where $V_{\rm inj} = 4\pi r_{\rm inj}^3/3$, and $\fcr$ is the fraction of the directly-injected SN energy that goes into CRs. The CR scalar is passively transported by solving a continuity equation with a gas velocity, while at the same time the CR scalar is depleted over a time scale $t_{\rm dec}$:
\begin{equation}
    \pderiv{\scr}{t}+\divergence{\scr\vel} = -\frac{\scr}{t_{\rm dec}}.
\end{equation}
The depletion of the CR scalar is compensated by additions in every cell to the CR energy density variable following $\dot{e}_{\rm c,SN} \Delta t = \scr(1-\exp(-\Delta t/t_{\rm dec}))$.

We adopt $\fcr=0.1$ \citep[e.g.,][]{2014ApJ...783...91C} and $t_{\rm dec}=40\;{\rm kyr}$, which is similar to the shell formation time of a single SN remnant exploded in a uniform medium with density $\nH=1\pcc$ \citep{2015ApJ...802...99K}. The tests for different injection \REV{parameters including varying $\fcr$ and $t_{\rm dec}$, as well as the case keeping the total SN energy fixed,} will be presented in a companion method paper.

\subsection{Models}\label{sec:model}

\begin{figure*}[ht!]
    \includegraphics[width=\linewidth]{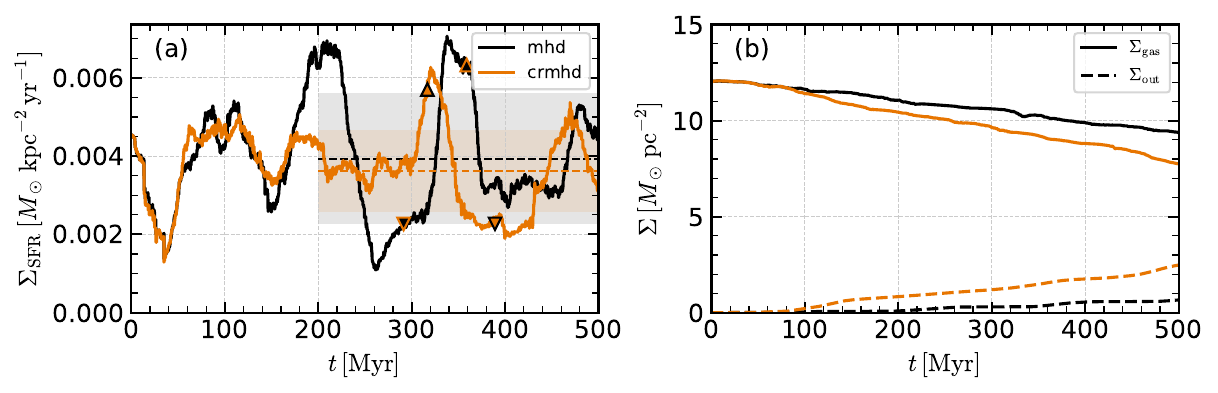}
    \caption{{\bf (a)} Time evolution of SFR surface density $\Ssfr$ calculated using the 40 Myr time bin (see \autoref{eq:sfr40}). The horizontal dotted lines with shaded area represent the time averaged SFR with the standard deviation after $t>200\Myr$. The downward and upward triangles respectively demark examples of quiescent (inflowing) and starburst (outflowing) epochs depicted  in \autoref{fig:snapshot_comp}.  {\bf (b)} Time evolution of the total gas surface density $\Sgas$ and mass loss per area $\Sigma_{\rm out}$. While mean SFRs are similar (within $\sim 10\%$) in the two models, the \crmhd\ model loses more mass (by a factor $\sim 4$) due to the CR driven winds. \label{fig:history}}
\end{figure*}

In this work, we present results from two models, with and without CRs, which we name \crmhd\ and \mhd, respectively. They are identical in all aspects other than inclusion of CR physics.
Our simulation domain is a vertically-elongated, uniform, Cartesian box with side lengths of $L_x=1024\pc$, $L_y=1024\pc$, and $L_z=8192\pc$ and number of zones of $N_x=128$, $N_y=128$, and $N_z=1024$, yielding a cubic cell of  resolution $\Delta x = 8\pc$. We use periodic boundary conditions in the $x$ and $y$ directions and outflow boundary conditions for vertical gas and CR velocities, with zero gradient extrapolation for other MHD variables, and  logarithmic extrapolation for CR energy density in the $z$ direction. The details of the vertical boundary conditions are presented in \autoref{sec:app_bc}.

The initial gas surface density is $\Sgas=12\Surf$. The initial vertical density profile assumes a form of
\begin{align}
\begin{split}
    \rho(z) =& \rho_1(z) + \rho_2(z)\\
            =&\rho_0\exp(-\tilde\Phi_{\rm tot}(z)/\sigma_1^2)+\\
                 &10^{-5}\rho_0\exp(-\tilde\Phi_{\rm tot}(z)/\sigma_2^2)\\
\end{split}
\end{align}
where for the approximate total gravitational potential we adopt $\tilde\Phi_{\rm tot}\equiv\Phi_{\rm ext}(z) + 2\pi G \Sgas h\ln\cosh(z/h)$ with $h = \Sigma_{\rm gas}/(2\rho_0)$. The two terms on the RHS represent warm and hot gas in rough hydrostatic equilibrium with $\sigma_1=10\kms$ for warm and $\sigma_2=10\sigma_1$ for hot gas. The midplane density normalization $\rho_0$ is determined by iteratively solving the constraint equation $\Sgas = \sum\rho(z) \Delta z$ for $\rho_0$. We find $n_0=\rho_0/(\mu_{\rm H}m_{\rm H})=1.48\pcc$. Given the density profile, the initial thermal pressure is set to $\Pth(z) = \rho_1(z)\sigma_1^2 + \rho_2(z)\sigma_2^2$. The azimuthal magnetic field $\Bvec=B_0(z)\yhat$ is initialized  as $B_0(z) = (8\pi \Pth(z)/\beta_0)^{1/2}$ with a constant initial plasma beta of $\beta_0=1$. The velocity field is initialized as a random realization of a turbulent velocity field with a power spectrum of $P_k(v)\propto k^{-2}$ and an equal mix of solenoidal and compressive power for a wavenumber range $1<kL_x/(2\pi)<8$. The amplitude of the initial turbulent velocity dispersion is set to have one dimensional velocity dispersion of $10\kms$.

For the \crmhd\ model, we initialize the CR energy density with a very small value, equal to the minimum thermal pressure. We adopt the maximum CR transport speed $\vmax=2\times10^4\kms$.
This value of $\vmax$ is selected as being large compared to the thermal and MHD signal speeds, while not being so large as to make the simulation timestep impractically small.

\subsection{Definitions of Thermal Phases, Outflow and Inflow, and Horizontal Averages}\label{sec:method_def}

We adopt the temperature cuts for the thermal phase definition identical to \classic\ \citep{2017ApJ...846..133K}:
\begin{itemize}
    \item \CNM\ -- $T<184\Kel$
    \item \UNM\ -- $184\Kel\leq T<5050\Kel$
    \item \WNM\ -- $5050\Kel\leq T<2\times10^4\Kel$
    \item \WHIM\ -- $2\times10^4\Kel\leq T<5\times10^5\Kel$
    \item \HIM\ -- $T\geq 5\times 10^5\Kel$.
\end{itemize}
For most analysis, we merge \CNM, \UNM, and \WNM\ into a single  ``warm-cold'' component (denoted by \twop)
and \WHIM\ and \HIM\ into a single hot component (denoted by \hot).
We also separate gas into outflowing and inflowing components using the outward vertical gas velocity $v_{\rm out}\equiv v_z{\rm sgn}(z)$
\begin{itemize}
    \item Outflow -- $v_{\rm out}>0$
    \item Inflow -- otherwise.
\end{itemize}
A Heaviside step function selecting for a given phase and flow direction, $\Theta^{\rm ph,dir}(T,v_{\rm out})$, is defined to return 1 for each cell satisfying specific temperature and velocity conditions, or return 0 otherwise.

The vertical profiles of interest are constructed using
a component-selected horizontal average, defined by
\begin{equation}\label{eq:fraction}
    \abrackets{q}^{\rm ph,dir}(z) = \frac{\sum_{x, y} q(x,y,z) \Theta^{\rm ph,dir}(T,v_{\rm out}) \Delta x \Delta y}{L_x L_y}.
\end{equation}
Note that this definition corresponds to the fractional contribution of the selected combined
thermal phase and velocity direction component to the volume-weighted average $\abrackets{q}$.
The mean value (or typical value) of $q$ for a given
component is denoted with an overline as
\begin{equation}\label{eq:mean}
\begin{split}
    \overline{q}^{\rm ph,dir}(z) &= \frac{\sum_{x, y} q(x,y,z) \Theta^{\rm ph,dir}(T) \Delta x \Delta y}{\sum_{x, y} \Theta^{\rm ph,dir}(T) \Delta x \Delta y} \\
    &= \frac{\abrackets{q}^{\rm ph,dir}}{f_A^{\rm ph,dir}},
\end{split}
\end{equation}
where $f_A^{\rm ph,dir} \equiv\abrackets{1}^{\rm ph,dir}$ is the area filling factor of the component.

\begin{figure*}[htb]
    \includegraphics[width=\linewidth]{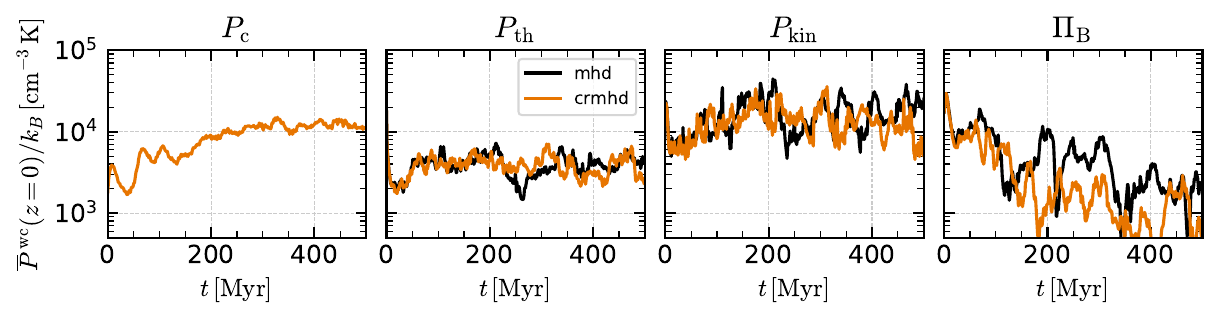}
    \includegraphics[width=\linewidth]{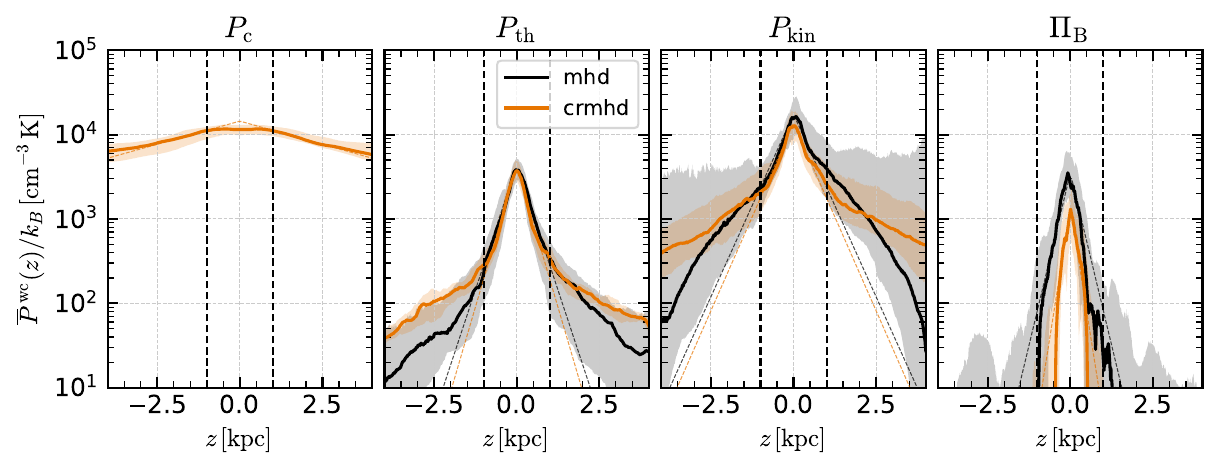}
    \includegraphics[width=\linewidth]{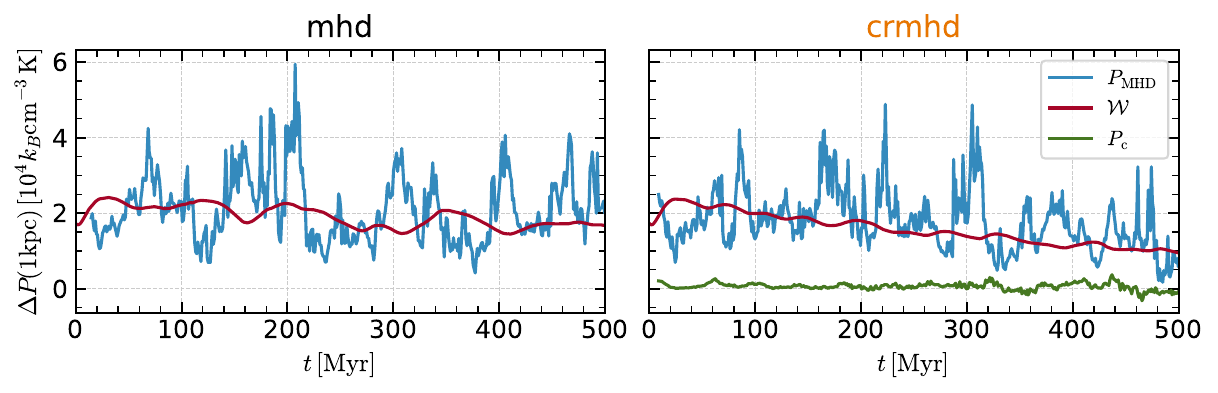}
    \caption{{\bf Top:} Time evolution of each stress component at the midplane.
    The evolution achieves a quasi-steady state after $t=200\Myr$ during which time-averaged profiles are constructed.
    {\bf Middle:} Time-averaged vertical profiles of each stress component. The solid lines represent the median after $t=200\Myr$, while the shaded areas denote the range between the 16th and 84th percentiles. The vertical dashed lines demark
    $|z|=1\kpc$. The exponential fit for each pressure ($|z|>1\kpc$ for $\Pcr$ and $|z|<1\kpc$ otherwise) is shown as thin dashed lines.
    {\bf Bottom:} Time evolution of each term in \autoref{eq:vde_tot} for $z_\mathrm{ref}=1\kpc$.
    Although $\Pcr$ reaches a quasi-steady state value comparable to other stress terms at the midplane (top row), its vertical gradient is so shallow (middle row) that the contribution in the offsetting the weight $\Delta\Pcr(1\kpc)$ is negligible (bottom row)  \label{fig:pressure_t}}
\end{figure*}

\begin{deluxetable}{lccccccccccccccc}
\tablecaption{Summary of key quantities in the PRFM theory \label{tbl:prfm}}
\tablewidth{0pt}
\tablehead{
    \colhead{Model} &
    \colhead{$\Ssfr$} &
    \colhead{$\Sgas$} &
    \colhead{$\tdep$} &
    \colhead{$\mathcal{W}$} &
    \colhead{$\Pcr$} &
    \colhead{$\Pth$} &
    \colhead{$\Pturb$} &
    \colhead{$\Pimag$} &
    \colhead{$\Ptot$} &
    \colhead{$\Ycr$} &
    \colhead{$\Yth$} &
    \colhead{$\Yturb$} &
    \colhead{$\Ymag$} &
    \colhead{$\Ytot$}
}
\startdata
\mhd   & $3.94\times10^{-3}$ & 10.3 & 2.61 & 21.0 & \nodata & 3.62 & 16.8 & 3.79 & 24.2 & \nodata & 1.92 & 8.90 & 2.01  & 12.8 \\
\crmhd & $3.62\times10^{-3}$ & 9.3  & 2.56 & 20.0 & 11.5    & 3.72 & 13.7 & 1.30 & 18.8 & 6.65    & 2.14 & 7.92 & 0.75 & 10.8 \\
\enddata
\tablecomments{All quantities are averaged over $t=200$--$500\Myr$. $\Ssfr$ is in units of $\sfrunit$. $\Sgas$ is in units of $\Surf$. $\tdep$ is in units of Gyr. Weight and pressures are in units of $10^3\,k_B\,\mathrm{cm^{-3}\,K}$. The weight is integrated over the entire box and all phases. The pressures are the mean values (or typical values) averaged over \twop\ gas at the midplane. Feedback yields ($\Upsilon\equiv P/\Ssfr$ for each pressure component) are in units of $10^2\,\mathrm{km\,s^{-1}}$.}
\end{deluxetable}

\section{Star Formation Rates and the PRFM Theory}\label{sec:sfr}

In this section, we first summarize the star formation rates (SFRs), which set the SN explosion rates and hence energy and momentum injection rates for outflow driving. We analyze the vertical dynamical equilibrium and the vertical stress components in each model, providing an explanation for the resulting SFRs in the context of the pressure-regulated, feedback-modulated (PRFM) star formation theory \citep{2022ApJ...936..137O}.

\autoref{fig:history}(a) plots the time evolution of the SFR surface density in both models, for which we sum up all stellar mass in sinks with mass-weighted mean age (see \autoref{sec:feedback}) in a time bin younger than a given value $t_{\rm bin}$ to define
\begin{equation}\label{eq:sfr40}
\Sigma_{{\rm SFR}} = \frac{\sum \Mstar (t_{\rm mage}<t_{\rm bin})}{L_x L_y t_{\rm bin}}.
\end{equation}
Here, we choose $t_{\rm bin}=40\Myr$, the time scale relevant for SN feedback from a single cluster given a single coeval stellar population \citep{1999ApJS..123....3L}.
The two models show very similar $\Ssfr$ in terms of both mean and standard deviation; $\Ssfr = (3.6\pm 1.0)\times10^{-3}\sfrunit$ in the \crmhd{} model and $(3.9\pm 1.7)\times10^{-3}\sfrunit$ in the \mhd{} model after $t>200\Myr$.
Despite the similar SFRs, \autoref{fig:history}(b) shows that the total integrated mass loss per area $\Sigma_{\rm out}$ in the \crmhd{} model is more than four times as large as that in the \mhd{} model.
With similar SFR in the \mhd\ and \crmhd\ models, the effects of thermal SN feedback on wind driving would also be similar, allowing us to isolate the effect of CRs (non-thermal SN feedback) in the wind analysis presented later.

We can understand from PRFM theory why CRs do not significantly alter SFRs.
By taking the horizontal average (as defined by \autoref{eq:fraction}), the vertical component of the momentum equation (\autoref{eq:mom_con}) in steady state can be written as
\begin{equation}\label{eq:vde}
    \deriv{}{z}\abrackets{\Pturb + \Pth + \Pmag -\frac{B_z^2}{4\pi}} =
    -\abrackets{\rho \deriv{\Phi}{z}} -\deriv{}{z}\abrackets{\Pcr} ,
\end{equation}
where $\Pturb = \rho v_z^2$ is the $zz$ component of the Reynolds stress tensor and $\Pmag = \Bvec\cdot\Bvec/8\pi$ is the magnetic pressure. The last term arises from \autoref{eq:CRflux} in the limit of steady-state and negligible collisional losses. We further define the $zz$ component of the Maxwell stress tensor $\Pimag \equiv \Pmag -B_z^2/4\pi$ and the total vertical stress $\Ptot \equiv \Pturb + \Pth + \Pimag$.

Vertical integration of \autoref{eq:vde} from the midplane ($z=0$) to a height $z$ and rearranging gives
\begin{equation}\label{eq:vde_tot}
\begin{split}
 \Delta \Ptot(z)  + \Delta\Pcr(z) =
    \int_0^{z} \abrackets{\rho \deriv{\Phi}{z}}dz \equiv \mathcal{W}(z)
\end{split}
\end{equation}
where $\Delta \Ptot(z) \equiv \abrackets{\Ptot}(0) - \abrackets{\Ptot}(z)$ and $\Delta \Pcr(z) \equiv \abrackets{\Pcr}(0)-\abrackets{\Pcr}(z)$.
Since a quasi-steady state is established in both models, we expect vertical dynamical equilibrium (i.e., \autoref{eq:vde_tot}) to hold on average.

In \autoref{eq:vde_tot}, the weight of the ISM (RHS)
demands a certain amount of total pressure support (LHS).  In our simulations (and in reality to a large degree), this support is provided by stellar feedback to the gas (through both SNe and FUV radiation in our models) as well as CRs (through SNe), with individual pressure terms in $\Ptot$ and $\Pcr$ all approximately linearly proportional to $\Ssfr$.
If the CR pressure support term $\Delta \Pcr(z)$ were non-negligible, the demand for support by gas ($\Delta\Ptot(z)$) would be reduced, such that the required SFRs could also be reduced.
However, we find that including CRs does not in fact affect $\Ssfr$ in our simulations (\autoref{fig:history}), implying that the CR contribution to vertical equilibrium within the disk must be negligible.  This does not, however, mean that the CR pressure is negligible. Instead, the implication is that $\Delta \Pcr(\zref)\ll \Delta \Ptot(\zref)$ for $z=\zref$ large enough to encompass the majority of the ISM mass.

To understand the above finding in more detail, we consider the vertical dynamical equilibrium of just the \twop\ gas component, which is both the relevant gas reservoir for star formation and the predominant contributor to the vertical ISM weight. To the extent that the total pressure is comparable in all phases (which is approximately true), $\overline{P}_{\rm MHD}^{\twop} = \overline{P}_{\rm MHD}^{\hot}$,  we have $\abrackets{\Ptot}=\sum_{\rm ph}\abrackets{\Ptot}^{\rm ph}=\overline{P}_{\rm MHD}^{\twop}\sum_{\rm ph}f_A^{\rm ph} = \overline{P}_{\rm MHD}^{\twop}$.

The top row of \autoref{fig:pressure_t} presents the time evolution of each stress component (including CR pressure) at the midplane as a function of time for the \twop{} phase. $\Pcr$ at the midplane increases quickly and reaches a saturated value as large as other terms, similar to what we have found in post-processing CR simulations \citep[see][]{2025ApJ...994...45H,2026ApJ...996...99L}.

The middle row of \autoref{fig:pressure_t} shows vertical profiles of the individual stress components for the \twop\ gas.
In our previous analyses \citep[e.g.,][]{2022ApJ...936..137O,2024ApJ...972...67K} for simulations that do not include CRs, we often simplify \autoref{eq:vde_tot} by taking $\Delta \Ptot(\zref)\rightarrow \Ptot(0)$, given that all MHD stress components are negligible at the top/bottom of the gas layer.
The middle row of \autoref{fig:pressure_t} shows that this assumption is valid for both the \mhd\ and \crmhd\ simulations, with an order of magnitude (or more) reduction in $\Pth$, $\Pturb$, and $\Pimag$ within $\sim 1\kpc$ of the midplane.
However, the first panel in the middle row of \autoref{fig:pressure_t} shows that the CR pressure profile is much flatter, especially within $|z|<1\kpc$.
The pressure scale heights for the \twop\ gas in the \crmhd\ model measured by an exponential fit to the time-averaged profiles within $|z|<1\kpc$ are $320$, $482$, and $175\pc$ for thermal, kinetic, and magnetic stresses, respectively, which are similar to the density scale height of $199\pc$ and to scale heights in the \mhd\ model.

In the bottom row of \autoref{fig:pressure_t}, we plot all of the terms in \autoref{eq:vde_tot} for $\zref=1\kpc$.
Vertical dynamical equilibrium is satisfied on average (with large temporal fluctuations in $\Delta \Ptot$) in both models, and the CR contribution \REV{$\Delta P_\mathrm{c}$ (green)} is always negligible for the \crmhd\ model.
\REV{Since $\Delta P_\mathrm{c}$ is negligible, \autoref{eq:vde_tot} implies that $\Delta P_\mathrm{MHD} \approx P_\mathrm{MHD} \approx \mathcal{W}$ will be similar in \crmhd\ and \mhd, which is evident from the two bottom panels of \autoref{fig:pressure_t}.  The very similar $P_\mathrm{MHD}$ implies the feedback rate that maintains the MHD pressure is similar, which explains why global SFRs remain the same irrespective of the presence of CRs. Fundamentally, this is because  the CR pressure within the warm-cold phase is uniform on large scales.} In detail,  the \crmhd\ model experiences slightly more secular decrease in the weight than the \mhd\ model because of the enhanced mass loss (see \autoref{fig:history}(b)) by CR-driven winds, as we shall analyze further in \autoref{sec:wind}.

\REV{It is worth noting that the large-scale average SFR is agnostic to potential effects of CRs. As delineated in \citet[][see their Section 2.4]{2022ApJ...936..137O}, any change in the small scale star formation efficiency primarily alters the time-averaged mass fraction of gravitationally bound clouds.  Studying potential effects on the mass fraction of dense gas due to changes induced by CRs in the interiors of gravitationally bound clouds \citep[e.g.,][]{2024ApJ...973...16F} would require higher resolution simulations than we have here (and proper chemistry modeling), an interesting potential direction for future research. }

The above results may be summarized by comparing the two models with respect to key quantities in the PRFM theory.  \autoref{tbl:prfm} includes the rate of star formation per unit area, the total weight integrated over the entire box, the component pressures at the midplane, and the ratios $\Upsilon \equiv P/\Ssfr$, which are the feedback yield parameters. The weight matches the total MHD pressure in both models.  The thermal, kinetic, and total MHD feedback yields are quite similar in the \mhd\ and \crmhd\ simulations.  The value of $\Upsilon_\mathrm{mag}$ is
lower in \crmhd\  than in \mhd; while the reason for this is not certain, we speculate that it is due to the greater outflow rate in \crmhd\ (see \autoref{sec:wind}).

Finally, we note that the scale height of CR pressure measured for $1\kpc<|z|<2\kpc$ is $4\kpc$. This value is much larger than that from the post-processing results ($\sim 0.5\kpc$ for R8, see \citealt{2022ApJ...929..170A,2025ApJ...994...45H}), although in the most comparable warm-wind CRMHD simulations of \citet{2024ApJ...964...99A} (see WW-ID and WW-HD models in Fig. 18 there), we also found that $\Pcr$ dropped by only a factor $\sim 2$ within $|z|=3\kpc$.
As we shall see in the next sections, even with the large CR scale height, the dynamical impact of CR pressure in the extraplanar region is quite substantial.
In the low-density extraplanar region, the thermodynamic properties of the gas are strongly affected by interactions with CRs, which in turn modify the resulting CR distribution and vertical profiles, leading to a larger CR scale height.

\begin{figure*}[ht!]
    \begin{center}
    \fig{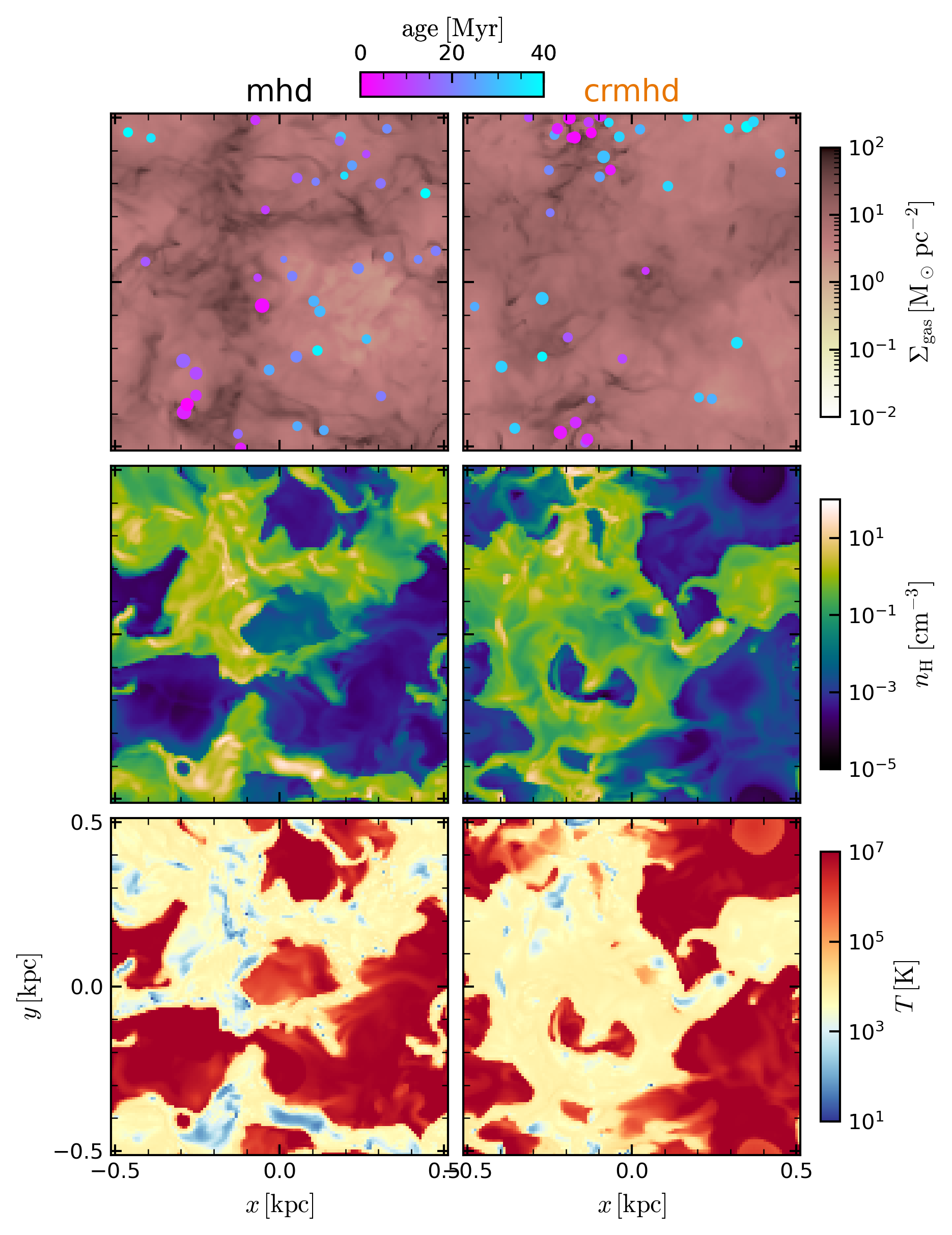}{0.41\textwidth}{(a) Quiescent Epoch, Face-on Views}
    \fig{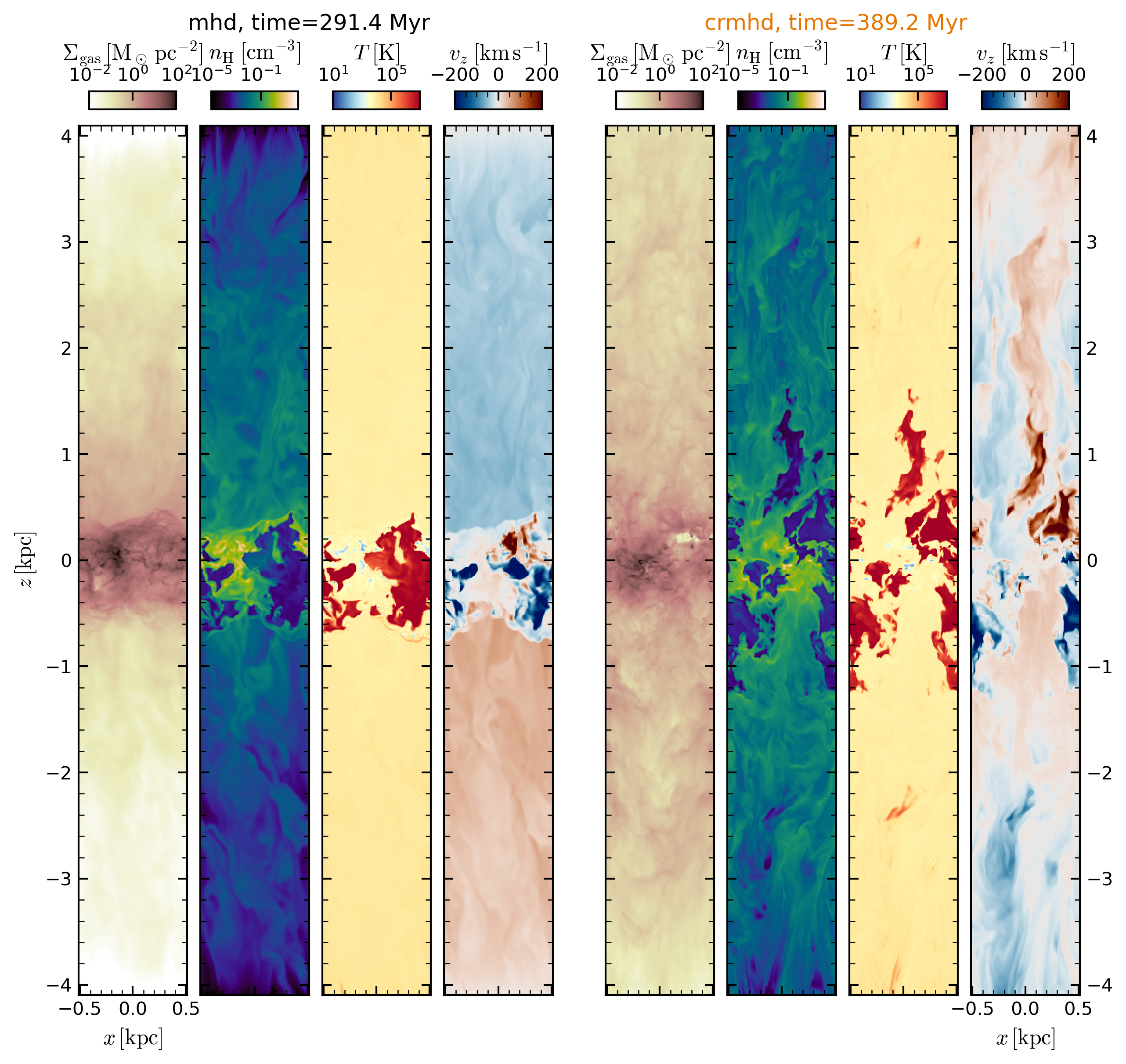}{0.57\textwidth}{(b) Quiescent Epoch, Edge-on Views}
    \end{center}
    \begin{center}
    \fig{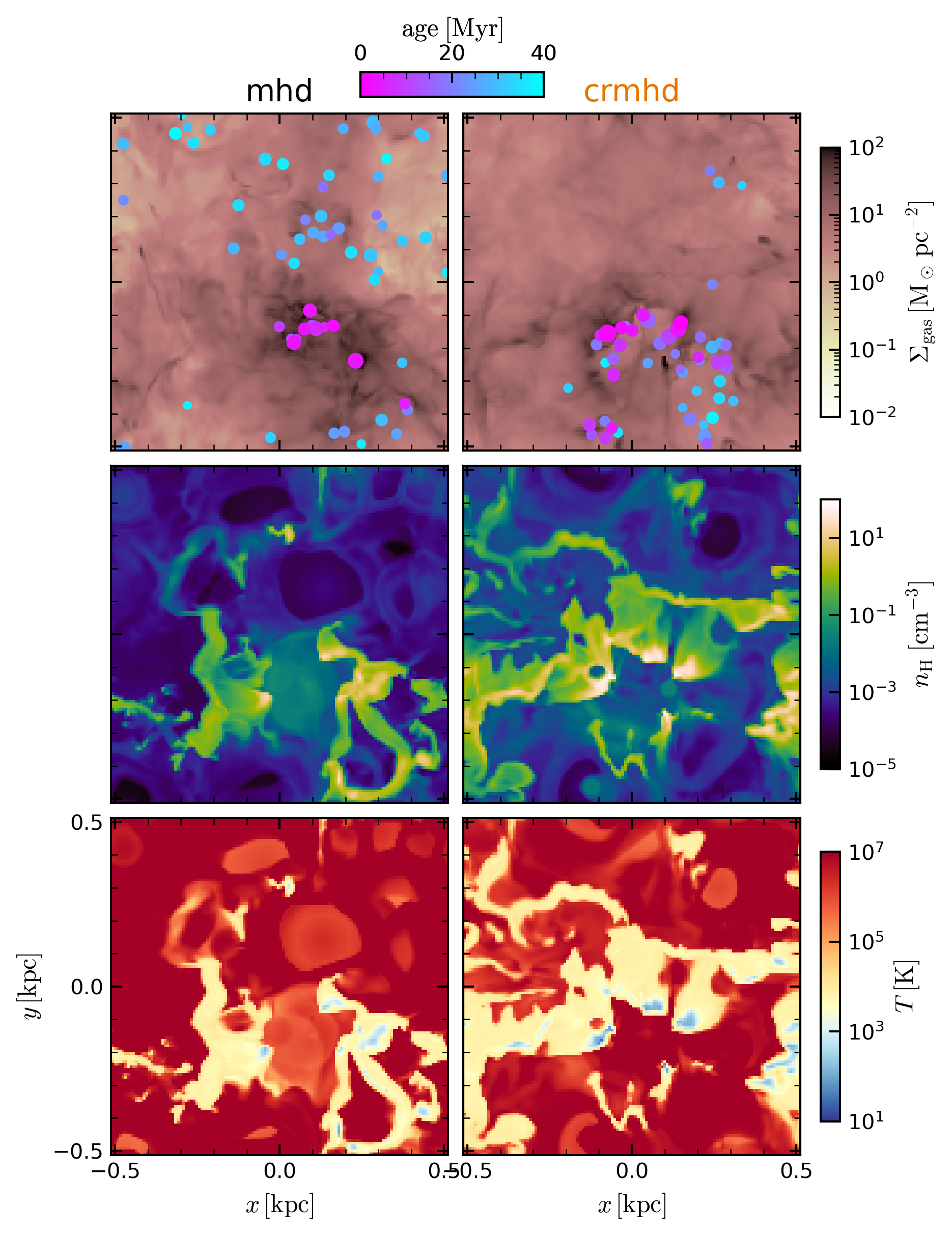}{0.41\textwidth}{(c) Starburst Epoch, Face-on Views}
    \fig{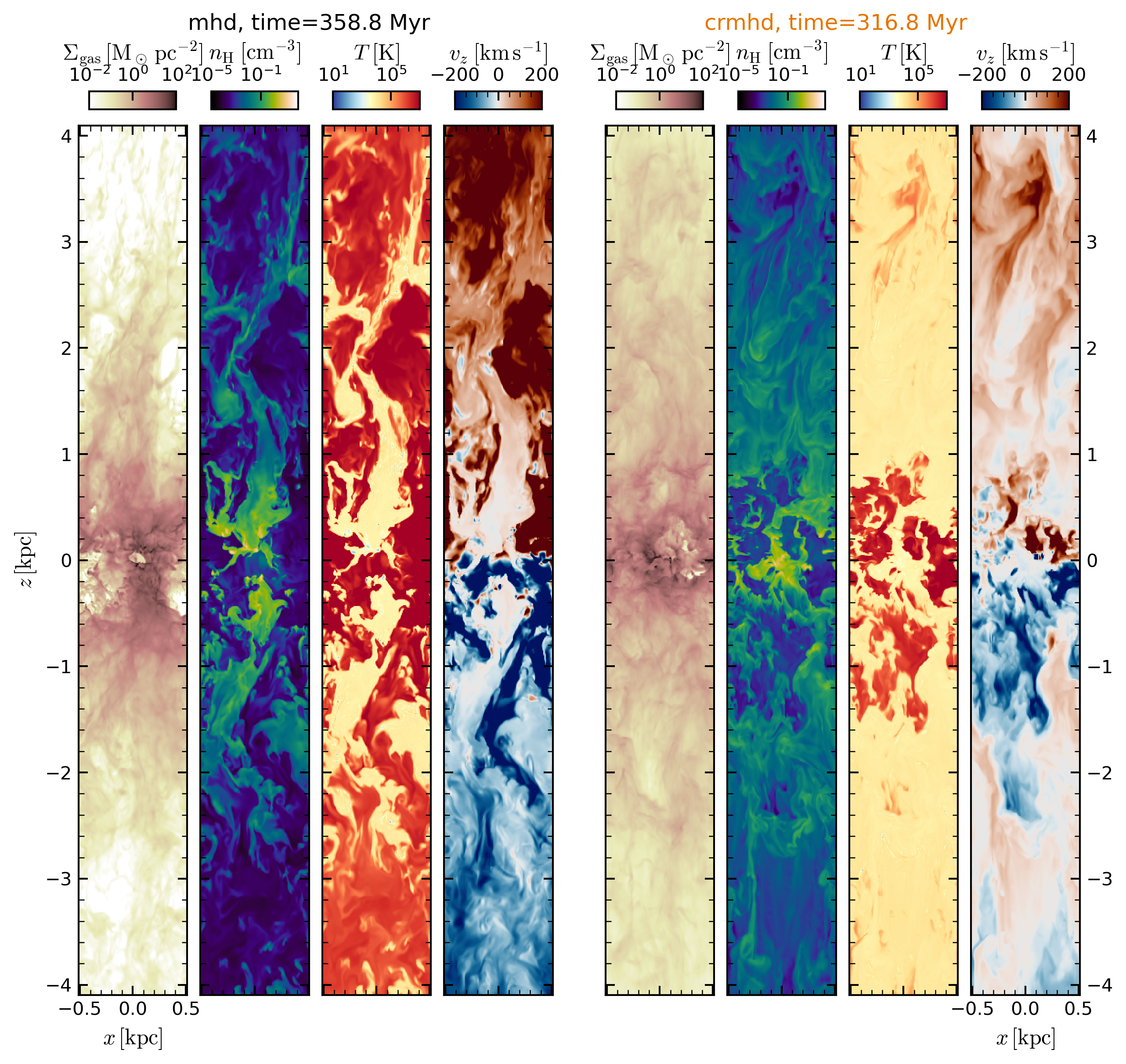}{0.57\textwidth}{(d) Starburst Epoch, Edge-on Views}
    \end{center}
        \caption{Comparison of thermal gas properties for example quiescent (top row; (a) and (b)) and starburst (bottom row; (c) and (d)) epochs as marked in \autoref{fig:history}. {\bf Left column, (a) and (c):} the face-on projection (along the $z$ axis) of gas surface density and midplane slices of hydrogen number density and temperature. Colored circles in the top row represent star clusters younger than $40\Myr$. {\bf Right column, (b) and (d):} the edge-on projection (along the $y$ axis) of gas surface density and $y=0$ slices of hydrogen number density, temperature, and vertical velocity. \label{fig:snapshot_comp}}
\end{figure*}

\begin{figure*}[htb]
    \includegraphics[width=\textwidth]{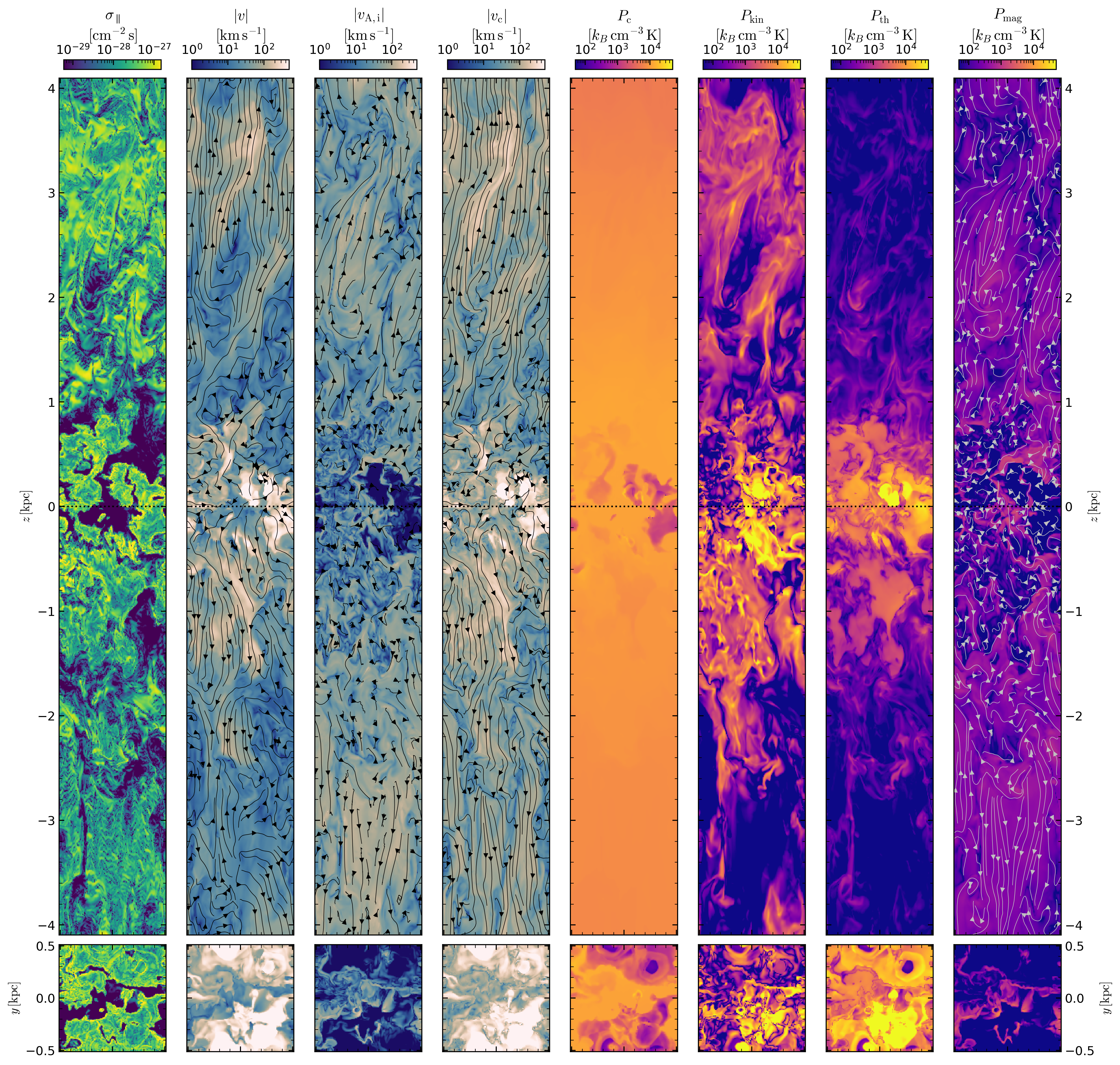}
    \caption{Physical quantities of interest for CR transport in slices through $y=0$ (top row) and $z=0$ (bottom row) from the \crmhd\ model at $t=316\Myr$ (the starburst epoch shown in \autoref{fig:snapshot_comp}). From left to right, we show the parallel scattering rate $\sigma_{\parallel}$, gas velocity magnitude $|v|$ with velocity streamlines, ion Alfv\'en velocity magnitude $|v_{A,i}|$ with its streamlines, effective CR transport speed $|v_{\rm c}| = |\Fcr|/(\ecr+\Pcr)$ with CR flux stream lines, CR pressure $\Pcr$, vertical component of turbulent pressure $\Pturb=\rho v_z^2$, thermal pressure $\Pth$, and magnetic pressure $\Pmag=B^2/(8\pi)$ with magnetic field lines.  \label{fig:snapshot}}
\end{figure*}

\section{Vertical Profiles of Gas and CR properties}\label{sec:ism_cr_trans}

In this section, we shall make extensive use of horizontally-averaged vertical profiles constructed  with phase separation as defined in \autoref{sec:method_def}.
In local simulations adopting horizontally periodic boundary conditions, flows of mass, momentum, and energy are mainly vertical such that the vertical profiles capture key physical processes of outflows. It is still important, however, to keep in mind the full complexity and the high degree of inhomogeneity in the ISM. We thus begin this section by examining example snapshots of physical quantities.

\subsection{ISM Structure Comparison}

The global SFRs examined in the previous section are more or less the same in the \mhd\ and \crmhd\ models, while here we can see that the ISM structure of the two models is significantly different, with larger differences farther from the midplane. \autoref{fig:snapshot_comp} displays the distribution of the thermal gas properties including density, temperature, and vertical velocity.
Sets of midplane (left) and vertical slices (right) are compared.
We choose  snapshots from similar evolutionary stages when both models are in a quiescent, inflow-dominated phase (top row; (a) and (b)) or in a starburst, outflow-dominated phase (bottom row; (c) and (d)).
Both epochs demonstrate the similarities near the midplane and stark differences in extraplanar regions.

For both \mhd\ and \crmhd, the surface density maps in the quiescent epoch of  \autoref{fig:snapshot_comp}(a) show a much more uniform gas distribution with fewer star clusters than those in the starburst epoch of \autoref{fig:snapshot_comp}(c).
In midplane slices of density and temperature,
the quiescent epoch shows more extended warm gas with pervasive cold gas and patchy, isolated hot gas filling less volume than in the starburst epoch. This is because the clusters younger than 10~Myr (shown in magenta) are the main contributors to FUV radiation and hence heating. Also, only significantly clustered SN explosions (as seen from the particle distribution in (c)) can create significant hot outflows  that break out into the extraplanar region. From the edge-on views, \autoref{fig:snapshot_comp}(b) and (d), it is evident that hot gas is almost completely missing in the extraplanar region during the quiescent epoch, while the hot outflow fills significant volume during the starburst epoch in the \mhd\ model. Consequently, for the \mhd\ model the vertical gas flows are mainly inflowing in (b) but outflowing in (d).

\begin{figure*}[ht!]
    \includegraphics[width=\linewidth]{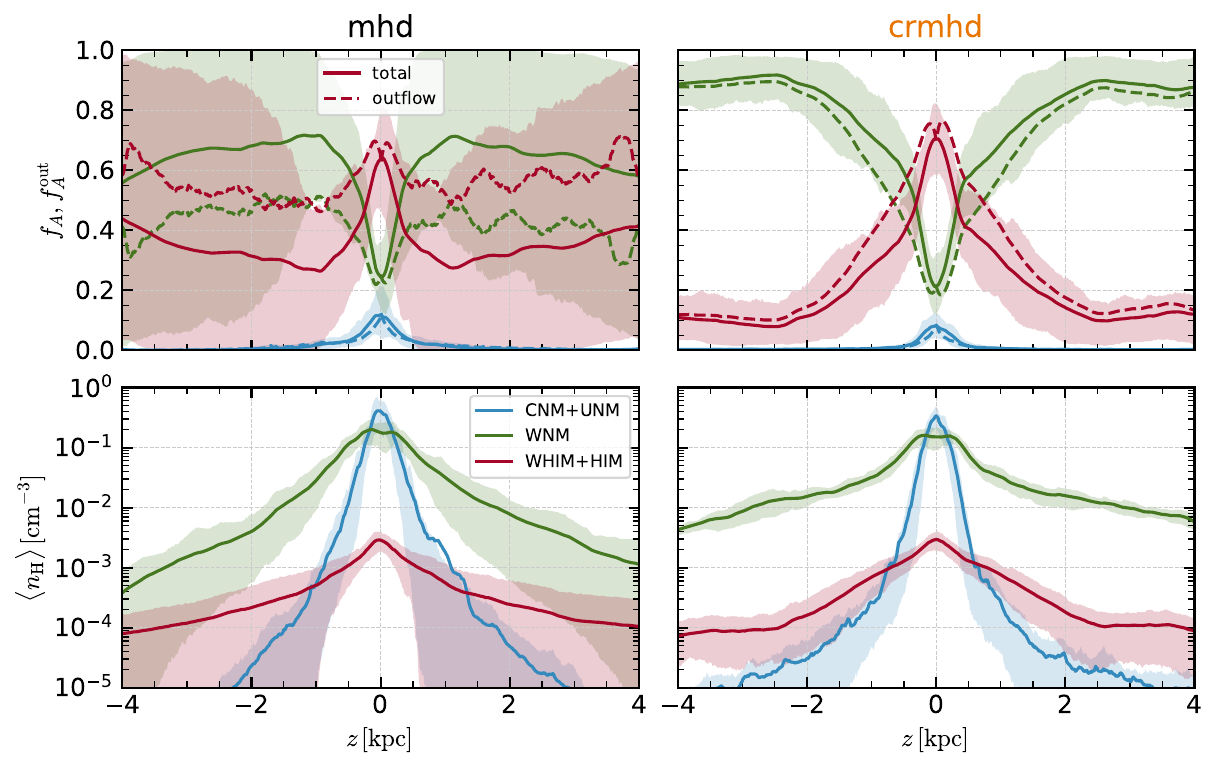}
    \caption{{\bf Top:} Vertical profiles of the volume filling factor of each phase. The solid line denotes the mean after $t=200\Myr$, while the shaded areas denote the range between the 16th and 84th percentiles. The dashed lines show the time-averaged filling factor of the outflowing component. {\bf Bottom:} Vertical profiles of the fractional density of each phase (i.e., the fractional contribution of each phase to the volume-weighted average). At $|z|>500\pc$, the gas in the \crmhd\ model is predominantly \WNM\ by volume and mass, even when selecting just the outflowing component. This is in contrast to the \mhd\ model, in which the hot component dominates the outflowing gas volume.
    \label{fig:mass_volume}}
\end{figure*}

There are also significant differences between the \crmhd\ and \mhd\ model. Two that are clearly evident in \autoref{fig:snapshot_comp} are (1)
during quiescent epochs, vertical flows remain bi-directional (inflowing and outflowing) in the \crmhd\ model; and (2)
during both epochs,
the denser, cooler gas fills a larger volume above $|z|\simgt 1\kpc$ in \crmhd\ compared to \mhd.
During starburst epochs, the hot, fast winds reach all the way to the vertical boundaries in the \mhd\ model, but this is not the case in the \crmhd\ model.
\REV{Clear hot ``breakout'' is lacking in the \crmhd\ model.  This is not due to lack of hot gas creation or burstiness of star formation; \crmhd\ and \mhd\ are similar in both respects  (see \autoref{fig:history}). Rather, in \crmhd\ the hot gas experiences more interaction with the volume-filling warm gas in the extraplanar region (\autoref{fig:mass_volume}), and quickly loses its energy through radiative cooling.
Although the above visual impressions of differences are } drawn from a particular snapshot cutting through a specific slice of the full volume, we shall show that these characteristics hold in general using the time series of the horizontally averaged vertical profiles (\autoref{sec:wind_flux}).

We now turn our attention to the distributions of additional physical quantities of interest for CR transport, focusing on the outflowing epoch of the \crmhd\ model. The first panel of \autoref{fig:snapshot} shows the result of
self-consistent modeling of the CR scattering rate under the assumption that the growth rate and damping rate of the streaming instability balance. Wherever there is high-density gas near the midplane (as seen in \autoref{fig:snapshot_comp}(d)), strong ion-neutral damping results in very low scattering rates, $\sigma_\parallel\ll 10^{-29}\cm^{-2}{\rm\, s}$. Most extraplanar regions where gas is low density (and also the regions of hot gas near the midplane) have weaker damping, and the scattering rate remains much higher, $\sigma_\parallel\sim10^{-28}-10^{-27}\cm^{-2}{\rm\, s}$.  As a consequence, CRs and gas are only weakly coupled in the \twop\ midplane gas, while in low-density warm extraplanar gas and hot gas, CRs and gas are well coupled.

The effective CR transport speed $|v_{\rm c}|\equiv |\Fcr|/4\Pcr$ is shown in the fourth column, and for GeV CRs is primarily set by advection and streaming (with their characteristic speeds, $|v|$ and $|v_\mathrm{A,i}|$ shown in the second and third columns, respectively).
In the hot gas (see temperature slice in \autoref{fig:snapshot_comp}(d)),
the CR transport speed is dominated by advection. The contribution of CR streaming
is more important in the warm, low-density gas, and especially notable for $z<-2\kpc$ in the snapshot shown. In the low scattering rate region near the midplane, CR diffusion is efficient, resulting in especially uniform CR pressure within the warm and cold gas. More generally, the efficient transport of CRs along magnetic fields makes the CR pressure (shown in the fifth column of \autoref{fig:snapshot}) quite smooth overall.  This is in stark contrast to all the other pressures, which have large fluctuations at all scales, as shown in the sixth to eighth columns.

\REV{When measured across horizontal directions at a given $z$, the quantitative level of pressure fluctuations with respect to the mean, $\delta P^2\equiv \abrackets{P^2}- \abrackets{P}^2$,  are $\delta P/\abrackets{P}\sim 1-2$ for turbulent, thermal, and magnetic pressures. For CRs, $\delta P_{\rm cr}/\abrackets{P_{\rm cr}}\lesssim 0.1$ in the extraplanar region, while ranging up to $\delta P_{\rm cr}/\abrackets{P_{\rm cr}}\sim 0.5$ in the midplane, mainly due to CR pressure variations across gas phases.
In the extraplanar region, density fluctuation are present within the warm gas of both \crmhd\ and \mhd, which are inherent in the turbulent, supersonic, SN-driven multiphase outflows. At the same time, in the extraplanar warm gas the magnetic pressure is generally larger than thermal pressure (i.e., $\beta\equiv \Pth/\Pmag<1$) for the \crmhd\ model (see \autoref{fig:snapshot}); $\beta<1$ for the \mhd\ model as well in the extraplanar warm gas. This condition is  favorable for CR acoustic instability \citep{1994ApJ...431..689B,2022MNRAS.513.4464T,2022MNRAS.510..920Q,2026ApJS..283...37Z} and the bottleneck effect \citep{2017MNRAS.467..646W,2021ApJ...913..106B} to develop. While space does not permit in the present work, it is of great interest to investigate further to what extent CR driven dynamical process may contribute to observable structure in outflows.  
}

\begin{figure*}[htb!]
    \includegraphics[width=\linewidth]{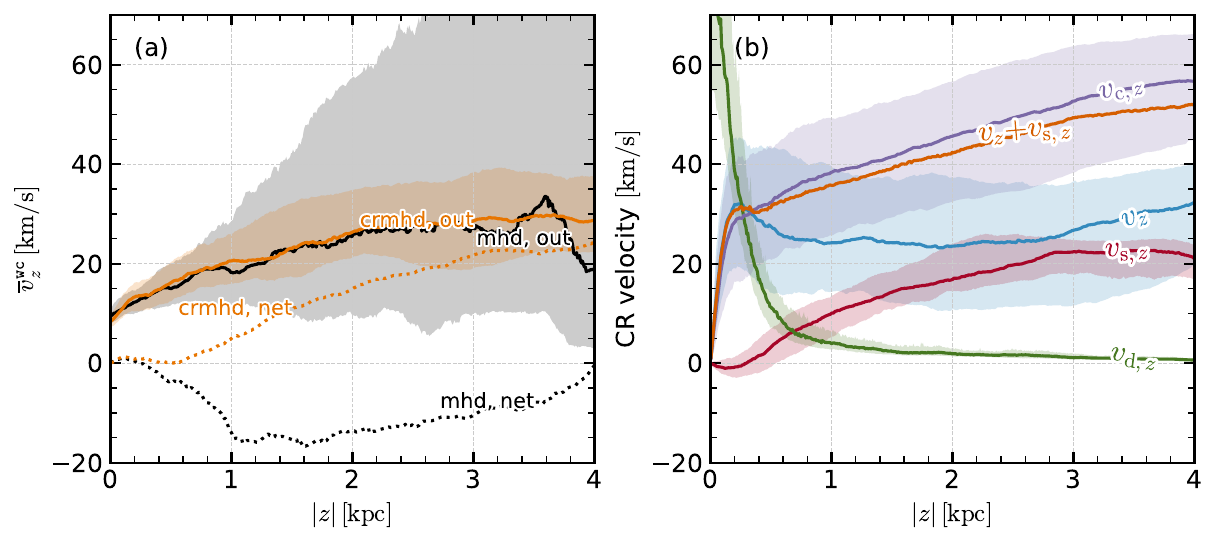}
    \caption{{\bf Left:} Vertical profiles of vertical gas velocities of the \twop\ phase based on both the upper and lower halves of the disk;
    the velocity in the lower half is multiplied by $-1$ before combining the distributions.
    We separately show the median velocities of only outflowing gas (solid lines) and all gas (dotted lines). {\bf Right:} Vertical profiles of vertical advection velocity (blue), streaming velocity (red), diffusion speed (green), and the effective CR transport velocity (purple). The sum of advection and streaming velocity is shown as vermilion.
    All horizontal averages are weighted by CR pressure and include all gas (without phase or inflow/outflow separation). In both panels, shaded regions denote the 16th to 84th percentiles of the temporal variations.
    \label{fig:cr_velocity}}
\end{figure*}

\begin{figure*}[htb]
    \includegraphics[width=\linewidth]{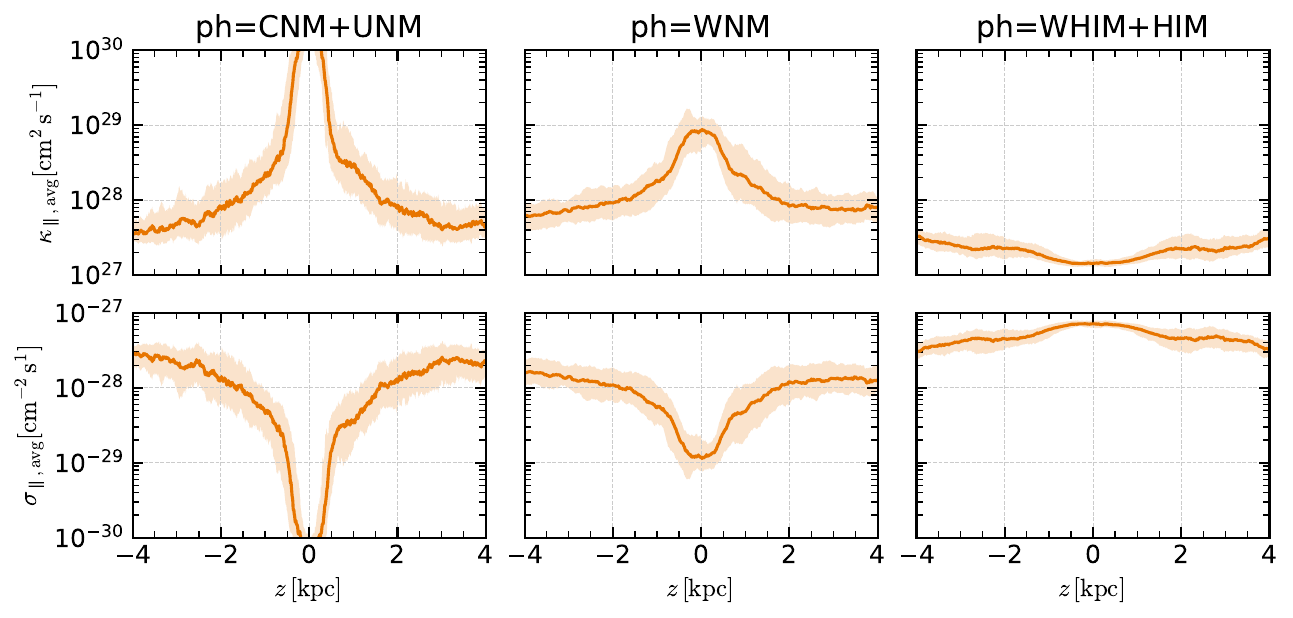}
    \caption{{\bf Top:} The effective CR diffusion coefficient, defined by the CR pressure gradient weighted average (\autoref{eq:kappa_avg}). Profiles represent horizontal averages, and we bin into three thermal components
    by separating the \twop\ phase into \CNM+\UNM\ (left) and \WNM\ (middle). {\bf Bottom:} The effective scattering rate, obtained from the inverse of $\kappa_{\parallel, \rm avg}$. \label{fig:cr_kappa}}
\end{figure*}

\begin{figure*}[htb]
    \includegraphics[width=\linewidth]{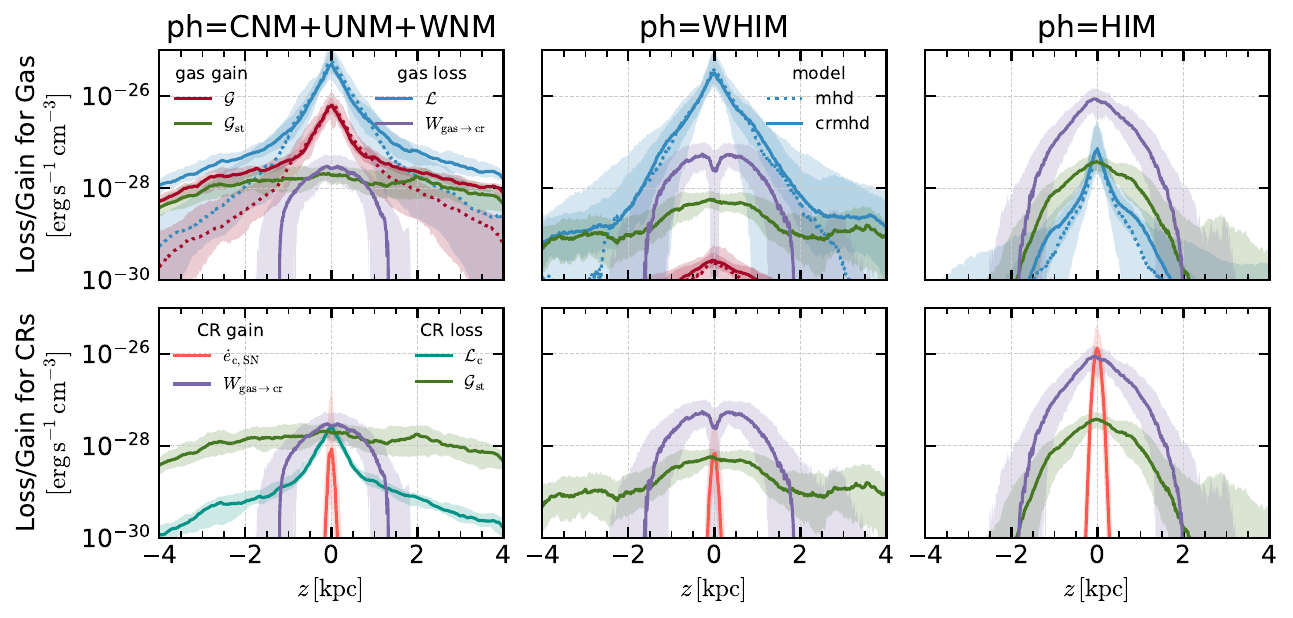}
    \caption{Decomposition of energy gain and loss terms in the \mhd\ (dotted) and \crmhd\ (solid) models. Profiles represent horizontal averages, and we bin into three thermal components
    by separating the \hot\ phase into \WHIM\ (middle) and \HIM\ (right).
     {\bf Top:} Energy gain and loss terms for the gas.   We show the radiative cooling and heating rate per volume in blue and red, respectively, while the gravity work term is omitted (and is small).  The CR streaming heating (\autoref{eq:cr_st_heating}; gain for the gas, and loss for the CR) and the gas work (\autoref{eq:gas_work}; loss for the gas, and gain for the CR) are shown in green and purple, respectively, in both rows. {\bf Bottom:} Energy gain and loss terms for the CRs. The CR energy injection rate by SNe $\dot{e}_{\rm c,SN} \equiv \scr/t_{\rm dec}$ is shown in coral, and the collisional CR loss term is shown in teal, which is only visible for the \twop\ phase (\CNM+\UNM+\WNM) in the left column.
    \label{fig:cr_coolheat}}
\end{figure*}

\subsection{Gas Phase}

To provide a view of the change in the overall properties of the multiphase ISM from midplane to extraplanar regions, \autoref{fig:mass_volume} plots vertical profiles of the phase-separated area filling factors $f_A^{\rm ph}=\abrackets{1}^{\rm ph}$ and contributions $\abrackets{\nH }^{\rm ph}$ from each phase to the total density. (Recall from \autoref{eq:mean} that the typical value of the density \textit{within} a given phase  is $\overline{n}^\mathrm{ph} = \abrackets{\nH }^{\rm ph}/f_A^{\rm ph}$.) In the \crmhd\ model, the \WNM\ phase dominates in the extraplanar regions $|z|>500\pc$ in terms of both mass and volume, while in the \mhd\ model, the \hot\ (\WHIM+\HIM) contribution to the volume is comparable to that of \WNM.\footnote{We note that the phase filling factors in the \mhd\ model are somewhat different from what we obtained in \classic\ and \ncr, shown in \citet{2020ApJ...897..143K} and \citet{2024ApJ...975..173L}, respectively. Two major physics elements not included
in the current \mhd\ model compared to the previous ones are runaway SNe and galactic differential rotation; there are also numerical differences in the way gas mass and energy are returned from star particles (see \autoref{sec:sink}). Tests show that including runaway SN does not significantly change $f_A^{\hot}$ for \mhd, while galactic differential rotation and varying mass return parameters can make more of a difference to $f_A^{\hot}$; this will be discussed in a future publication.
We find, however, that the \crmhd\ model is less sensitive to these physical and numerical elements.
}
These differences are amplified when comparing the filling factor of the outflowing component. However, the \crmhd\ model shows only a small difference between $f_A$ and $f_A^{\rm out}$ (higher \hot\ phase filling factor in the outflows), while the majority of the outflow volume is filled with the \hot\ gas in the \mhd\ model.
The temporal fluctuations of the filling factors are also much smaller in the \crmhd\ model.
The \WNM\ dominates the density of extraplanar gas in both models, with a steeper decrease of the density profile in the \mhd\ model at large $|z|$. For the \crmhd\ model, density profiles of \WNM\ taper off slowly at large $|z|$ due to CR-driven warm outflows.
The density profiles of \twop\ gas similarly drop off slowly with $|z|$ in the CRMHD simulations of \citet[][see Fig. 18 and 19 there]{2024ApJ...964...99A}, although the overall normalization there is lower by a factor $\sim 3-6$, due to the lower CR momentum flux injection rate (see further discussion in \autoref{sec:transfer}).

\subsection{Vertical Velocities}\label{sec:cr_velocity}

\autoref{fig:cr_velocity}(a) shows the vertical velocity in the \twop\ phase for outflowing gas only (solid) and all gas (dotted). The vertical gas velocity of outflowing gas is more-or-less the same for both models,
increasing with $z$ in a similar fashion.
When considering the net velocity, the alternating inflows and outflows in the \mhd\ model during the chosen time window ($t=200-500\Myr$)
do not cancel out, leading to net negative velocity (dotted black)
and large temporal fluctuations in the outflowing velocity (shaded gray). In contrast, outflows are persistent in the \crmhd\ model, leading to
net positive velocity at $|z|\gtrsim0.5$kpc (dotted orange) with modest temporal fluctuations of the outflow (shaded orange).
The net outflow velocity profiles of \twop\ gas found in the most comparable CRMHD simulations of \citet[][see models WW-ID and WW-HD in Figs. 18 and 19 there]{2024ApJ...964...99A} are similar to the net velocity profile in the \crmhd\ model here.

Although the velocity profile of outflowing gas is similar for both models, the physical reason for the velocity increase is different, as we shall explain via quantitative analysis in
\autoref{sec:transfer}.
The mean velocity increase in the \mhd\ model is mainly due to dropout of the low velocity portion of the distribution, as has been shown in our previous work \citep{2018ApJ...853..173K,2020ApJ...900...61K,2020ApJ...894...12V}. In the \crmhd\ model, however, the \twop\ outflow is actually accelerated (or deceleration by gravity is greatly reduced, as argued in \citealt{2018MNRAS.479.3042G}). In \autoref{fig:cr_velocity}, additional acceleration can be deduced from the reduced difference in velocities between outflowing and all gas in the \crmhd\ model.

\autoref{fig:cr_velocity}(b) shows the characteristic velocities for CR transport.
The  steady state CR flux is obtained as the solution of \autoref{eq:steady_flux}, such that $\Fcr$ is decomposed as
\begin{equation}\label{eq:steady_flux2}
    \Fcr \approx \Fadv + \Fst + \Fd
\end{equation}
where the advection, streaming, and diffusion fluxes are $\Fadv\equiv (\ecr + \Pcr)\vel$, $\Fst\equiv(\ecr+\Pcr)\vel_s$, and $\Fd \equiv (\ecr+\Pcr)\vel_d$
with the diffusion velocity
\begin{equation}\label{eq:vdiff_def}
    \vel_d  \equiv - \frac{1}{\ecr + \Pcr}
   \rttensor{\sigma}^{-1}
    \cdot{\nabla \Pcr},
\end{equation}
respectively; $ \ecr + \Pcr=4\Pcr=(4/3)\ecr$ for our relativistic equation of state.  The effective local
CR transport velocity can be defined as
\begin{equation}\label{eq:vcr_def}
\velcr\equiv \frac{\Fcr}{\ecr+\Pcr},
\end{equation}
and we therefore have $\velcr\approx \vel +\vel_s +\vel_d$
in regions close to steady state. In regions out of steady state
(with very low scattering rates), the sum of the characteristic velocities exceeds $\velcr$.
For \autoref{fig:cr_velocity}(b), we locally calculate the vertical component of each velocity and take $\Pcr$-weighted averages in the horizontal direction. Note that we take the average of the diffusion speed $|v_{d,z}|$ rather than velocity.

For vertical CR transport out of the disk, a critical quantity is the effective velocity at the interface where the CR-gas coupling is stronger \citep{2025ApJ...989..140A, 2025ApJ...994...45H}. For CRs with $\sim$GeV energy, a transition from diffusion-dominated to dynamically-controlled transport occurs
where the density drops enough that ion-neutral damping is significantly reduced. With only nonlinear Landau damping, wave amplitudes remain sufficiently high that diffusion speeds (green) become smaller than dynamical speeds (blue and red).  In the present simulation, the transition to the strongly-coupled regime (with sub-dominant $v_d$) occurs at $z\sim 0.5-1\kpc$. The measured effective vertical transport velocity for CRs
at $|z|=1$kpc is
$v_{{\rm c},z}\sim 35\kms$ averaged over all gas (purple),
which is indeed comparable to the sum (vermillion) of
advection (blue) and streaming (red), $v_{z}\sim 25\kms$ and $v_{{\rm s},z}\sim 10 \kms$.

While streaming is quite important to transport within \twop\ gas, \autoref{fig:snapshot} makes clear that the CR transport in the hot gas is completely dominated by advection (and CRs are always well coupled).

\subsection{CR Diffusion Coefficients}\label{sec:diffusion}

The CR gradient weighted diffusion coefficient at a given $z$ is defined by
\begin{equation}\label{eq:kappa_avg}
    \kappaavg \equiv \frac{\abrackets{|F_{d,\parallel}|}}{\abrackets{|\nabla_\parallel \Pcr|}}
\end{equation}
where  $F_{d,\parallel} \equiv - \nabla_\parallel\Pcr/\sigma_\parallel$ is the steady-state diffusive flux with $\nabla_\parallel\Pcr \equiv \hat \Bvec\cdot\nabla\Pcr$.
\autoref{fig:cr_kappa} shows the profiles of the mean diffusion coefficient (top row) for three different phases \CNM+\UNM, \WNM, and \hot\ (\WHIM+\HIM) along with the inverse of it (bottom row) to show the mean scattering rate.
The mean diffusivity is highest at the midplane in both \CNM+\UNM\ and \WNM\ because this is where damping is strongest. The peak value in the \WNM\ is $\kappaavg\sim 10^{29}\cm^2{\rm \,s}^{-1}$ and decreases outward, approaching $\kappaavg\approx 10^{28}\cm^2{\rm \,s}^{-1}$. In the colder phase, the diffusion coefficient gets much larger, implying complete decoupling of CRs from the gas. In the \hot\ phase, $\kappaavg$ slightly increases with altitude reaching values of a few $\times 10^{27}\cm^2{\rm \,s}^{-1}$. Overall, values of $\kappaavg \sim 10^{27}-10^{28}\cm^2{\rm \,s}^{-1}$ in the extraplanar region are consistent with the findings from our previous post-processing simulations \citep[e.g.,][]{2022ApJ...929..170A,2025ApJ...989..140A}.

\subsection{Energy Gain and Loss}\label{sec:source_sink}

We now investigate the spatial distribution of the individual
energy source and sink terms appearing on the RHS of \autoref{eq:energy_con} and \autoref{eq:CRenergy}. The term arising from the interaction between gas and CRs appears with opposite signs in equations. Omitting the flux divergence term on the LHS, while explicitly including a term for CR energy injection from SNe, the gas and CR energy equations with source and sink terms can be written as
\begin{eqnarray}\label{eq:source_terms}
    \dot{\mathcal{E}}_{\rm tot}+\cdots &=& \mathcal{G} - \mathcal{L} - W_{\rm grav} - W_{\rm gas\rightarrow cr} + \mathcal{G}_{\rm st}\\
    \dot{e}_{\rm c}+\cdots &=& \dot{e}_{\rm c,SN}- \mathcal{L}_{\rm c} + W_{\rm gas\rightarrow cr} - \mathcal{G}_{\rm st}.
\end{eqnarray}
The sign of each term above is chosen so that each term is mostly positive. That is, we find that the work of gravity on the gas, as well as the gas-CR interaction, results in net energy losses from the gas.  The sign of the streaming velocity is such that CR streaming represents an energy gain for the gas. For the thermal gas, we have radiative cooling $\mathcal{L}=\nH^2\Lambda$ and heating $\mathcal{G}=\nH\Gamma$ terms as well as the work term due to gravity
\begin{equation}\label{eq:grav_work}
W_{\rm grav} \equiv \rho \vel \cdot\nabla\Phi.
\end{equation}
We explicitly include the CR gain term by SN injection using the passive scalar
\begin{equation}\label{eq:cr_inj}
\dot{e}_{\rm c,SN}\equiv \scr/t_{\rm dec}
\end{equation}
while the CR loss term is
\begin{equation}\label{eq:cr_cool}
\mathcal{L}_{\rm c} \equiv {\Lambda_{\rm coll}\nH\ecr}.
\end{equation}

The gas and CR interaction terms are separated into the work done by gas on CRs (if positive, this means loss for the gas and gain for the CR)
\begin{equation}\label{eq:gas_work}
W_{\rm gas\rightarrow cr} \equiv{ -\vel \cdot
    \sigmatot \cdot
    \rbrackets{  \Fcr - 4\Pcr\vel }},
\end{equation}
and the gas heating mediated by the damping of Alfv\'en waves excited by streaming instability (if positive, this means gain for the gas and loss for the CR)
\begin{equation}\label{eq:cr_st_heating}
\mathcal{G}_{\rm st} \equiv {
    \vel_\mathrm{s} \cdot
    \sigmatot \cdot
    \rbrackets{  \Fcr - 4\Pcr\vel }}.
\end{equation}
We note that these work terms are calculated based on the actual source terms implemented in the code, per \autoref{eq:CRenergy}.  For reasons of numerical stability \citep[see discussion in][]{2018ApJ...854....5J}, $\sigmatot$ includes both a physical scattering term and a term that ensures that in steady state, \autoref{eq:steady_flux} holds.  As a result, the CR energy source term in steady state is given by \autoref{eq:source_steady}, such that $W_{\rm gas\rightarrow cr} \rightarrow \vel \cdot\nabla \Pcr$ and $\mathcal{G}_\mathrm{st}\rightarrow-\vel_\mathrm{s}\cdot \nabla\Pcr$. In practice, we find that \autoref{eq:gas_work} and \autoref{eq:cr_st_heating} agree with their steady-state values to within $\sim 10\%$.

\autoref{fig:cr_coolheat} plots vertical profiles of the energy gain and loss terms for the gas (top) and CRs (bottom). Here, we separate the \hot\ phase into the \WHIM\ and \HIM\ phases \REV{(middle and right columns, respectively)}, as radiative cooling in the \hot\ phase is completely dominated by the \WHIM. Terms for the \mhd\ model are shown with dotted lines and only appear in the top row. For both models, we omit the gravity work term as it is small compared to the radiative cooling and heating term at all heights.

We first compare radiative cooling and heating terms between the two models. The cooling rates in \WHIM\ are more-or-less the same in both models at $|z|<2\kpc$, but higher at $|z|>2\kpc$ in the \crmhd\ model, where it is balanced by the CR heating.
For the \twop\ phase, cooling is significantly enhanced in the extraplanar region of the \crmhd\ model compared to \mhd, mainly due to the increased density of the \twop\ phase.  The enhanced cooling in the \crmhd\ model is compensated by correspondingly enhanced heating; heating details are discussed below.

For the \crmhd\ model, the two terms providing in-situ gas heating are CR streaming heating (green) and radiation heating (red).  For the \twop\ (\CNM+\UNM+\WNM) component, the radiation heating far exceeds streaming heating in the midplane region, while the two terms become comparable at $|z|>2\kpc$. Radiative gas cooling (blue) largely compensates these heating terms (especially at large $|z|$), with the excess of cooling over heating  ultimately attributable to energy originally from SNe \citep[see][]{2023ApJ...946....3K}.
The cooling in the \WHIM\  is comparable to the cooling in the \twop\ phase, especially near the midplane. In the \WHIM, radiative heating is negligible, mainly because we assume that the photoelectric heating is turned off in the fully ionized gas. Cooling of the \WHIM\ is comparable to streaming heating at $|z|>2\kpc$; a similar result was found in \citet[][see Fig. 22 there]{2024ApJ...964...99A}.  In the midplane region, the excess of cooling over heating in the \WHIM\ ultimately comes from energy deposited by SNe in the ISM, as in the \twop\ gas.
In the \HIM\ phase, radiative cooling is weak due to the low density, while transfer of energy to the CRs by work is quite significant.
The \HIM\ gas loses  energy via adiabatic expansion and mixing with lower temperature gas, followed by radiative cooling (after phase transition to the cooler phases); the result of this is seen as net cooling in \twop\ and \WHIM\ gas.

\REV{In the bottom row, we show two gain and two loss terms for CRs: the CR energy injection rate from SNe 
(red; \autoref{eq:cr_inj}) and the work done by gas on CRs (purple; \autoref{eq:gas_work}); and the CR collisional (blue; \autoref{eq:cr_cool}) and streaming losses (green; \autoref{eq:cr_st_heating}). The majority of the CR energy injection by SNe occurs in the \HIM\ phase near the midplane (bottom right), by design of our algorithm. Gas work adds to the CR energy within $|z|<2\kpc$, again primarily within the \HIM. For both \HIM\ and \WHIM\ the CR energy loss is much smaller than the gain.
In the \twop\ phase (bottom left), on the contrary, the loss terms are much larger than the gain terms. However, the total CR loss across all phases is still much smaller than the total gains from work, leading to outward CR energy fluxes that are significantly larger than what is injected by SNe.}

\REV{
Within the neutral gas layer  $|z|<500\pc$, the total work term contribution is about twice as large as the direct SN injection energy to CRs.  As we shall show in the next section, this enhances
the equilibrium CR pressure at a given star formation rate (CR feedback yield $\Ycr\equiv\Pcr/\Ssfr$).
Additional work at larger $|z|$ further enhances the CR energy loading factor (details will be provided in \autoref{fig:loading_z} and \autoref{fig:cumul_flux}).
Integrating over the whole computational volume, the total work done by gas on the CRs exceeds the CR streaming losses by a factor of 6. However, the flat profile of streaming losses suggests that in reality CR energy and momentum would continue to be transferred from CRs to gas at scales exceeding our computational domain.}

\section{Multiphase Outflows}\label{sec:wind}

Having established that CRs significantly impact the state of the extraplanar gas, in this section we present detailed analysis of multiphase outflows.

\begin{figure*}[htb]
    \includegraphics[width=\linewidth]{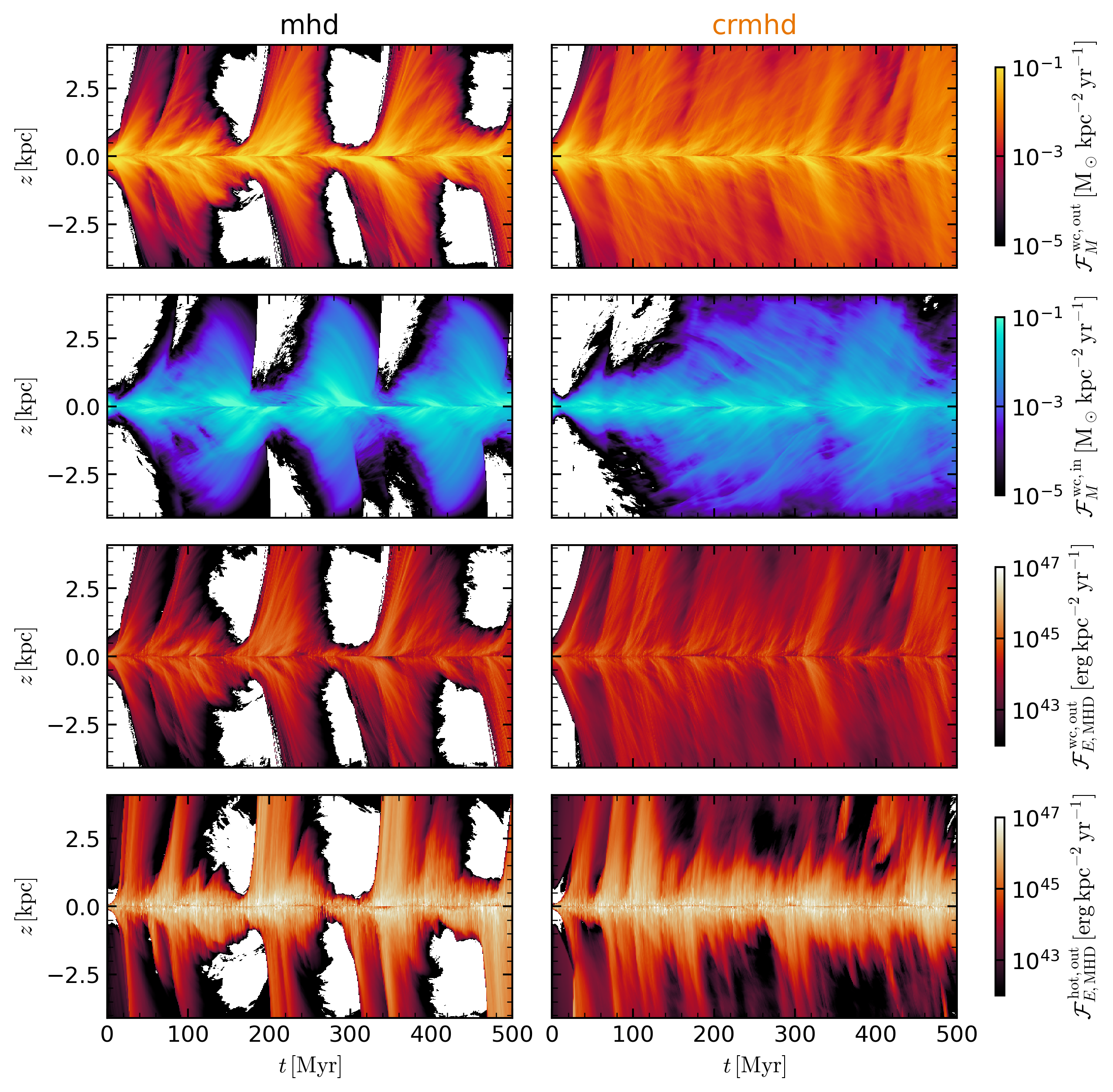}
    \caption{Space-time plots of selected horizontally-averaged fluxes from the \mhd\ (left) and \crmhd\ (right) models. The top two rows respectively show the outward and inward mass fluxes of the \twop\ phase. The bottom two rows are the outflow energy fluxes carried by the \twop\ and \hot\ phases, respectively. The \crmhd\ model shows pervasive simultaneous outflows and inflows of \twop\ gas, while the \mhd\ model is more stochastic, with alternating periods of outflow/inflow. \label{fig:flux_tz}}
\end{figure*}
\begin{figure*}[htb]
    \includegraphics[width=\linewidth]{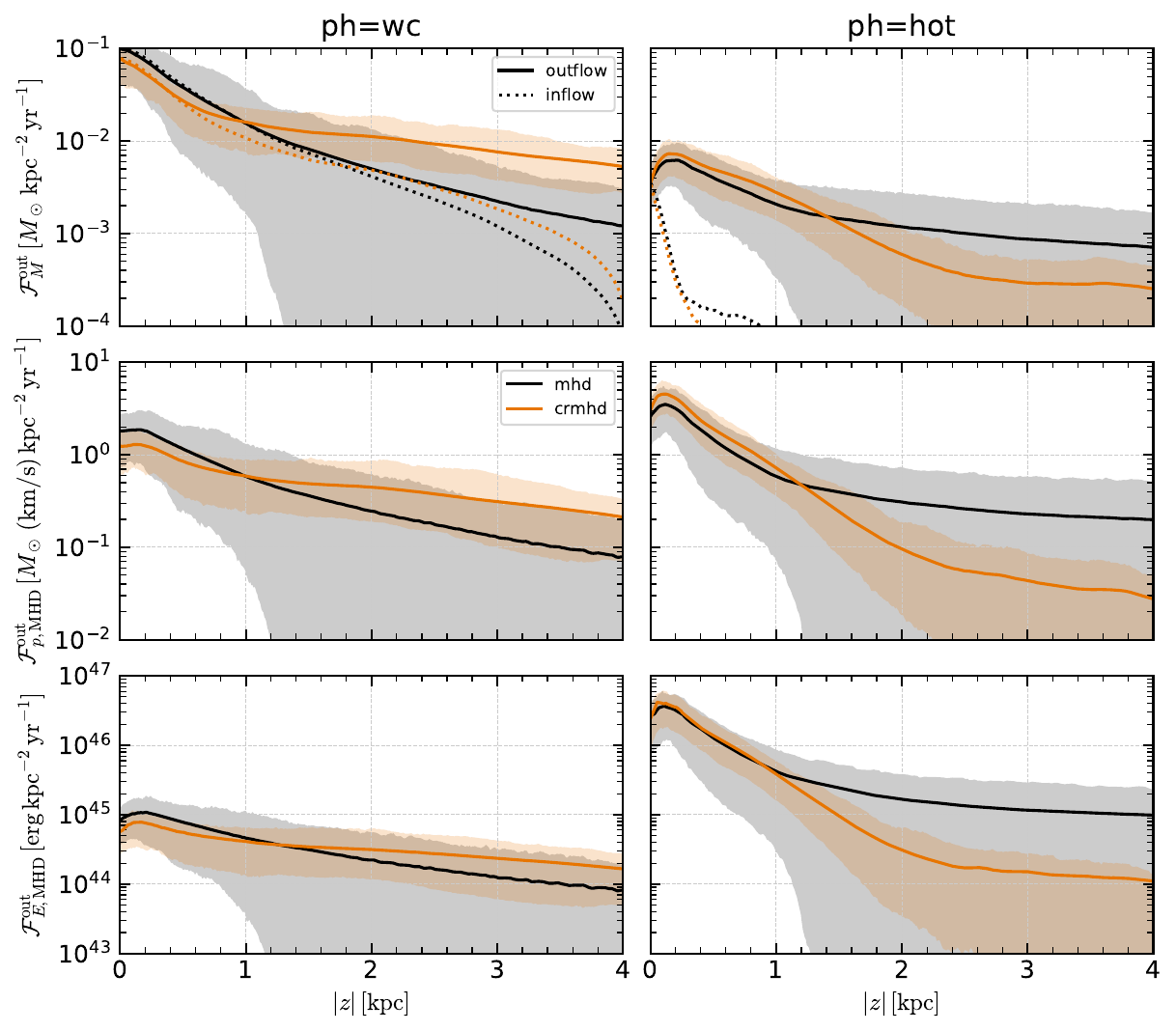}
    \caption{Vertical profiles of outflow fluxes (through both upper and lower halves combined) in the simulation domain. The solid lines denote the mean MHD fluxes (\autoref{eq:mflux}-\autoref{eq:eflux_mhd}) after $t=200\Myr$, while the shaded areas denote the range between the 16th and 84th percentiles. In the top row, we also show the mean inflow mass fluxes with dotted lines.
    \label{fig:flux_out_z}}
\end{figure*}

\subsection{Outflow Fluxes}\label{sec:wind_flux}
We define the vertical mass, momentum, and energy fluxes for the thermal gas as
\begin{equation}
    \mathcal{F}_M =  \abrackets{\rho v_\mathrm{z}}\;,
\label{eq:mflux}
\end{equation}
\begin{equation}
    \mathcal{F}_{p,\mathrm{MHD}} =   \abrackets{\Pturb + \Pth + \Pimag}=\abrackets{\Ptot}\;,
\label{eq:pflux_mhd}
\end{equation}
\begin{equation}
    \mathcal{F}_{E, \mathrm{MHD}} =    \abrackets{(\etot + \Pth + \Pmag) v_z - \frac{(\mathbf{B}\cdot\vel)B_\mathrm{z}}{4\pi}}\;.
\label{eq:eflux_mhd}
\end{equation}
For the CR fluid, the momentum and energy fluxes are simply
\begin{equation}\label{eq:pflux_cr}
    \mathcal{F}_{p,\mathrm{c}} = \abrackets{\Pcr}
\end{equation}
\begin{equation}\label{eq:eflux_cr}
    \mathcal{F}_{E,\mathrm{c}} = \abrackets{F_{{\rm c},z}}.
\end{equation}
All fluxes are separately calculated for each phase and velocity component as in \autoref{sec:method_def}.

\autoref{fig:flux_tz} displays the space-time plots of selected MHD fluxes; we show outward and inward mass fluxes of the \twop\ phase (top two rows), and outward energy fluxes of the \twop\ and \hot\ phases (bottom two rows). On large scales, the mass fluxes in the \mhd\ model are characterized by alternating outflows and inflows, while the \crmhd\ model shows pervasive simultaneous outflows and inflows of the \twop\ phase. In both models, quasi-periodic bursts of star formation (as shown in \autoref{fig:history}(a)) induce large outflow mass fluxes near the midplane, which quickly decrease within the disk scale height. In the \mhd\ model, only the stronger outflows break into the extraplanar region above $|z|>500\pc$,  followed by inflows with returning flux comparable to the recent outflow flux. In the \crmhd\ model, however, there are continuous outflows with no distinct breakout events. Steady inflows are established after $\sim150\Myr$, but the mass flux is smaller than that of the outflows. Overall, in the \crmhd\ model CR feedback results in  steady outflows of the \twop\ phase, leading to the extended distributions of \twop\ gas  consistently seen in the maps and profiles of the previous section (e.g., \autoref{fig:snapshot_comp} and \autoref{fig:mass_volume}).

In the two models, the  energy fluxes injected near the midplane are similar, due to the similar SFRs. In the \mhd\ model, strong bursts of star formation create episodes of hot outflows that break out into the extraplanar region, with energy flux roughly constant with $z$. Breakout events in the \hot\ gas are accompanied by periods of high \twop\ energy and mass outflows, followed by mass inflows of the \twop\ phase. Strong inflows of the \twop\ phase often suppress hot outflows before they reach the boundaries.
The situation is quite different in the \crmhd\ model: the energy flux of hot outflows gradually decreases outward, mostly vanishing above $|z|>2\kpc$. This is mainly due to the interaction between \hot\ phase outflows and the extraplanar \twop\ gas populated by steady outflows that fill significant extraplanar volume (\autoref{fig:mass_volume}). For the \crmhd\ model, the \twop\ phase continuously delivers energy fluxes that only slowly decline with distance, dominating the energy flux from the \hot\ phase far away from the midplane.

\autoref{fig:flux_out_z} plots the time-averaged vertical profiles of outflow fluxes. We combine the fluxes through both the upper and lower horizontal surfaces at a fixed height $|z|$ using the outward velocity $v_{\rm out} =v_z{\rm sgn}(z)$ and outward CR flux $F_{\rm c, out} = F_{{\rm c},z}{\rm sgn}(z)$ in the flux calculations. In the top row, we show both outflow (solid lines) and inflow (dotted lines) mass fluxes. The middle and bottom rows show the momentum and energy fluxes, respectively, of only the outflowing gas.

Within the disk ($z<500\pc$), the outflow and inflow mass fluxes of the \twop\ gas are similar to each other in both models; this reflects the turbulent cycling of gas with a vertical crossing time of a few 10s of Myr. In the \mhd\ model the outflow mass flux keeps dropping at large $|z|$, while in the \crmhd\ model the outflow mass flux flattens out above $|z|\sim 1\kpc$. The inflow mass flux profiles are similar in both models, dropping sharply near the boundaries due to the outflowing boundary condition (see \autoref{sec:app_bc}). The net outgoing mass flux (the difference between outflow and inflow mass fluxes) is thus much larger in the \crmhd\ model than in the \mhd\ model, with five times the rate of mass leaving the box.
The large, steady mass flux of \twop\ gas in the \crmhd\ model is
made possible by the continuous acceleration by CR momentum transfer in the extraplanar region, as we shall discuss further in \autoref{sec:transfer}.  The momentum and energy fluxes of the \twop\ gas are also relatively flat above  $|z|\sim 1\kpc$ for the \crmhd\ model, while they drop off slightly more rapidly for the \mhd\ model.

In the  \mhd\ model,  all \hot\ outflow MHD fluxes flatten out above $|z|\sim 1\kpc$, while in the \crmhd\ model they continue to decrease with the same logarithmic slope
up to $|z|\sim 2\kpc$. The decrease of the \hot\ MHD fluxes in the \crmhd\ model is due to the enhanced interaction with the abundant extraplanar \twop\ gas.
It is also noteworthy that the temporal variation represented by the shaded area (the 16th to 84th percentile over 300 snapshots) is smaller in the \crmhd\ model than the \mhd\ model, reflecting the steadier outflows seen in \autoref{fig:flux_tz}.
Similar to mass fluxes below $|z|<1\kpc$,  MHD momentum and energy fluxes  in this region are more-or-less the same for both phases and both models.

\subsection{Loading Factors}

\begin{figure}[htb]
    \includegraphics[width=\linewidth]{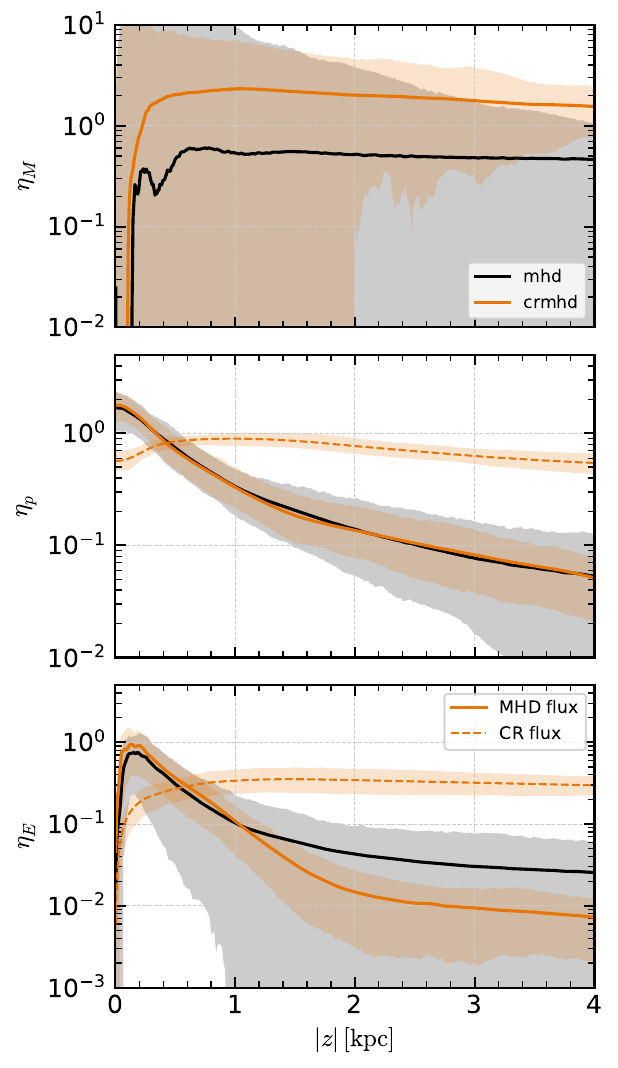}
    \caption{Vertical profiles of loading factors summed up over both upper and lower halves of the disk, all thermal phases, and both inward and outward vertical velocities. Solid lines denote mean values after $t=200\Myr$, while the shaded areas denote the range between the 16th and 84th percentiles. The dashed lines in the momentum and energy loading factor panels (second and third rows) denote the corresponding CR loading factors.
    \label{fig:loading_z}}
\end{figure}

We define the outflow loading factors as the fluxes normalized by the areal star formation rate and momentum and energy injection rates \citep[e.g.,][]{2020ApJ...900...61K}. The reference fluxes are defined using the SN rate
\begin{equation}\label{eq:flux_ref}
    \mathcal{F}_{q, {\rm ref}} \equiv q_{\rm ref}\frac{\dot{N}_{\rm SN}}{L_xL_y}
\end{equation}
where $q_{\rm ref}$ is the reference value of mass, momentum, and energy for each SN event. It is customary to use $M_{\rm ref} = m_* = 95\Msun$, the total mass of new stars formed per SN (determined from the adopted IMF and population synthesis models), making $\mathcal{F}_{M, {\rm ref}}=\Sigma_{\rm SFR,40}$ for the reference mass flux. In practice, we use $\Sigma_{\rm SFR,40}/m_*$ for the areal SN rate $\dot{N}_{\rm SN}/L_xL_y$ rather than the actual measured SN rate; the two are almost identical.
The resulting reference mass fluxes for the \mhd\ and \crmhd\ models are $3.9\times10^{-3}\sfrunit$ and $3.6\times10^{-3}\sfrunit$, respectively.
For the energy flux, the reference energy is the adopted SN explosion energy, $E_{\rm ref} = E_{\rm SN}= 10^{51}\erg$. The resulting reference energy fluxes for the \mhd\ and \crmhd\ models are $4.2\times10^{46} \erg\ \kpc^{-2}\yr^{-1} $ and $3.8\times10^{46}\erg\ \kpc^{-2}\yr^{-1}$, respectively. For the momentum, we use a characteristic momentum at the end of the energy conserving stage of a SN remnant evolution, $p_{\rm ref} = E_{\rm SN}/v_{\rm cool} = 1.25\times10^5\Msun\kms$ with $v_{\rm cool}=200\kms$, which is close to the actual momentum injected to the ISM \citep[e.g.,][]{2015ApJ...802...99K}. The resulting reference momentum fluxes for the \mhd\ and \crmhd\ models are $5.2 \Msun \kms \kpc^{-2}\yr^{-1} $ (equivalent to $2.5\times10^4 k_B \pcc \Kel$ in pressure units) and $4.8\Msun\kms \kpc^{-2}\yr^{-1}$ (equivalent to $2.3\times10^4 k_B \pcc \Kel$ in pressure units), respectively. For the CR loading factors, we use the same reference momentum and energy fluxes as for the thermal gas.

We adopt the time averaged SFR with a 40~Myr time bin to define the reference fluxes.
In principle, one can define the reference fluxes and loading factors in a fully time dependent manner, but connecting the instantaneous injection (or star formation) rates with the resulting fluxes is non-trivial due to time delays.\footnote{To first order, the fluxes measured farther from the midplane where star formation and feedback injection occur are more delayed, but there is no clear one-to-one correspondence as there are multiple complications, e.g., failed outflow launching, interaction with fountain flows, and extended star formation and feedback that launches one big outflow. See the Appendix of \citealt{2020ApJ...900...61K}. Given the large temporal fluctuations in both measured and reference fluxes, taking ratios without proper accounting of time delays sometimes causes erroneous results.} Since our simulations are local, SFRs and associated injection rates do not experience significant secular evolution, as seen in \autoref{fig:history}.

In \autoref{fig:loading_z}, we plot the net mass, momentum, and energy loading factors using the combined fluxes through both the upper and lower sides (without the thermal phase and inflow/outflow distinctions of \autoref{fig:flux_out_z}).
The \crmhd\ model shows on average $\eta_M \sim 1.5-2.2$ beyond $|z|=1\kpc$, dominated by the \twop\ phase. The mass loading factor in the \mhd\ model also shows net outflows with $\eta_M\sim 0.5$.
The MHD momentum loading factors are almost the same for both models.
The energy loading factor is $\eta_{\rm E} \sim 0.1$ at $|z| \sim 1\kpc$ in both models, decreasing more slowly in the \mhd\ model to
$\eta_{\rm E}\sim 0.026$
at $|z|=4\kpc$, and more steeply in the \crmhd\ model to
$\eta_{\rm E}\sim 0.007$ at $|z|=4\kpc$.

In the \crmhd\ model, the CR momentum flux $\mathcal{F}_{p,\mathrm{c}}=\Pcr$ has a local minimum at the midplane when averaged over all phases, because of the low CR pressure within the interiors of superbubbles, which are concentrated in the midplane.  When considering just the \twop\ gas, the CR pressure profile is essentially flat at $|z|\lesssim 0.5\kpc$ (see the $\Pcr$ panel of \autoref{fig:pressure_t}).
Given the steady decline in the MHD momentum flux, the CR momentum flux becomes larger than that of the thermal gas above $z\sim 0.5\kpc$, with a maximum momentum loading factor of $\eta_{p,\rm{CR}}\sim 0.9$, declining very slowly to $\sim 0.5$ at $|z|=4\kpc$.

The CR energy flux increases at first steeply with $|z|$, and then more gradually, surpassing the MHD energy flux at $|z|\sim 0.5\kpc$. As shown in \autoref{fig:cr_coolheat}, the work term appears as a gain for the CRs (and loss for the thermal gas) within $|z|<2\kpc$; this is the origin of the gradual increase in $\mathcal{F}_{E,\mathrm{c}}$.   Beyond
$|z|\sim 1\kpc$, $\mathcal{F}_{E,\mathrm{c}}$ is quite flat, with a loading factor $\eta_{E,CR} \sim 0.3$.
Thus, even though only 10\% of the SN energy is injected as CRs, due to significant transfers from  the thermal gas as well as cooling of the gas, the  majority of the outflowing energy and momentum flux  beyond $|z|=0.5\kpc$ is carried by CRs.

\begin{figure*}[htb]
    \includegraphics[width=\textwidth]{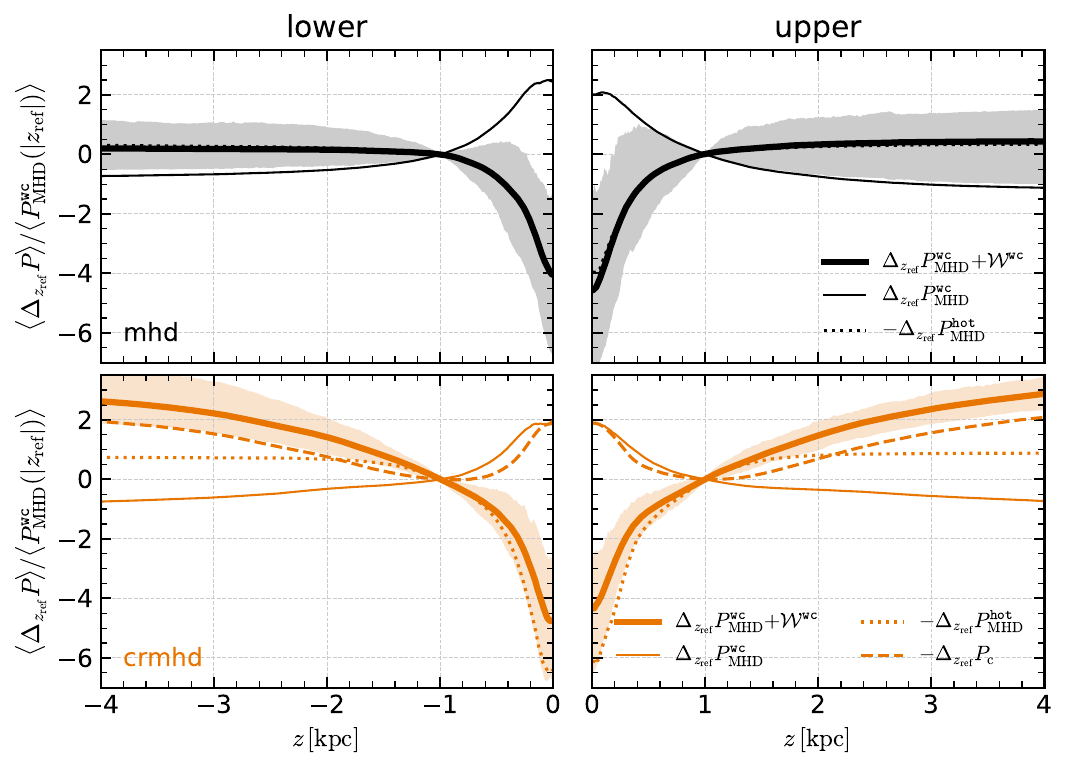}
    \caption{Momentum flux difference profiles of the lower (left) and upper (right) half of the disk
    for both the \mhd\ and \crmhd\ models (top and bottom, respectively). Thin solid lines show the MHD flux itself (i.e., total MHD stress), which decreases at increasing $|z|$ in both models. Thick solid lines show the net momentum gain of the \twop\ phase obtained by adding the weight contribution (LHS of \autoref{eq:mom_transfer}).  Dotted lines show the transfer to the \twop\ phase of MHD flux lost by the \hot\ phase. The momentum flux transfer from CRs is shown as dashed lines for the \crmhd\ model; this significantly boosts the momentum added to \twop\ gas in the extraplanar region.
    All profiles are normalized by the mean MHD flux of the \twop\ gas at $z_{\rm ref}=
    1\kpc$, $\abrackets{\Ptot}^{\twop}(\zref)=2.7\times10^3$ and $2.0\times10^3k_B\pcc\Kel$ for \mhd\ and \crmhd, respectively.
    \label{fig:dflux}}
\end{figure*}

\begin{figure*}[htb]
    \includegraphics[width=\linewidth]{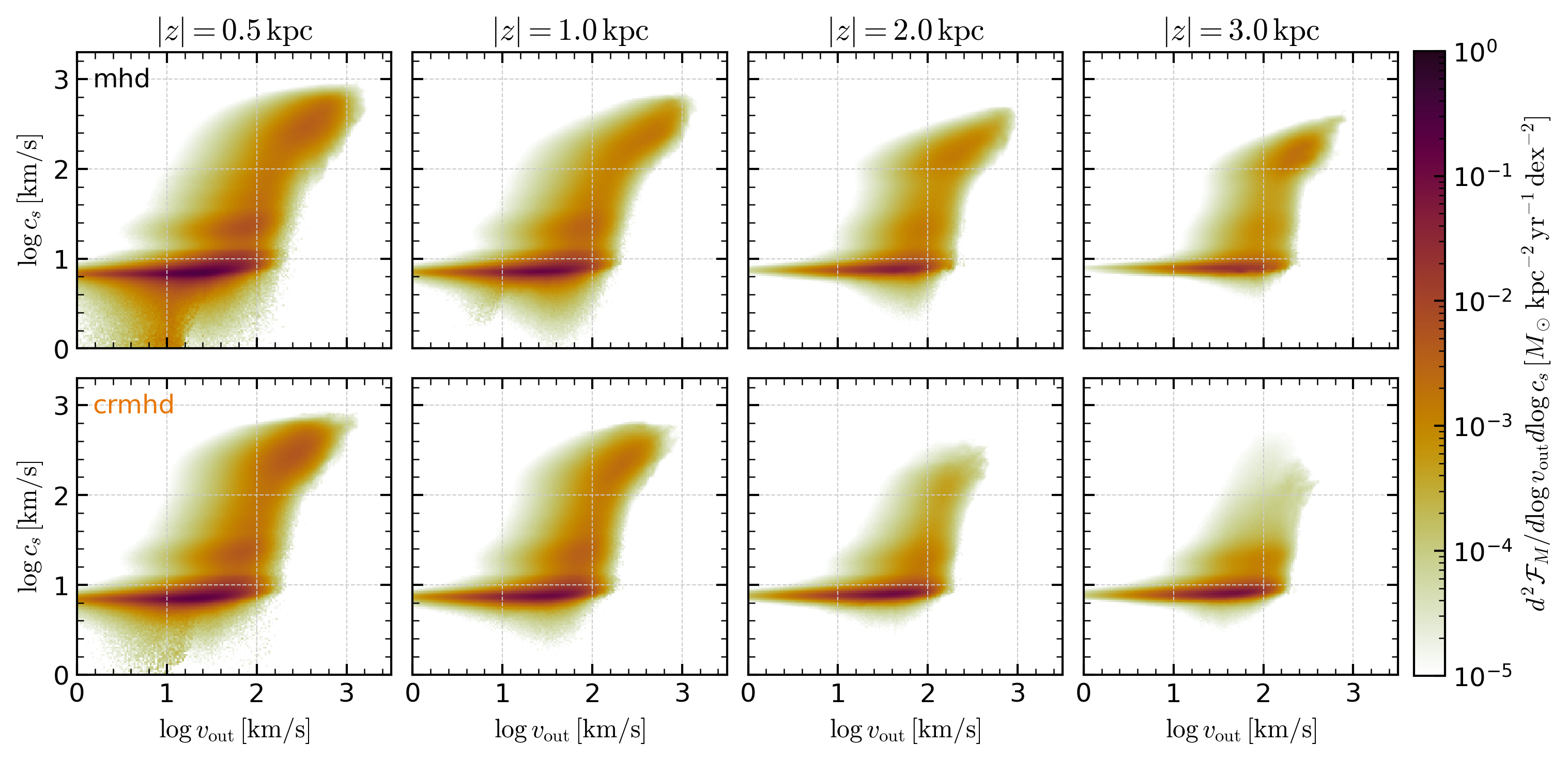}
    \caption{Time averaged mass flux distributions (joint PDFs) of outflow velocity and sound speed, at different distances $|z|$ from the midplane. Top row shows \mhd\ model and bottom row shows \crmhd\ model.
    \label{fig:jointpdf}}
\end{figure*}

\begin{figure*}[htb]
    \includegraphics[width=\linewidth]{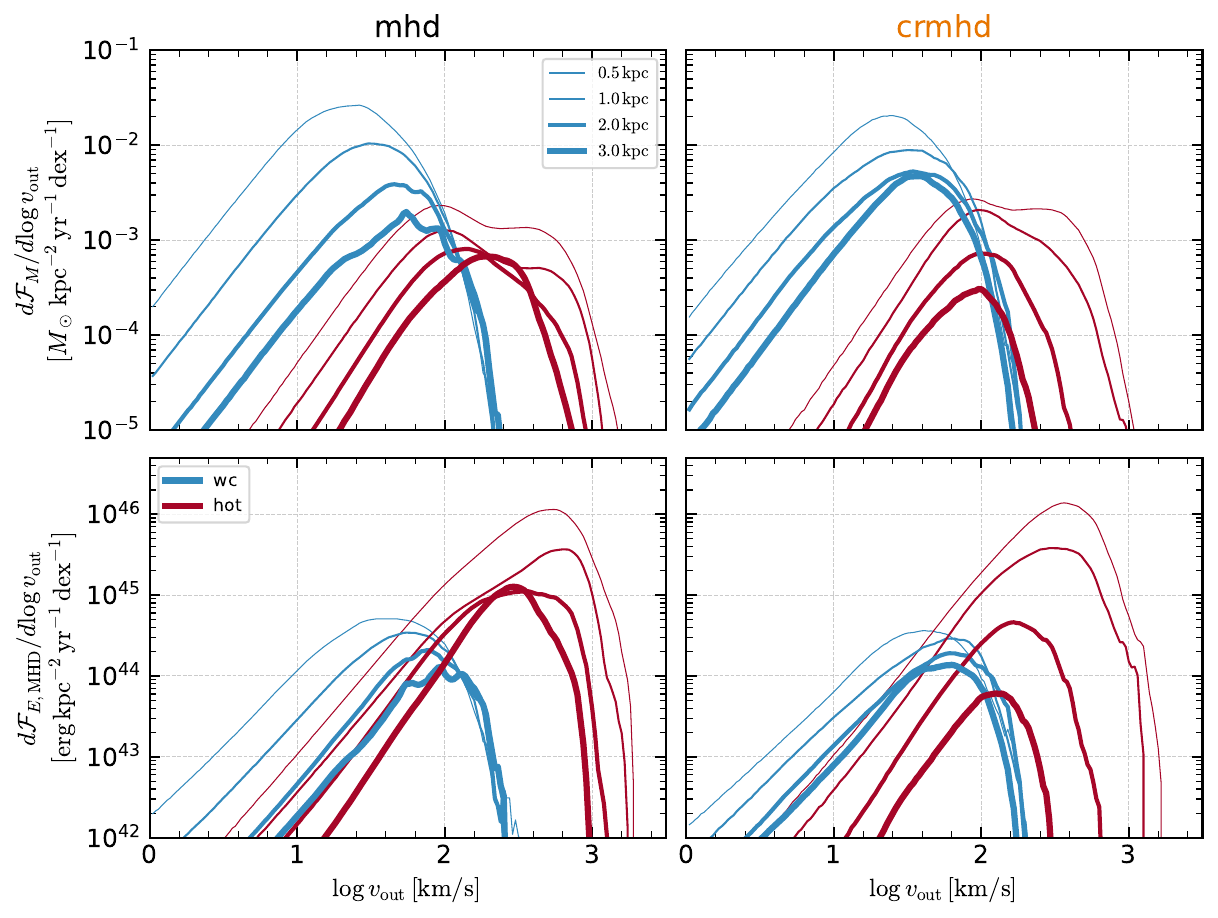}
    \caption{Marginalized {\bf mass (top)} and {\bf energy (bottom)} flux distributions as a function of the outflow velocity for the {\bf wc (blue)} and {\bf hot (red)} phases. \label{fig:voutpdf}}
\end{figure*}

\subsection{Momentum Transfer}\label{sec:transfer}
In this section, we analyze momentum transfer as a function of $z$
to understand how the \twop\ phase gains or loses momentum flux by interacting with the \hot\ gas, with CRs, and with the gravitational potential. We decompose the momentum flux of the thermal gas (i.e., the total vertical stress $\Ptot$) into the \twop\ and \hot\ phases and rearrange the integrated steady state momentum equation (similar to \autoref{eq:vde_tot}, but integrated from $\zref$ to $z$) such that
\begin{equation}\label{eq:mom_transfer}
    \Delta_{\zref} P_{\rm MHD}^{\twop}(z) + \Delta_{\zref}\mathcal{W}(z) =
    -\Delta_{\zref} P_{\rm MHD}^{\hot}(z) - \Delta_{\zref} \Pcr(z)
\end{equation}
Here, the pressure difference at a height $z$ with respect to a reference height $\zref$ is $\Delta_{\zref} P_{\rm MHD}^{\rm ph}(z) \equiv \abrackets{\Ptot}^{\rm ph}(z)-\abrackets{\Ptot}^{\rm ph}(\zref)$; note the difference in signs compared to \autoref{eq:vde_tot}. The weight difference is $\Delta_{\zref}\mathcal{W}(z) = \mathcal{W}(z) - \mathcal{W}(\zref)$
with the weight definition in \autoref{eq:vde_tot}. We adopt $z_{\rm ref}=\pm1\kpc$ for the upper and lower half of the disk, respectively.
We confirm that the steady state condition is satisfied on average at every $z$ (i.e., LHS is equal to RHS).
\autoref{eq:mom_transfer} explains that in a steady state, any net momentum flux transfer to the \twop\ phase (\textit{including} what is used in climbing out of the gravitational potential well) is at the expense of momentum flux losses from the \hot\ phase and CR fluid.

\autoref{fig:dflux} presents the flux difference profiles term-by-term for both the lower and upper half of the disk. The profiles are normalized by the mean of total stress of the \twop\ phase at $\zref=\pm1\kpc$, $\abrackets{\Ptot}^{\twop}(\zref)=2.7\times10^3$ and $2.0\times10^3k_B\pcc\Kel$ for \mhd\ and \crmhd, respectively.
The results from the \mhd\ model (top row) are consistent with our previous analysis in \citet{2020ApJ...894...12V}. The MHD momentum flux of the \twop\ phase itself (the first term of LHS; thin solid line) secularly decreases with increasing $|z|$,  mainly due to gravitational deceleration. By adding in the gravitational weight contribution, the \emph{net} flux difference (the entirety of LHS; thick solid line) increases outward, i.e., this shows the momentum flux that the \twop\ would have gained in the absence of gravity. The momentum flux loss from the \hot\ phase (RHS in the absence of the CR term), shown as the dotted line, is consistent with the thick solid line; for \mhd\ the net momentum flux gain of the \twop\ phase can be fully explained by momentum flux loss of the \hot\ phase.
For the \mhd\ model, the majority of the transfer of momentum flux out of the \hot\ gas occurs at $|z|\lesssim 0.5\kpc$, since this is where the interactions between phases is most extreme.
The level of the flux transfer between phases in the extraplanar region $|z|>1\kpc$ is about 30\% of the reference flux at $|\zref|=1\kpc$.

For the \crmhd\ model, the cumulative net momentum flux gain (thick solid line) for the \twop\ gas in the extraplanar region $|z|>\zref$ is much larger than for the \mhd\ model, $\sim3$ times the flux at $\zref$.
The extraplanar momentum transfer from the \hot\ to the \twop\ phase is similarly larger than that of the \mhd\ model.
Most notably, the momentum transfer out of the CRs (dashed line) exceeds the \hot\ MHD flux transfer above $|z|=2\kpc$. Both transfer channels are necessary to explain the net gain in the MHD momentum flux of the \twop\ phase.
The rate of momentum transfer from the CRs to the gas is roughly constant as a function of $|z|$, leading to a roughly linear $\Delta_{\zref} \Pcr$ profile above $|z|=1\kpc$.
A similar roughly linear profile of  momentum flux transfer from CRs to \twop\ outflows is evident in Fig. 16 of \citet{2024ApJ...964...99A}.
In contrast, the transfer of momentum from \hot\ to \twop\ in the \crmhd\ model occurs mainly at $|z|=1-2\kpc$ (as seen in \autoref{fig:flux_tz}), with very little hot gas present beyond this (see \autoref{fig:mass_volume}). The result is an initial steep rise and then flat profile for $\Delta_{\zref} P_\mathrm{MHD}^\mathrm{hot}$ in \autoref{fig:dflux}.

Similar to the situation for the \mhd\ model, the majority of momentum flux transfer out of the \hot\ phase occurs in the disk region at $|z|\lesssim 0.5\kpc$.  The net momentum transfer to the \twop\ gas (heavy line) is very similar in the \crmhd\ and \mhd\ model within this disk region, and the $\Ptot$ profiles for \twop\ gas (thin lines) are also quite similar.
For the \crmhd\ model, the \hot\ MHD flux loss at $|z|\lesssim 1 \kpc$ slightly exceeds the gain by the \twop\ gas; the excess goes into energizing CRs
through adiabatic work, as shown in \autoref{fig:cr_coolheat}. This is also responsible for the increase in $\Pcr$ from $z=0$ to $\sim 1\kpc$, as seen in the momentum loading panel of \autoref{fig:loading_z}. The shape of $-\Delta_{\zref}\Pcr$ in  \autoref{fig:dflux}, with a local minimum at $|z|\sim 1\kpc$, is the mirror image of the momentum loading profile for CRs in \autoref{fig:loading_z}.

\subsection{Joint PDFs}

Galactic outflows are inherently multiphase, comprising gas at a wide range of flow velocities.  It is of great interest to analyze how the presence of CRs alters the thermo-kinetic distribution of the outflowing gas.
\autoref{fig:jointpdf} shows, for both the \mhd\ and \crmhd\ model, the time-averaged joint distributions of outflow velocity and sound speed weighted by the mass flux in the extraplanar region. At each reference height ($|z|=0.5$, 1, 2, and 3~kpc; both sides of the disk are summed up), we take a $\Delta z =100\pc$ thick slab to calculate the mean fluxes within the slab.
The integral of the distribution in each panel shows the total outward mass flux at a given $|z|$.
Both models exhibit characteristic shapes similar to those seen in previous works \citep[e.g.,][]{2020ApJ...903L..34K,2024ApJ...960..100S}. The narrow, nearly horizontal portion at around $c_s\sim 10\kms$ represents warm outflows that carry the majority of mass flux. The hot outflows, especially above $c_s>100\kms$, show correlated distributions in the outflow velocity and sound speed plane.

The mass flux distributions at $|z|=0.5\kpc$ are similar in \mhd\ and \crmhd, in both magnitude and shape. Both models lose outflowing mass flux at higher $|z|$ (as seen in \autoref{fig:flux_out_z}), but the \crmhd\ model has much greater reduction in the \hot\ phase.
The distribution of the \twop\ gas is similar in shape, but the magnitude is smaller in the \mhd\ model at large $|z|$.

To visualize the evolution in the distribution more quantitatively, \autoref{fig:voutpdf} presents the marginalized distributions for the \twop\ and \hot\ phases as a function of height. We integrate the joint distribution along the $\log c_s$ axis with the boundary between the two phases at $\log c_s = 1.18$. In the top row, the mass flux distributions of the \twop\ phase (blue) show a gradual shift of the most probable velocity (the peak of the distribution) toward a higher velocity at larger $|z|$, while the maximum velocity remains more or less the same. This shift corresponds to the increase in the vertical velocity shown in \autoref{fig:cr_velocity}, which can be due either to acceleration or to raining out of low velocity material as fountains. As shown in \citet{2020ApJ...894...12V}, a ballistic fountain model explains the overall shift in the distribution for the \mhd\ model, while still requiring some momentum transfer from \hot\ to \twop\ (via mixing), as seen here in \autoref{fig:dflux}.
In the \crmhd\ model, the distribution drops less overall, and has little change in the peak from 2 to 3 kpc
keeping the majority of the moderate velocity component $v_z>50\kms$ from 1 kpc. This is possible because in the \crmhd\ model the CR pressure gradient provides additional acceleration, as shown in \autoref{fig:dflux}.
Especially at $|z|<2\kpc$ the characteristic outflow velocities shown in \autoref{fig:cr_velocity} are quite similar between the two models, but the distribution reveals that the reasons differ.

For the \hot\ phase, the evolutionary trends are again quite distinct in the two models.
The mass flux distribution in the \mhd\ model decreases from 0.5 to 2 kpc, but the change from 2 to 3 kpc is less significant. In the \crmhd\ model, in contrast, the overall decrease in the distribution continues and is mainly driven by the highest velocity component. As can be inferred from \autoref{fig:jointpdf}, the highest velocity component is from \HIM, while the component near $\vout=100\kms$ is from \WHIM. The \HIM\ phase in the \crmhd\ model cannot survive at higher $|z|$, mainly because the increased filling factor of the \WNM\ phase enhances energy loss from the \HIM\ phase through mixing.

\section{Discussion}
\subsection{Comparison with Post-Processing Results}\label{sec:diss_pp}
The present work confirms the major conclusions reported on CR transport from our previous post-processing simulations  \citep{2021ApJ...922...11A,2022ApJ...929..170A,2024ApJ...964...99A,2025ApJ...989..140A,2025ApJ...994...45H}. In particular, we find a characteristic two-zone vertical profile for CR pressure in \twop\ gas, with a flat region in the midplane where diffusion dominates transport in the dense, neutral gas, and an exponential region at high altitude where
transport is dynamical in the low-density gas where scattering rates remain high.

Within the high-scattering region above $|z|\sim 1\kpc$, \autoref{fig:cr_velocity} shows that the effective CR vertical velocity $v_{\mathrm{c},z}$ is nearly equal to the dynamical CR transport speed $v_{\rm dyn,z}= v_z+v_{s,z}$. (Note that throughout this subsection, we omit angle brackets $\langle ...\rangle$ indicating horizontal averages, to simplify notation.) In steady state, the horizontally-averaged CR energy equation becomes $dF_{{\rm c},z}
/dz \approx d\rbrackets{v_{\rm dyn,z} 4\Pcr}/dz \approx  v_{\rm dyn,z}d \Pcr/dz$, such that the CR pressure profile is expected to follow $\Pcr\propto v_{\rm dyn,z}^{-4/3}$. Equivalently, $H_c=(3/4)H_a$, where $H_c\equiv|d\ln  \Pcr/dz|^{-1}$ is the CR scale height, and $H_a\equiv|d\ln v_{\rm dyn,z}/dz|^{-1}$ is the acceleration length scale.

The normalization of the relation is set at the interface where there is a transition from diffusion-dominated to dynamically-controlled CR transport. At the interface $|z|\sim z_t$, the vertical CR flux through both upper and lower sides is
\begin{equation}
\begin{split}
    F_{\rm out}(z_t) \equiv& F_{{\rm c},z}(z_t)-F_{{\rm c},z}(-z_t)\\
    =& 8v_{{\rm c},z}(z_t)\Pcr(z_t)\\
    \equiv& 8\veff \overline{P}_{\rm c}^{\twop}(0),
\end{split}
\end{equation}
where $\veff \approx v_\mathrm{dyn,z}(z_t)$ for GeV CRs and
$\Pcr(z_t)\approx \overline{P}_{\rm c}^{\twop}(z_t)\approx\overline{P}_{\rm c}^{\twop}(0)$.
This total flux should balance the \emph{net} CR energy injection rate obtained by integrating the RHS of \autoref{eq:CRenergy} and evaluating at $z=z_t$. For the latter, we can define a cumulated vertical profile integrating sources and sinks as
\begin{equation}
    F_{\rm in}\equiv F_{\rm in, SN} + F_{\rm in, W} - L_{\rm coll}-G_{\rm st},
\end{equation}
where the spatial distribution contributing to the individual terms is analyzed in \autoref{sec:source_sink}.  The  direct SN source term is defined as $F_{\rm in,SN}(z)\equiv(L_xL_y)^{-1}\int_{-z}^{z} \dot{e}_{\rm c,SN} dV$ using $\dot{e}_{\rm c,SN}$ (see \autoref{eq:cr_inj}), with the work gain, collisional loss, and streaming losses similarly defined as integrals of
$W_{\rm gas\rightarrow cr}$,  $\mathcal{L}_{\rm c}$, and $\mathcal{G}_{\rm st} $, using \autoref{eq:gas_work}, \autoref{eq:cr_cool}, and \autoref{eq:cr_st_heating}, respectively.

\begin{figure}
    \includegraphics[width=\linewidth]{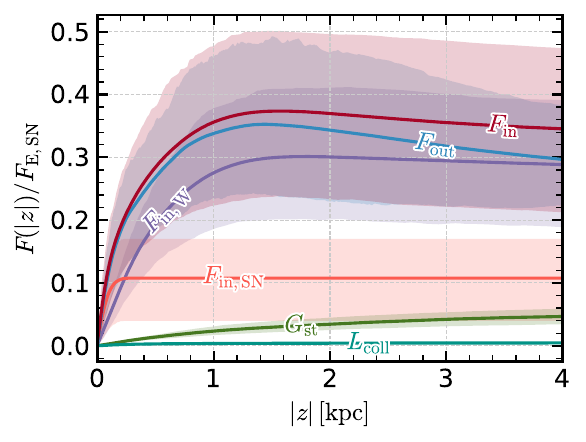}
    \caption{\label{fig:cumul_flux}
    Vertical profiles of cumulative CR energy source terms. In steady state, the CR flux through a height $\pm z$, $F_{\rm out}(|z|) \equiv F_{{\rm c},z}(z)-F_{{\rm c},z}(-z)$, matches the net CR  source term integrated within $[-z,z]$, $F_{\rm in} \equiv F_{\rm in, SN} + F_{\rm in ,W} - G_{\rm st} - L_{\rm coll}$. In the figure, all terms are normalized by the total SN energy injection rate per area $F_{\rm E, SN}\equiv \Esn\Ssfr/m_*$.
    }
\end{figure}

\autoref{fig:cumul_flux} shows the contribution from each term to the cumulative CR energy source profile,
normalized by the total SN energy injection rate per area $F_{\rm E,SN}\equiv E_{\rm SN}\Ssfr/m_*$.
As also seen in \autoref{fig:cr_coolheat}, the CR energy gain by adiabatic work exceeds direct injection from SNe, the streaming loss is about half of the direct SN injection, and the collisional loss term is negligible. Direct SN injection is fully accounted for within $|z|\sim 200\pc$, converging to $\fcr=0.1$. The work term increases up to $|z|\sim1.5\kpc$, while streaming losses continue to accumulate throughout the domain.
\REV{The slight mismatch between $F_{\rm out}$ and $F_{\rm in}$, with $F_{\rm in}>F_{\rm out}$, means that the total CR energy in the simulation domain is gradually increasing over time.  Over the time interval we considered for the time average, $\int\ecr dV$ grows about 50\%, but this growth does not seem to continue beyond $500$~Myr.}

The net source term contribution may be characterized by $\fcnet\equiv F_{\rm in}/F_{\rm E,SN}$, which is the sum of the direct injection $\fcr$ and the additional contribution from volumetric terms $(F_{\rm in,W}-L_{\rm coll}-G_\mathrm{st})/F_{\rm E,SN}$.  Since the collisional and streaming losses are small, $\fcnet\sim \fcr + f_\mathrm{W}$ where $ f_\mathrm{W}\equiv F_{\rm in,W}/F_{\rm E,SN}$.
From \autoref{fig:pressure_t}, the $\Pcr$ profile within \twop\ gas transitions from flat to exponentially declining between $z_t=500\pc$ and $1\kpc$, where we evaluate
$\fcnet=0.27$ and $0.34$, respectively.
By equating $F_{\rm in}(z_t) = F_{\rm out}(z_t)$, we have
\begin{align}\label{eq:veffz}
  \veff =& \frac{\fcnet E_{\rm SN}\Ssfr/m_*}
  {8 \overline{P}_{\rm c}^{\twop}(0)}\\
  =&38\kms\rbrackets{
  \frac{\fcnet}{0.3}}
  \rbrackets{\frac{\Ssfr}{4\times10^{-3}\sfrunit}}\nonumber\\
  &\rbrackets{\frac{\overline{P}_{\rm c}^{\twop}(0)}{10^4k_B\pcc\Kel}}^{-1}.
\end{align}
We use our fiducial parameters for $E_{\rm SN}=10^{51}\erg$ and $m_*=95\Msun$.
In the \crmhd\ model, we have
$\overline{P}_{\rm c}^{\twop}(0)/k_B=(1.14\pm0.14)\times10^4\pcc\Kel$ averaged over $|z|<50\pc$ and $\Ssfr=(3.6\pm1.0)\times10^{-3}\sfrunit$.
With $\fcnet\sim0.3$,
this implies an effective transport velocity $\veff\sim 30\kms$.  This value is indeed in good agreement with the dynamical value we find at $z_t\sim 0.5-1\kpc$, $v_{\mathrm{dyn},z}\sim 35\kms$.

Equivalently, in \citet{2025ApJ...994...45H} it was argued that  \autoref{eq:veffz} can be used to predict the total midplane CR pressure of the \twop\ gas if the SFR is known, and $\veff$ for GeV CRs may be estimated as $\sim v_{\mathrm{dyn},z}(z_t)$.
This prediction for CR pressure may be given via the CR feedback yield
\begin{align}
    \Upsilon_{\rm c} &\equiv \frac{\overline{P}_{\rm c}^\twop(0)}{\Ssfr}
    = \frac{\fcnet
    (E_\mathrm{SN}/m_*)}{8 \veff}\\
    &= 664\kms
    \rbrackets{\frac{\fcnet}{0.3}}
    \rbrackets{\frac{\veff}{30\kms}}^{-1},
\end{align}
when expressed in terms of $\veff$ and the total fraction of SN energy that goes into CRs.
With $\fcnet\sim0.3$
and $\veff\sim v_{\mathrm{dyn},z}(z_t)= 35\kms$,
$\Upsilon_{\rm c}\sim 570\kms$, slightly smaller than the directly-measured value $\Upsilon_{\rm c}\sim 665\kms$ listed in \autoref{tbl:prfm}.
We note that in \citet{2025ApJ...994...45H}, a smaller value $\Upsilon_\mathrm{c}\sim 400\kms$ was obtained from post-processing an MHD-only model similar to the present one. The main reasons for the difference are that in post-processing,
$\fcnet\sim \fcr$,
and in the absence of CR-pressure driven acceleration the value of the dynamical velocity was lower,
$v_{\mathrm{dyn},z}\sim 16\kms$.
In \citet{2025ApJ...994...45H}, several different \tigress\ MHD simulations
were post-processed with CR transport to obtain a fit
$\veff\propto\Ssfr^{0.20}$, similar to the fitted \twop\ outflow velocity  ($v_{\rm out}\propto\Ssfr^{0.23}$; see \citealt{2020ApJ...900...61K}), which leads to $\Upsilon_{\rm c}\propto\Ssfr^{-0.23}$.
Extending the current fully coupled dynamical simulation to other galactic conditions will be of great interest for exploring the environmental dependence of CR transport.

In \autoref{sec:sfr} we noted that the exponential scale height of the CR pressure in the region at $1\kpc<|z|<2\kpc$ is $H_c\sim 4\kpc$, significantly larger than the scale height of $H_c\sim 0.5\kpc$ found from post-processing simulations.  We can understand the reason for this difference based on the analytic model for vertical CR transport  presented in \citet{2025ApJ...989..140A}.  There, it was argued that the CR diffusivity, the acceleration scale of gas, and the velocity in the extraplanar region where CR-gas coupling becomes strong all combine in determining $H_c$ (see Eq. 20 in that paper).  In particular, there is a larger acceleration scale in the present simulations, which include the back-reaction of CR pressure gradients on the gas, than was present in the post-processing simulations, where gas acceleration was provided only by MHD stresses. A detailed analysis will be presented in a separate publication (F. Yu et al, in preparation), but the extended scale of momentum transfer in the \crmhd\ simulation compared to the \mhd\ simulation is evident in \autoref{fig:dflux}.

\subsection{Comparison with Other Simulations}\label{sec:diss_sim}
There have been several simulations utilizing a similar local, tall-box setup including CR feedback combined with a range of ISM physics and CR transport treatments.  Earlier simulations often adopt constant CR diffusion coefficients without CR streaming or explicit modeling of star formation \citep{2016ApJ...816L..19G, 2018MNRAS.479.3042G,2016ApJ...827L..29S,2023MNRAS.520.4621S}. \citet{2021MNRAS.504.1039R} use the SILCC framework to model the ISM physics (non-equilibrium chemistry but fixed UV radiation), star formation, and feedback, and include CR streaming but still adopt a constant CR diffusion coefficient (see also \citealt{2023MNRAS.522.1843R} for varying environments).

Overall, the common conclusion in the literature is that CR-mediated dynamics drives stronger, more mass-loaded, and cooler outflows. However, quantitative results are sensitive to the CR transport model and the diffusion coefficient \citep{2017MNRAS.467..906W,2021MNRAS.501.4184H}.
Some work has suggested that low diffusion coefficients $\kappa_{\rm CR}\le10^{28}\cm^{2}{\,\rm s}^{-1}$  can lead to suppression of the SFR
\citep[e.g.,][]{2019MNRAS.488.3716C,2020A&A...638A.123D,2021ApJ...910..126S}. In global galaxy simulations with a variable scattering rate \citep{2025A&A...698A.104T}, however, CRs have little effect on the SFR, similar to what our present \mhd\ vs. \crmhd\ models demonstrate.

Most recently, \citet{2025ApJ...987..204S} present a set of local simulations using the moving-mesh {\tt Arepo} code coupled with a two-moment Alfv\'en wave regulated CR transport scheme \citep{2019MNRAS.485.2977T,2021MNRAS.503.2242T} and the CRISP framework. They also calculate the CR scattering rate self-consistently considering both non-linear Landau and ion-neutral damping, but the actual formalisms differ quite a bit from the present work.
For a quantitative comparison, we focus on the CR-NL-IN model of \citet{2025ApJ...987..204S}, which includes both wave damping mechanisms and is thus the most similar to our \crmhd\ model.
Their simulation lasts $250\Myr$ and time-averaged values are reported for $112-250\Myr$; both intervals are shorter by a factor of 2 than ours. They found overall higher SFRs in both MHD and CRMHD models than ours (by a factor $\sim 2.5$)
and slightly higher SFRs ($\sim$50\%) in their MHD model than their CR-NL-IN model.

The mass loading factors (not separated by phase) are $\sim 0.1$ and $0.4$ at $z=1\kpc$ for their MHD and CR-NL-IN models, respectively, decreasing to $<0.01$ and $\sim 0.1$ at $z=3\kpc$.
Their higher (and more sustained with distance) mass loading with CRMHD is
similar to our finding, but our mass loading factor is $\sim 1$. The significant drop of the total mass loading factor in their MHD model is also quite different from what we find;
our \mhd\ model maintains $\eta_{M}\sim 0.5$ at $z>3\kpc$ when including both \twop\ and \hot\ outflows (see also \citealt{2020ApJ...900...61K}).

Energy loading factors (without ``boosting'' by the inclusion of galactic rotation) of the thermal gas show even more dramatic differences from our results: in their MHD model, $\eta_E$ decreases from $10^{-3}$ at $z=1\kpc$ to $10^{-5}$ at $z=3\kpc$, while their CR-NL-IN model maintains $\eta_E\sim10^{-4}$ at all heights. Including previous \tigress\ simulations \citep{2020ApJ...900...61K}, $\eta_E>0.01$ at all heights without CRs in solar neighborhood conditions (excluding galactic rotation), regardless of whether runaway SNe are included \citep{2018ApJ...853..173K}. Similar simulations with resolved multiphase outflows commonly show $\eta_E\ge 0.01$ (dominated by the hot wind) \citep{2020ApJ...890L..30L}. It is noteworthy that the low energy loading factors $\eta_E\le10^{-3}$ of \citet{2025ApJ...987..204S} are also seen in other Arepo simulations \citep{2020MNRAS.491.2088K}. A potential difference stems from the hydrodynamics schemes, i.e., Eulerian vs. (semi-)Lagrangian. Although high-resolution global dwarf galaxy simulations using SPH \citep{2019MNRAS.483.3363H,2024ApJ...960..100S} also obtain energy loading factors\footnote{The energy associated with galactic rotation is not removed in these global simulations.
But their mass loading factors are $\eta_M>1$ at $1\kpc$, much higher than that of \citet{2025ApJ...987..204S}, and the rotation velocity in dwarf galaxies is less than the $220\kms$ assumed in \citet{2025ApJ...987..204S}. Both imply that the genuine hot gas energy loading factors are higher in the dwarf simulations.}  higher than $0.01$, $\eta_E$ can drop significantly at low mass resolution ($\sim100\Msun$) from \citet{2019MNRAS.483.3363H}. Since \citet{2025ApJ...987..204S} adopted the target resolution of $10\Msun$ with an additional limit to the cell volume, lack of resolution is unlikely the reason for low $\eta_E$. Given that the ISM pressure near the midplane in \citet{2025ApJ...987..204S} is about an order of magnitude lower than our simulations ($\Pth\sim4\times10^{3}k_B\Kel\pcc$; see \autoref{fig:pressure_t}), there are evidently significant differences in cooling and heating processes, and these may also affect properties of outflows.

Similar CRMHD simulations employing a variable scattering rate model have been run for Milky-Way-like galaxies, either in isolation \citep{2025A&A...698A.104T,2025arXiv251016125T} or in cosmological zoom-ins \citep{2021MNRAS.501.4184H,2021MNRAS.501.3663H,2022MNRAS.517.5413H}.
In particular, \citet{2025A&A...698A.104T} explore the effect of CR feedback in a global, Milky Way like galaxy considering similar physical ingredients to ours; their ISM physics produces a multiphase ISM (albeit with very low thermal pressure, as seen in \citealt{2025ApJ...987..204S}) and they include two-moment CR transport with non-linear Landau and ion-neutral damping. From comparisons between their MHD and CRMHD simulations, similar to this paper, they find SFRs are barely changed ($\sim20\%$ lower in CRMHD). Outflows are warm-gas dominated, and the mass loading factor measured at $z=10\kpc$ is enhanced by a factor of $\sim 4$ with CRs (similar to the enhancement measured at $z>1\kpc$ in our simulations). The energy loading factor remains at a level of $\sim 0.01$ in both models. The CR energy loading factor at $z=10\kpc$ is $\sim 0.04$ which is similar to their injection rate of $0.05$, while we get $\sim 0.3$ at $z=1\kpc$, which is 3 times larger than the injection rate of $0.1$, mainly due to additional CR energy gain through adiabatic work (\autoref{fig:cumul_flux}).

\section{Conclusion}\label{sec:conclusion}

\subsection{Summary}

We present results from our newly implemented \tigresspp\ simulation framework for the star-forming ISM, including CR feedback.
This paper is our first demonstration of dynamically coupled CRMHD simulations.
These simulations consist of multiphase MHD with gravitational collapse creating star clusters that produce SNe and radiative heating \citep[as in \classic,][]{2017ApJ...846..133K}, now extended with CR injection and two-moment CR transport in which the scattering rate for source terms is computed based on the self-confinement paradigm \citep[as in][]{2021ApJ...922...11A}.

We compare our CRMHD simulation (\crmhd) with an MHD-only simulation (\mhd) having otherwise identical background conditions and physics.  The environment selected for this demonstration is representative of the solar neighborhood.
\REV{The current CRMHD simulation confirms the salient features from our previous post-processing simulations of CR transport in the multiphase ISM. 
At the same time, it reveals and quantifies mutual dynamical impacts of the CRs and the gas on each other.}

Our main findings are as follows:
\begin{itemize}
    \REV{
    \item The dynamically-coupled CRMHD simulation in this paper recovers our main
    previous findings regarding CR transport. For a realistic, multiphase ISM, CR transport 
    results in a two-zone CR system \citep{2025ApJ...989..140A,2025ApJ...994...45H}.
    In high-density, neutral gas near the midplane, Alfv\'en waves that could scatter CRs are strongly suppressed by ion-neutral damping, resulting in a diffusion-dominated CR transport region with
    $\kappa_\parallel\geq 10^{29}\cm^2{\rm \,s}^{-1}$ for GeV CRs (\autoref{fig:cr_kappa}). Strong diffusion leads to uniform CR pressure extending up to $|z|\sim 1\kpc$ for GeV CRs (\autoref{fig:pressure_t}). In gas that is low density and/or well ionized,
    the nonlinear Landau mechanism mediates less efficient wave damping, enabling stronger coupling between CRs and gas.
    In the extraplanar region at $|z|\gtrsim 1\kpc$, $\kappa_\parallel\leq 10^{28}\cm^2{\rm \,s}^{-1}$ for GeV CRs, and CR transport is dynamically controlled by advection and streaming, with characteristic velocity $\gtrsim 30\kms$ (\autoref{fig:cr_velocity}).}

    \REV{\item SFRs in the CRMHD simulation remain essentially unchanged compared to the MHD simulation (\autoref{fig:history}), with the slightly lower value at later times in \crmhd\ due to the reduction in the disk's gas surface density (given its higher mass loss rate). The values for the thermal, turbulent, and magnetic feedback yield (\autoref{tbl:prfm}) are also essentially unchanged by the presence of CRs. The nearly-uniform CR pressure within the
    main gas disk provides minimal vertical support against gravity (\autoref{fig:pressure_t}).  Thus, the presence of CRs does not quantitatively alter the level of star formation and feedback demanded to maintain the thermal, turbulent, and magnetic pressure that supports the ISM's weight.}

    \item The midplane CR pressure $\Pcr\sim 1.1\times10^4k_B\pcc\Kel$
    is similar to the other pressures in the gas (totaling $\Ptot\sim 1.9\times10^4k_B\pcc\Kel$; \autoref{fig:pressure_t}), in line with our finding of approximate equipartition based on post-processing in a wide range of galactic environments \citep{2025ApJ...994...45H,2026ApJ...996...99L}, and comparable to what is observed in the solar neighborhood.
    The extraplanar CR pressure follows an exponential profile with a scale height of $\sim 4\kpc$, much larger than that of the post-processing results, but similar to results in the CRMHD simulations of \citet{2024ApJ...964...99A}.
    This can be understood based on the increase in the acceleration scale for the gas in the region where CRs transition from diffusive to dynamical transport (see \autoref{sec:diss_pp}).  The larger scale for extraplanar acceleration owes, in turn, to the spatially extended transfer of CR momentum to the gas (\autoref{fig:dflux}).

    \item The extraplanar CR pressure gradient gradually accelerates warm-cold gas, leading to more persistent, volume-filling outflows (\autoref{fig:mass_volume}, \autoref{fig:flux_tz}), similar to those seen in the more idealized CRMHD simulations of \citet{2024ApJ...964...99A}.
    Inside of $|z|<2\kpc$, acceleration of \twop\ gas is achieved via momentum transfer from the hot phase, but at larger $|z|$ the momentum transfer from CRs to \twop\ gas is twice that from the \hot\ phase (\autoref{fig:dflux}).
    The mass loading factor (summed over both sides of the disk) in the \crmhd\ simulation is $\sim 1.5$ above $|z|=1\kpc$, which is $\sim 5$ times larger than in the \mhd\ simulation (\autoref{fig:loading_z}). However, the hot wind energy loading factor is significantly reduced in the \crmhd\ simulation, due to interaction with \twop\ gas.
    Nonetheless, the total energy loading factor for outflows in the \crmhd\ simulation is an order of magnitude larger than that in the \mhd\ simulation, with the vast majority carried in CRs.

    \item In addition to the CR-gas momentum exchange that drives warm gas out of the disk at a high rate, there are significant energy exchanges between gas and CRs. In the low-density outflowing \twop\ gas at  $|z|\gtrsim1\kpc$, streaming CRs heat the gas at a rate comparable to the heating by radiation (\autoref{fig:cr_coolheat}). CRs are deposited in SN feedback regions in our simulations at a level of 10\% of the SN energy, which takes place within $\sim 100\pc$ of the midplane. However, the flow of CRs down pressure gradients in the hot gas does substantial additional work on the CR fluid within $|z|\lesssim 2\kpc$. This work leads to $\sim 3$ times more energy gain in the CRs than what is directly injected by SNe.  A very interesting question for future investigation will be to characterize the properties of the regions of high CR pressure gradients between $200\pc\lesssim |z| \lesssim 1\kpc$, where the majority of the CR energy flux leaving the galaxy is accumulated (see \autoref{fig:cumul_flux}).
\end{itemize}

\subsection{Future Perspectives}
In this first paper, we have limited the scope to direct comparisons of key physical outcomes between the \mhd\ and \crmhd\ models for solar neighborhood conditions, considering CRs with $\sim$GeV energy. The fiducial \crmhd\ model we present here can also serve as a baseline for several other studies. In particular, one very interesting set of questions concerns the impact of simplifications that have often been made in other galactic CRMHD simulations:  adopting constant diffusion coefficients, assuming gas is fully ionized when setting the Alfv\'en velocity for streaming (rather than realistically computing the ionization state of different gas phases), and in some cases ignoring CR streaming altogether.  We have conducted a set of controlled simulations of CRMHD models with varying assumptions, testing the impacts of each of these simplifications; results of this investigation will be presented in upcoming work. This study also includes exploration of the effect of initial magnetic field strengths.

It is also important to extend to galactic conditions beyond the solar neighborhood.  In previous work, we explored differences in CR transport in a set of different \tigress\ simulations via a post-processing approach \citep{2022ApJ...929..170A,2025ApJ...994...45H}.  With our new \tigresspp\ framework, we will be able to investigate the back-reaction of CRs on gas dynamics, which is of particular interest for assessing potential changes in outflow loading factors relative to those driven solely by SN feedback, as quantified from a set of \classic\ simulations in \citet{2020ApJ...900...61K}.   For most direct comparison, these studies of varying galactic environments will include galactic differential rotation, as in our \classic\ simulation suite.

Most of the physics modules adopted for the present simulations are based on the \classic\ framework of \citet{2017ApJ...846..133K} and simply ported from \athena\ into \athenapp, but we have also introduced a few changes. One of these is considering sink particles as a combined star-forming entity, consisting of both new stars and a gas reservoir (rather than having 100\% star formation efficiency). Currently, the gas reservoir is dispersed ``manually'' with a prescribed mass return into a fixed volume.  The impact of the parameter choices for the present model will be explored in an upcoming publication.  In the future, we envision improving upon this (somewhat rudimentary) approach with a subgrid model for cloud disruption that captures the detailed physics of early feedback \citep[e.g.,][]{2021ApJ...911..128K,2025ApJ...989...42L,2025ApJ...989...43L} with greater veracity.

A more detailed view of the thermal and chemical state of the multiphase ISM, as well as greater realism in the physics of early feedback, can be achieved by combining the CRMHD model of the present work with the \ncr\ framework \citep{2023ApJ...946....3K}. In \ncr\ \citep{2023ApJS..264...10K}, both ionizing and non-ionizing UV radiation from star particle sources is followed via adaptive ray tracing, and coupled with a photochemistry module to set abundances of hydrogen species as well as carbon and oxygen species that are key coolants \citep[see also][]{2015MNRAS.454..238W,2021ApJ...920...44H,2023ApJ...952..140H,2022MNRAS.512..348K,2026arXiv260221790R}.  In addition to non-equilibrium cooling and heating of warm and cold atomic and molecular gas and nebular cooling of warm and hot ionized gas, \ncr\ provides for realistic spatial distribution of photoionized gas throughout the multiphase ISM, including in the fountain/outflow region when ionizing photons are able to escape from the midplane region.  Explicit CR transport will allow for realistic spatial and temporal variation in the photochemistry module, while at the same time the possibility of photoionized gas in the extraplanar region may potentially alter CR transport through its effect on the ion Alfv\'en speed and the ion-neutral damping rate.

Finally, much of our empirical understanding of CRs owes to phenomenological modeling of CR transport at varying energy, distinguishing between CR primaries and secondaries, and between species with long and short lifetimes \citep[e.g.][]{Grenier+15}.  Although the present work considers only CRs of $\sim$ GeV energy, the numerical methods we have implemented are straightforwardly applied to multiple energy groups of both hadrons and electrons \citep[see also][]{2020MNRAS.491..993G,2022MNRAS.516.3470H,2026arXiv260221147D}.  Initial studies of both protons and electrons in the $1-100$ GeV range in post-processing mode have been very informative about the relative roles of diffusive vs. dynamical transport vs. losses as a function of energy \citep{2025ApJ...988..214L,2025ApJ...989..140A}, as well as how to infer properties of CR populations and  magnetic fields from synchrotron emission \citep{2026ApJ...996...99L}. Extending these initial investigations into full CRMHD simulations using the \tigresspp\ framework will be of great interest.

\begin{acknowledgements}
We are grateful to the referee for their constructive comments, which helped improve the manuscript.
Support for this work was provided by grant 80NSSC22K0717 from NASA ATP to CGK, grant AST-2407119 from the NSF to LA and ECO, and grants 510940 and  10013948 from the Simons Foundation to ECO. LA was supported in part by the Program “Rita Levi Montalcini” of the Italian MUR.
SM acknowledges support from the EACOA Fellowship awarded by the East Asian Core Observatories Association.
J.-G.K acknowledges support from KIAS Individual Grant QP098701 at Korea Institute for Advanced Study.
Resources supporting this work were provided in part by the NASA High-End Computing (HEC) Program through the NASA Advanced Supercomputing (NAS) Division at Ames Research Center, and in part on computational systems managed and supported by Princeton University’s Research Computing.
This research has made use of NASA's Astrophysics Data System.
 \end{acknowledgements}

\software{{\tt Athena++} \citep{2020ApJS..249....4S},
{\tt astropy} \citep{astropy:2013,astropy:2018,astropy:2022},
{\tt scipy} \citep{2020SciPy-NMeth},
{\tt numpy} \citep{vanderWalt2011},
{\tt IPython} \citep{Perez2007},
{\tt matplotlib} \citep{Hunter:2007},
{\tt xarray} \citep{hoyer2017xarray},
{\tt pandas} \citep{mckinney-proc-scipy-2010},
{\tt CMasher} \citep{CMasher},
{\tt adstex} (\url{https://github.com/yymao/adstex}),
{\tt FFTW} \citep{FFTW05},
{\tt fftMPI} (\url{https://lammps.github.io/fftmpi})
}

\bibliography{ref}
\bibliographystyle{aasjournalv7}

\appendix

\section{Effects of the Vertical Boundary Conditions}\label{sec:app_bc}

In our previous simulations using the \tigress\ framework \citep[with methods presented in][]{2017ApJ...846..133K,2020ApJ...898...35K,2023ApJ...946....3K}, we used the {\tt diode} boundary conditions (BCs) for MHD variables (see below for explicit prescription) and found that the BCs do not affect the results. However, the same {\tt diode} BCs cannot be straightforwardly applied to CRMHD simulations.
The CR energy density profile near the boundaries is nearly flat, and the effective CR transport velocity $v_{{\rm c},z}=F_{{\rm c},z}/(4\Pcr)\sim 60\kms$ is smaller than the maximum wave speed in the extraplanar region, $\sim (\Pcr/\rho)^{1/2}\sim 100\kms$.
At the same time, warm outflows always occupy most of volume near boundaries, also with moderate or low velocities. When using simple {\tt diode} BCs, these conditions make the dynamical effects of reflected waves from boundaries non-negligible. We find empirically that CR energy density profiles and dynamics are strongly affected by the choice of BCs.
To  optimize the balance of computational cost and robustness, we therefore test how outcomes vary under different possible choices of BCs, comparing our standard domain size with simulations using larger vertical domains.

The present simulation, like other numerical models of stratified local disks, remains global in the vertical direction.
Generally, in such cases
one can employ an extrapolation from the active zones into the ghost zones at the $z$ boundaries. For example, the following are three possible choices for the variable $q$ in the $k$-th ghost zone of the lower and upper boundaries, with the last active zone index $\ks$ and $\ke$, respectively:
\begin{itemize}
    \item Zero gradient extrapolation:
    \begin{equation}
        q(\ks-k) = q(\ks), \quad q(\ke+k) = q(\ke)
    \end{equation}
    \item Constant logarithmic gradient extrapolation:
    \begin{eqnarray}
        q(\ks-k) &=& q(\ks)\sbrackets{\frac{q(\ks)}{q(\ks+1)}}^k, \\
        q(\ke+k) &=& q(\ke)\sbrackets{\frac{q(\ke)}{q(\ke-1)}}^k
    \end{eqnarray}
    \item Constant linear gradient extrapolation:
    \begin{eqnarray}
        q(\ks-k) &=& q(\ks)-k\sbrackets{q(\ks)-q(\ks+1)}, \\
        q(\ke+k) &=& q(\ke)+k\sbrackets{q(\ke)-q(\ke-1)}
    \end{eqnarray}
\end{itemize}
We refer these specifications as {\tt zero},  {\tt lingrad}, and {\tt lngrad}, respectively.

Additional constraints at the boundaries are sometimes desirable to ensure specific properties. For example, with a closed, stratified box like ours, we desire no inflows and density decreasing outward. To ensure the former property, a common approach is to prescribe an extrapolation (usually with zero gradient) for the normal velocity only when gas is outflowing
and set $v_z=0$ in the ghost zones otherwise.
Formally, this prescription (which we term {\tt outflow}) can be written
\begin{eqnarray}
    v_z(\ks-k) = v_z(\ks) {\rm max} \rbrackets{\frac{v_z(\ks)}{|v_z(\ks)|}\frac{z(\ks)}{|z(\ks)|},0},\\
    v_z(\ke+k) = v_z(\ke) {\rm max} \rbrackets{\frac{v_z(\ke)}{|v_z(\ke)|}\frac{z(\ke)}{|z(\ke)|},0}.
\end{eqnarray}
For the latter property, we introduce the limited logarithmic extrapolation (called {\tt lngrad-lim})
\begin{eqnarray}
    q(\ks-k) = q(\ks){\rm min}\rbrackets{\sbrackets{\frac{q(\ks)}{q(\ks+1)}}^k,1},\\
    q(\ke+k) = q(\ke){\rm min}\rbrackets{\sbrackets{\frac{q(\ke)}{q(\ke-1)}}^k,1}.
\end{eqnarray}

Having explicitly defined five possible extrapolation choices, which can be separately applied to each variable, \autoref{tbl:mhd_bcs} lists three sets of BCs we have tested.
In the first row, we give the name we adopt here to refer to a given set of choices: {\tt open}, {\tt diode}, or {\tt lngrad-outflow}.

In ZEUS-2D \citep{1992ApJS...80..753S}, the term {\tt outflow} was used for BCs that apply {\tt zero} extrapolation for all variables. This convention was subsequently used in several widely used public codes,
e.g., Athena and \athenapp\footnote{\url{https://github.com/PrincetonUniversity/athena}},
Ramses\footnote{\url{https://bitbucket.org/rteyssie/ramses}},
FLASH\footnote{\url{https://flash.rochester.edu/site/flashcode}},
and Pluto\footnote{\url{https://plutocode.ph.unito.it}}.
However, this choice is only guaranteed to produce ``outflowing'' results if the flows have supersonic outward velocities. For subsonic outflows, this choice can  cause the reflection of waves. If the velocity at the boundary is inward, this choice can cause inflows, acting as if there exists an infinite reservoir of matter. This choice of BCs is therefore often referred to alternatively as  {\tt open} or {\tt inflow/outflow} BCs.  We adopt the term {\tt open} for zero-gradient BCs, as listed as the header for the second column of the table.

For the reasons described above, additional numerical conditions help to achieve a desired physical state when flows may be subsonic or inward at a boundary.  Thus, in  \tigress\ we adopted the set of BCs listed under the heading {\tt diode} in the third column.

In the fourth column, under the heading {\tt lngrad-outflow}, we list a set of BCs  in which variables are extrapolated with a range of different gradients. The logarithmic extrapolation for density is exact when
combined with a linear gravitational potential (or constant gravity) and an isothermal equation of state. Therefore, for this choice, we adopt {\tt lngrad-lim} for $\rho$ with {\tt lingrad} for $\Phi$ and {\tt zero} for $T$ (i.e., isothermal for $P$).

The last three rows in \autoref{tbl:mhd_bcs} denote the sets of BCs for CR variables, which are motivated by the corresponding MHD BCs. Note that we apply extrapolation for the effective CR transport velocities, not the CR fluxes.

We caution that the names we adopt in \autoref{tbl:mhd_bcs}, and more generally  terms for sets of BCs, are not used consistently in the literature.  For example, in FLASH \citep{2000ApJS..131..273F}, the term {\tt diode} is consistent with our present usage, and the same is true in \citet{2015ApJ...809..187S,2017ApJ...851...93K}.  However, \citet{2017ApJ...838...99C} refer to the set we label as {\tt diode} as ``closed boundary conditions.''
\citet{2025MNRAS.539.1706V} use the term {\tt diode} to refer to a choice of BCs with {\tt zero} extrapolation for outflowing gas and {\tt reflective} BCs for inflowing gas.

\begin{figure*}
    \includegraphics[width=\linewidth]{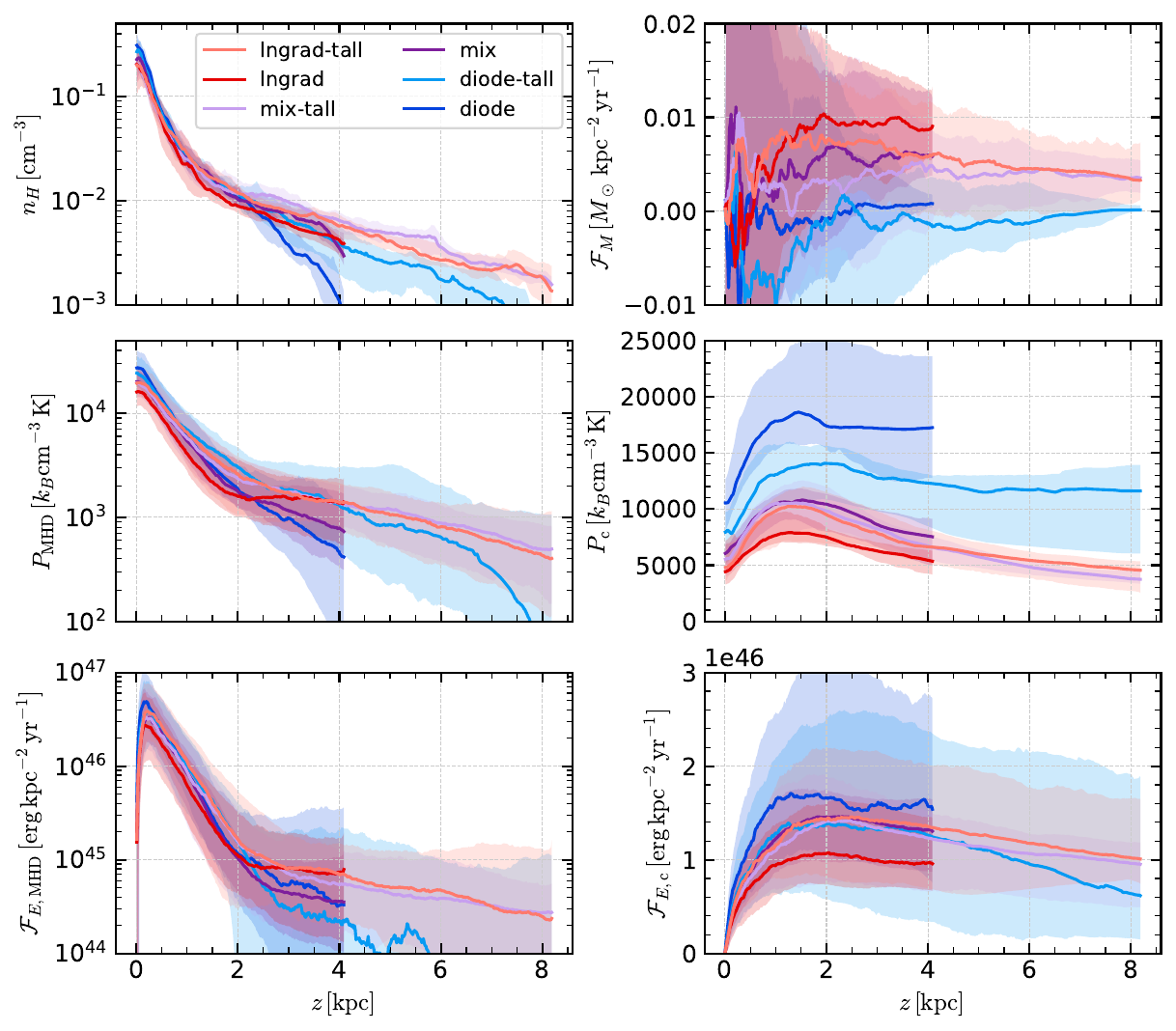}
    \caption{Time averaged vertical profiles of selected quantities for tests of BCs. From top to bottom, we show hydrogen number density and mass flux, vertical stress of the thermal gas and CR pressure, and vertical energy flux of the thermal gas and CR flux. Note that the panels in the right column use linear scales while those in the left column use log scales. The solid lines denote medians while the shaded areas depict the 16th to 84th percentiles over $t=200-500\Myr$.\label{fig:bc_test}}
\end{figure*}

We have run several \crmhd\ simulations using standard ($z\pm4096\pc$) and taller ($z\pm8192\pc$) boxes at low resolution ($\Delta x=16\pc$), to test how results depend on the choice of BCs.  We consider three combinations of BCs:
\begin{enumerate}
    \item {\bf mix}: {\tt diode} for MHD and {\tt lngrad-outflow} for CR
    \item {\bf diode}: {\tt diode} for both MHD and CR variables
    \item {\bf lngrad}: {\tt lngrad-outflow} for both MHD and CR variables.
\end{enumerate}
Our fiducial choice is {\bf mix}.
In \autoref{fig:bc_test}, we show time averaged vertical profiles of selected quantities, with model names in the key based on the choice of BCs and the box size.

The taller box runs show overall more consistent results across different BCs.
However, the {\bf diode} BCs produce a steep (and presumably unphysical) gradient in $\Ptot$ near the outer boundary in the tall box.
Also, {\bf diode} produces the largest inconsistency in CR pressure profile when the box size is changed.  Physically, this is because the CR pressure is not naturally decreasing outward as steeply as the MHD variables, so CRs accumulate near the vertical boundaries when the zero gradient is imposed.
The flat CR pressure in {\bf diode} reduces the impact of CRs, resulting in lower mass fluxes at high $|z|$.
Both {\bf mix} and {\bf lngrad} are
equally
valid choices in terms of consistency of most variables between the resulting vertical profiles in the standard boxes and the taller boxes.
However, {\bf mix} produces better agreement across box sizes for CR energy flux and MHD mass flux.  Also, the MHD portion of the  BCs in {\bf mix} are the same as chosen for our previous \tigress\ simulations. For these reasons, we have adopted the choice {\bf mix} for the \crmhd\ simulation presented in the main text.

\begin{table}[]
    \centering
    \begin{tabular}{cccc}
        variable      & {\tt open}    & {\tt diode}  & {\tt lngrad-outflow} \\
        \hline
        $\rho$        & zero    & zero     & lngrad-lim    \\
        $v_x$, $v_y$  & zero    & zero     & zero         \\
        $v_z$         & zero    & outflow  & outflow   \\
        $P$           & zero    & zero     & isothermal   \\
        $r$           & zero    & zero     & zero       \\
        $B$           & zero    & zero     & zero       \\
        $\Phi$        & zero    & zero     & lingrad      \\
        \hline
        $\ecr$         & zero    & zero     & lngrad-lim    \\
        $F_{{\rm c},x}/\Pcr$, $F_{{\rm c},y}/\Pcr$
                      & zero    & zero     & zero          \\
        $F_{{\rm c},z}/\Pcr$ & zero    & outflow  & outflow       \\
    \end{tabular}
    \caption{Boundary condition prescriptions tested here.  See text for details.}
    \label{tbl:mhd_bcs}
\end{table}

\end{document}